\documentclass[12pt,a4paper,final]{iopart}

\usepackage{iopams}  
\usepackage{graphicx}
\usepackage[breaklinks=true,colorlinks=true,linkcolor=blue,urlcolor=blue,citecolor=blue]{hyperref}
\usepackage{braket}
\usepackage{xcolor}
\usepackage{sidecap}
\usepackage{mathrsfs}
\usepackage{verbatim}

\begin{document}

\title[]{The uniform electron gas at high temperatures: \emph{ab initio} path integral Monte Carlo simulations and analytical theory}

\author{Tobias Dornheim}
\address{Center for Advanced Systems Understanding (CASUS), 03581 G\"orlitz, Germany}
\address{Helmholtz-Zentrum Dresden-Rossendorf (HZDR), 01328 Dresden, Germany}
\ead{t.dornheim@hzdr.de}

\author{Jan Vorberger}
\address{Helmholtz-Zentrum Dresden-Rossendorf (HZDR), 01328 Dresden, Germany}

\author{Zhandos Moldabekov}
\address{Center for Advanced Systems Understanding (CASUS), 03581 G\"orlitz, Germany}
\address{Helmholtz-Zentrum Dresden-Rossendorf (HZDR), 01328 Dresden, Germany}

\author{Gerd R\"opke}
\address{Institut f\"ur Physik, Universit\"at Rostock, 18041 Rostock, Germany}

\author{Wolf-Dietrich Kraeft}
\address{Institut f\"ur Physik, Universit\"at Rostock, 18041 Rostock, Germany}

\begin{abstract}
We present extensive new \emph{ab initio} path integral Monte Carlo (PIMC) simulations of the uniform electron gas (UEG) in the high-temperature regime, $8\leq\theta=k_\textnormal{B}T/E_\textnormal{F}\leq128$. This allows us to study the convergence of different properties towards the classical limit. In particular, we investigate the classical relation between the static structure factor $S(\mathbf{q})$ and the static local field correction $G(\mathbf{q})$, which is only fulfilled at low densities. Moreover, we compare our new results for the interaction energy to the parametrization of the UEG by Groth \emph{et al.}~[PRL \textbf{119}, 135001 (2017)], which interpolates between PIMC results for $\theta\leq8$ and the Debye-H\"uckel limit, and to higher order analytical virial expansions. Finally, we consider the momentum distribution function $n(\mathbf{q})$ and find an interaction-induced increase in the occupation of the zero-momentum state even for $\theta\gtrsim32$. All PIMC data are freely available online, and can be used as input for improved parametrizations and as a rigorous benchmark for approximate methods.
\end{abstract}

\vspace{2pc}

\section{Introduction\label{sec:introduction}}

The uniform electron gas~\cite{loos,review,quantum_theory} (UEG, also known as \emph{jellium}, or quantum one-component plasma) is one of the most fundamental model systems in a number of research fields such as statistical physics and quantum chemistry. Having originally been introduced as a simplified model for the conduction electrons in metals~\cite{mahan2012many}, the UEG has given new insights into a number of interesting effects such as superconductivity~\cite{BCS2,BCS}, Wigner crystallization of electrons~\cite{Drummond_Wigner_2004,PhysRevB.100.235116}, and the theory of collective excitations~\cite{pines,bonitz_book}.

Despite its apparent simplicity, the accurate description of the UEG had remained a most formidable challenge for decades, and facilitated a number of methodological innovations such as dielectric theories~\cite{stls_original,vs_original,Iyetomi_PRB_1981,stls,stls2,stolzmann,schweng,dynamic_ii,arora,tanaka_hnc,Tolias_JCP_2021,castello2021classical} and quantum-to-classical mappings~\cite{perrot,Dutta_PRE_2013,Sandipan_PRE_2013,doi:10.1063/1.4865935}. The first rigorous results at $T=0$ have been presented in the seminal work by Ceperley and Alder~\cite{Ceperley_Alder_PRL_1980} on the basis of numerical quantum Monte Carlo (QMC) simulations. These data have subsequently been used as input for parametrizations of the exchange--correlation (XC) energy of the UEG~\cite{Perdew_Zunger_PRB_1981,vwn,Perdew_Wang_PRB_1992}, which has been pivotal for the potentially unrivaled success of density functional theory (DFT) regarding the description of real materials~\cite{PBE_1996,Burke_Perspective_JCP_2012,Jones_RevModPhys_2015}. In the mean time, different authors have used ground-state QMC methods to study and parametrize a number of additional properties of the UEG, such as the momentum distribution $n(\mathbf{q})$~\cite{Ortiz_Ballone_PRB_1994,Holzmann_PRL_2011}, the static density response function $\chi(\mathbf{q})$ and local field correction $G(\mathbf{q})$~\cite{moroni,moroni2,bowen2,cdop,Chen2019}, and the pair correlation function $g(r)$~\cite{Perdew_Wang_PDF_1992,Ortiz_Ballone_PRB_1994,Spink_PRB_2013}.

In addition, the last decades have witnessed a remarkable surge of interest in the study of matter at extreme conditions~\cite{fortov_review}. Of particular importance is the so-called warm-dense matter (WDM) regime~\cite{wdm_book,new_POP}, which is defined by two characteristic parameters that are of the order of unity at the same time~\cite{Ott2018}: a) the density parameter (also known as Wigner-Seitz radius in the literature) $r_s=\overline{r}/a_\textnormal{B}$ (with $\overline{r}$ and $a_\textnormal{B}$ being the average interparticle distance and first Bohr radius) and b) the degeneracy temperature $\theta=k_\textnormal{B}T/E_\textnormal{F}$ (with $E_\textnormal{F}$ being the usual Fermi energy). Often the coupling parameter $\Gamma=e^2/r_s\langle K \rangle$ is used in addition to characterize the deviation from the ideal gas ($\langle K \rangle$ is the mean kinetic energy, $e$ is the elementary charge). These conditions are ubiquitous throughout our universe, and naturally occur in astrophysical objects such as giant planet interiors~\cite{Militzer_2008,saumon1}. Moreover, WDM is crucial for technological applications such as the discovery of novel materials~\cite{Kraus2016,Kraus2017}, and is predicted to occur on the compression pathway of a fuel capsule towards inertial confinement fusion~\cite{hu_ICF}.

Evidently, previous ground-state results for the UEG are insufficient for an accurate theoretical description of these extreme states, which has sparked a series of new developments in the field of fermionic QMC simulations of the UEG in the WDM regime~\cite{Brown_PRL_2013,PhysRevE.91.033108,Malone_PRL_2016,Schoof_PRL_2015,Dornheim_NJP_2015,dornheim_jcp,dornheim_prl,Lee_JCP_2021,Yilmaz_JCP_2020,dornheim_POP,Filinov_CPP_2021}. These efforts have culminated in the first accurate parametrizations of the XC free energy $f_\textnormal{xc}(r_s,\theta,\xi_\sigma)$ [with $\xi_\sigma=(N^\uparrow-N^\downarrow)/N$ being the spin-polarization] of the UEG~\cite{review,groth_prl,ksdt,status}. This makes it possible to perform explicitly thermal DFT simulations~\cite{Mermin_DFT_1965} of WDM on the level of the local density approximation~\cite{kushal,karasiev_importance}. This has been followed by the rigorous investigation of a number of other properties of warm dense electrons, including the static density response~\cite{dornheim_pre,groth_jcp,dornheim_ML,dornheim_HEDP,dornheim_electron_liquid,Tolias_JCP_2021,Dornheim_PRL_2020_ESA,Dornheim_PRB_2021}, the momentum distribution function~\cite{Hunger_PRE_2021,Dornheim_PRB_nk_2021,Dornheim_PRE_2021}, and the static structure factor~\cite{dornheim_cpp,dornheim_electron_liquid,dornheim_prl,Lee_JCP_2021}.
Moreover, it has even been possible to use path integral Monte Carlo (PIMC)~\cite{cep,Berne_JCP_1982,Berne_JCP_1983} results for the imaginary-time version of the intermediate scattering function $F(\mathbf{q},\tau)$ [cf.~Eq.~(\ref{eq:F}) below] as a starting point for an analytic continuation~\cite{Jarrell_Gubernatis_PhysRep_1996} to the dynamic structure factor $S(\mathbf{q},\Omega)$~\cite{dornheim_dynamic,dynamic_folgepaper,Dornheim_Vorberger_finite_size_2020}---the key property in X-ray Thomson scattering experiments~\cite{siegfried_review,kraus_xrts}---and related quantities~\cite{Hamann_PRB_2020,Hamann_CPP_2020}.
Finally, we mention recent advances in the description of nonlinear effects in the warm dense electron gas~\cite{Dornheim_PRL_2020,Dornheim_PRR_2021,Dornheim_CPP_2021,Dornheim_JPSJ_2021,Dornheim_JCP_ITCF_2021}, which cannot be neglected in many situations of experimental relevance, and might give rise to an improved way of diagnostics of WDM experiments~\cite{moldabekov2021thermal}.

Yet, these remarkable achievements have been limited to $\theta\leq8$ in the case of $f_\textnormal{xc}$, and $\theta\leq4$ for the static density response and the momentum distribution function. We stress that these temperature parameters are not so high that quantum degeneracy effects can be completely neglected~\cite{review}. Therefore, there remains a gap in our understanding of the UEG between the WDM regime ($\theta\sim1$) and the classical limit ($\theta\gg1$). 
This regime of moderate and  weak degeneracy is highly relevant for astrophysics as well as inertial confinement fusion.
For example, the transition from strongly degenerate quantum regime to so called classical plasmas takes place on the way from the core of white dwarfs to their atmosphere \cite{Koester_1990, doi:10.1063/1.5097885}. A further example of a moderately to weakly degenerate plasma is given by the outermost layer of neutron stars~\cite{Haensel}, which are often being referred to as their atmosphere. In particular, understanding the properties of the plasma of the atmospheres of neutron stars is important for the interpretation of X-ray observations from pulsars \cite{AJ2021, AJ2_2021}. Helioseismology and asteroseismology need reliable equation of state (EOS) data at low densities in  order to explain the multitude of solar oscillation modes that are observed~\cite{Daeppen1988,aerts2010,hecker2018}.
From a practical point of view, the relevance of this study is associated with the generation and use of hot dense plasmas at the National Ignition Facility \cite{Hu2010, NIF_2020} and at the “Z” pulsed power facility at Sandia National Laboratories~\cite{Z_Sandia}.

For the warm dense matter region, analytical results for the EOS are scarcely available. Numerical simulations should meet exact analytical results where they are known to be accurate. 
Concerning the linear direct term of the EOS, i.e., the term of the order $\xi$, we find the term $\xi/6$ in
Eq. (2.53) of Ref.~\cite{kraeft1986quantum} and elsewhere. In contrast, the mean value of the Coulomb potential does not deliver a linear direct term (like $\xi/6$)~\cite{WDK2005,Kraeft_2015}. In Sec. \ref{sec:xi/6} of the
present paper, the nonexistence of the linear direct term is confirmed by comparison to PIMC data.
However, the exchange contribution to the EOS contains a linear term, Eq. (2.54) of Ref.~\cite{kraeft1986quantum}. Efforts to get results beyond the second virial coefficient are reported by DeWitt {\it et al.}~\cite{RIEMANN1995,riemann_1997,DeWitt1998,Dewitt1998c}.
In contrast, Brown $\&$ Yaffe \cite{BY2001} and Alastuey $\&$ Perez 
\cite{AP1996} achieve $n^{5/2}$ expressions under the assumption of the existence of the linear direct term in the EOS, $\xi/6$, and thus end up with wrong results.

In the present work, we remedy this unsatisfactory situation by presenting extensive new \emph{ab initio} PIMC simulations of the UEG in the high-temperature regime, $8\leq\theta\leq128$. This allows us unambiguously answer a number of open questions, such as the validity of the classical relation between the static structure factor and the static local field correction, the range of validity of low-degeneracy density and fugacity expansions for the EOS~\cite{RIEMANN1995,riemann_1997,Kraeft_2015}, the absence of the direct $\xi/6$ term, and the reliability of the interpolation of the parametrization of $f_\textnormal{xc}$ by Groth \emph{et al.}~\cite{groth_prl} between PIMC data for $\theta\leq8$ and the Debye-H\"uckel (DH) limit at high temperature. All our PIMC results are freely available online~\cite{repo} and can be used as input for improved models~\cite{status}, or to benchmark new approximations.

The paper is organized as follows: In Sec.~\ref{sec:theory}, we introduce the relevant theoretical background, including the idea behind the PIMC method (Sec.~\ref{sec:pimc}), how to overcome finite-size effects in the interaction energy (Sec.~\ref{sec:FSC}), and some basic relations of linear response theory (Sec.~\ref{sec:LRT}). In Section~\ref{analy}, we present the appropriate virial expansion of the equation of state of the electron gas in the high temperature case. Sec.~\ref{sec:results} is devoted to the presentation of our new simulation results, starting with a brief discussion of the fermion sign problem~\cite{dornheim_sign_problem} in Sec.~\ref{sec:FSP}. In Sec.~\ref{sec:static}, we analyse the static structure factor and static linear response function, with a particular emphasis on the validity of the classical relation between these two properties. Sec.~\ref{sec:potential} contains the discussion of the interaction energy of the UEG, and we compare our new finite-size corrected PIMC results to different models. Then follows the discussion of the direct linear term in the virial expansion in Sec.~\ref{sec:xi/6}. Lastly, we consider the momentum distribution function in Sec.~\ref{sec:momentum}.
The paper is concluded by a brief summary and outlook in Sec.~\ref{sec:summary}.

\section{Theory\label{sec:theory}}

We consider the standard UEG model as it has been described in detail e.g. in Refs.~\cite{review,loos}. In particular, we take into account the interaction of the $N$ electrons in the main simulation cell with the infinite array of periodic images via the Ewald summation technique, see the paper by Fraser \emph{et al.}~\cite{Fraser_PRB_1996} for an extensive yet accessible discussion. Furthermore, we restrict ourselves to the fully spin-unpolarized (paramagnetic) case, i.e., $N^\uparrow=N^\downarrow=N/2$.

Note that we assume Hartree atomic units throughout the remainder of this work.

\subsection{Path integral Monte Carlo\label{sec:pimc}}

The PIMC method~\cite{cep} allows for the exact solution of the full quantum many-body problem in the canonical ensemble, where the volume $\Omega=L^3$, the inverse temperature $\beta=1/T$, and the number density $n=N/\Omega$ are being kept constant.
As a starting point, we evaluate the corresponding partition function in coordinate space,
\begin{eqnarray}\label{eq:Z}
Z_{\beta,N,\Omega} = \frac{1}{N^\uparrow! N^\downarrow!} \sum_{\sigma^\uparrow\in S_{N^\uparrow}} \sum_{\sigma^\downarrow\in S_{N^\downarrow}} \textnormal{sgn}(\sigma^\uparrow,\sigma^\downarrow)\int d\mathbf{R} \bra{\mathbf{R}} e^{-\beta\hat H} \ket{\hat{\pi}_{\sigma^\uparrow}\hat{\pi}_{\sigma^\downarrow}\mathbf{R}}\ ,
\end{eqnarray}
where the two sums are carried out over all elements $\sigma^i$ ($i\in\{\uparrow,\downarrow\}$) of the permutation groups $S_{N^i}$, which is required for proper antisymmetrization to take into account the fermionic nature of the electrons. Further, the operators $\hat{\pi}_{\sigma^\uparrow}$ and $\hat{\pi}_{\sigma^\downarrow}$ realize the respective permutation of the $\sigma^i$, and the vector $\mathbf{R}=(\mathbf{r}_1,\dots,\mathbf{r}_N)^T$ contains the coordinates of both spin-up and spin-down electrons.
Unfortunately, the straightforward evaluation of Eq.~(\ref{eq:Z}) is precluded by the fact that the kinetic ($\hat{K}$) and potential ($\hat{V}$) contributions to the full Hamiltonian $\hat{H}$ do not commute,
\begin{eqnarray}\label{eq:com}
e^{-\beta\hat{H}} \neq e^{-\beta\hat{K}} e^{-\beta\hat{V}}\ .
\end{eqnarray}
To overcome this obstacle, we employ the exact semi-group property of the density operator
\begin{eqnarray}\label{eq:group}
e^{-\beta\hat{H}} = \prod_{\alpha=0}^{P-1} e^{-\epsilon\hat{H}}\ ,
\end{eqnarray}
with $P\in\mathscr{N}$ and the definition $\epsilon=\beta/P$. 
In this way, we have transformed the evaluation of the matrix element of the density operator $\hat{\rho}=e^{-\beta\hat{H}}$ into the product of $P$ matrix elements, but at $P$ times the original temperature. Specifically, the commutator error in the primitive factorization Eq.~(\ref{eq:com}) vanishes as $P^{-2}$
when the number of factors is increased, and eventually becomes negligible in practice.
The resulting expression for the partition function in the path integral picture is given in the literature~\cite{cep,Dornheim_permutation_cycles,review}, and need not be repeated here.

\begin{figure}\centering
\hspace*{-0.016\textwidth}\includegraphics[width=0.5\textwidth]{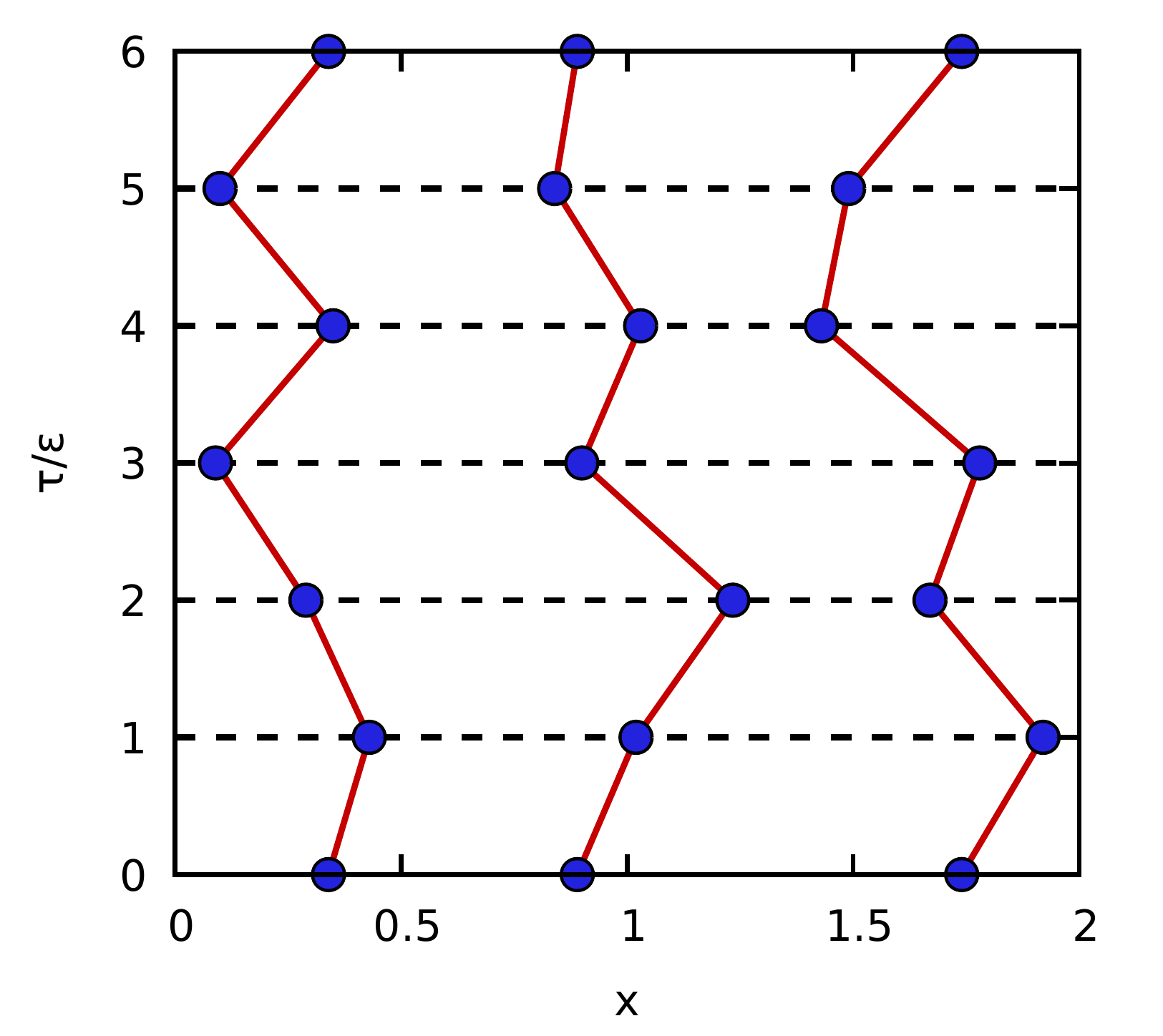}\includegraphics[width=0.5\textwidth]{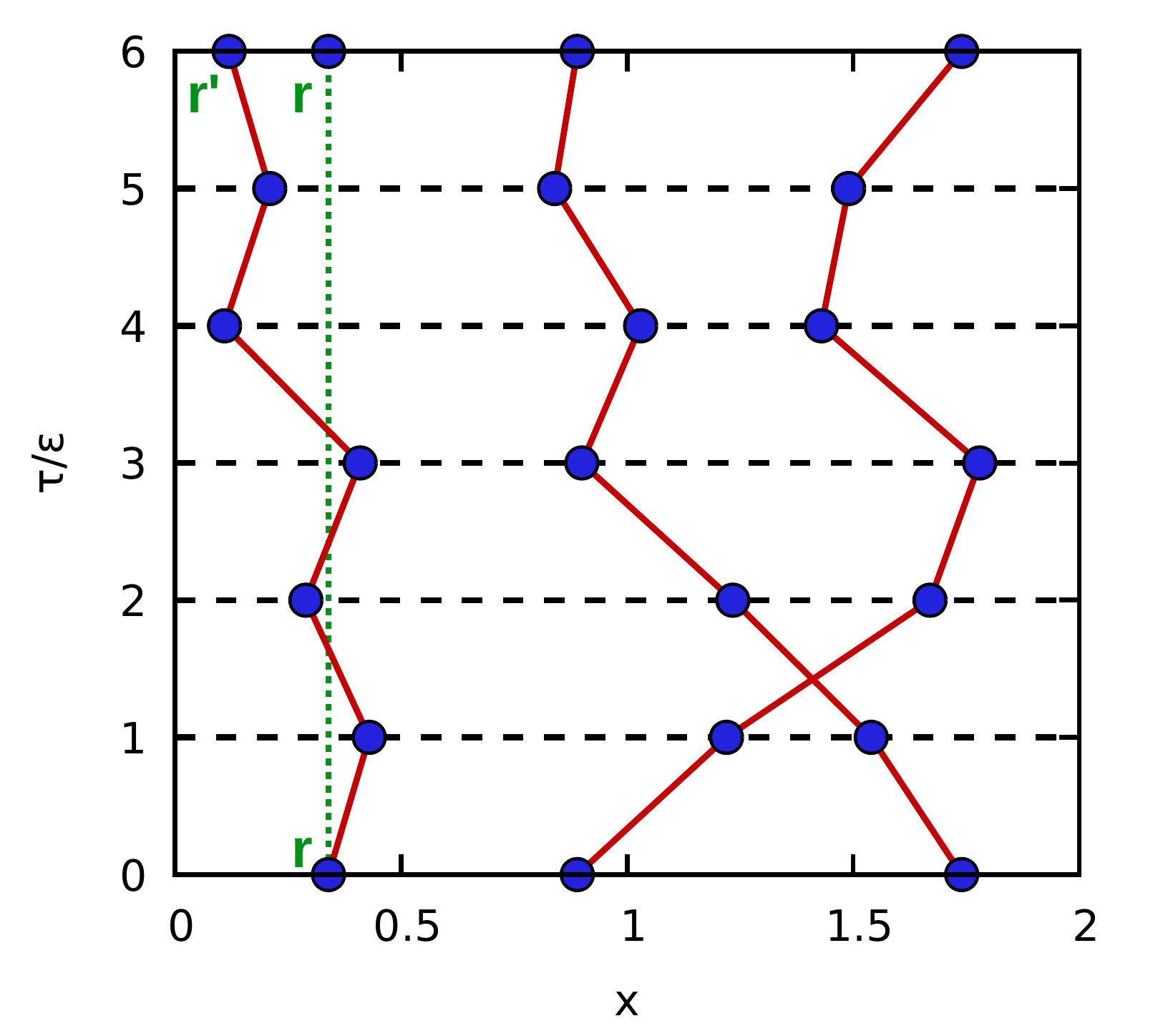}
\caption{\label{fig:PIMC}
Illustration of the path integral Monte Carlo configuration space. Left: Configuration of $N=3$ particles in the $x$-$\tau$-plain. Right: \emph{Open} configuration contributing to the estimation of the momentum distribution, cf.~Eq.~(\ref{eq:nk_formula}). We note the presence of a permutation cycle of the two particles on the right, leading to a negative configuration weight, $W(\mathbf{X})<0$ in this case. Taken from Ref.~\cite{Dornheim_PRB_nk_2021}.
}
\end{figure}

A graphical illustration of the path integral formalism is given in the left panel of Fig.~\ref{fig:PIMC}, where we show a configuration of $N=3$ particles in the $x$-$\tau$-plain. Specifically, $\tau\in[0,\beta]$ denotes the \emph{imaginary time}, which has been discretized by using $P=6$ high-temperature factors in this example. Further, each particle is represented by an entire path of coordinates, and the coordinates on the first and last slice are equal due to diagonal nature of the trace in Eq.~(\ref{eq:Z}), which can be rewritten as
\begin{eqnarray}\label{eq:PIMC_Z}
Z=\int\textnormal{d}\mathbf{X}\ W(\mathbf{X})\ .
\end{eqnarray}
Here $\mathbf{X}$ denotes a so-called path configuration, which contributes to the total partition function with the weight $W(\mathbf{X})$.
The basic idea of the PIMC method is then to use the Metropolis~\cite{metropolis} Monte Carlo algorithm to generate a Markov chain of configurations $\{\mathbf{X}_i\}$ that are distributed according to the probability
\begin{eqnarray}\label{eq:prob}
P(\mathbf{X}) = \frac{W(\mathbf{X})}{Z}\ .
\end{eqnarray}

An additional difficulty arises from the fact that Eq.~(\ref{eq:Z}) contains the sum over all possible permutations of particle coordinates of electrons of the same spin-orientation. In particular, each additional pair exchange leads to a factor of minus one in $Z$, which means that the configurations weights $W(\mathbf{X})$ can be negative. Such a case is illustrated in the right panel of Fig.~\ref{fig:PIMC}, where the two particles on the right form an \emph{exchange-cycle}~\cite{Dornheim_permutation_cycles}, i.e., a single trajectory with two particles in it. Consequently, it is $W(\mathbf{X})<0$, which, in turn, precludes the interpretation of Eq.~(\ref{eq:prob}) as a proper probability. To overcome this obstacle, we generate a Markov chain of configurations that are distributed according to
\begin{eqnarray}\label{eq:modified}
P'(\mathbf{X}) = \frac{1}{Z'} |W(\mathbf{X})| \ ,
\end{eqnarray}
with the modified normalization
\begin{eqnarray}\label{eq:Z_bose}
Z' = \int \textnormal{d}\mathbf{X}\ |W(\mathbf{X})| \ .
\end{eqnarray}
The exact fermionic expectation value of interest is then given by the ratio
\begin{eqnarray}\label{eq:ratio}
\braket{\hat A} = \frac{\braket{\hat A \hat S}'}{\braket{\hat S}'}\ ,
\end{eqnarray}
where $\braket{\dots}'$ indicates that the expectation value is computed with respect to Eq.~(\ref{eq:modified}). The denominator of Eq.~(\ref{eq:ratio}) is the so-called \emph{average sign}, which is defined as
\begin{eqnarray}
S = \braket{\hat S}' = \left< \frac{W(\mathbf{X})}{|W(\mathbf{X})|} \right>'
\end{eqnarray}
and constitutes a straightforward measure for the amount of cancellation of positive and negative terms in the fermionic PIMC simulation.
We note that, in first order, the statistical uncertainty of an observable $\hat{A}$ scales as~\cite{binder}, 
\begin{eqnarray}\label{eq:sqrt}
\frac{\Delta A}{A}\sim \frac{1}{S\sqrt{N_\textnormal{MC}}}\ .
\end{eqnarray}
The exponential decay of $S$ with important system parameters like $N$ or $\beta$ thus directly leads to an exponential increase in the Monte Carlo error bars and, consequently, to a vanishing signal-to-noise ratio. This is the origin of the notorious fermion sign problem, which severely limits the application of the PIMC method to the simulation of fermions~\cite{dornheim_sign_problem,Dornheim_JPA_2021}. Yet, the sign problem is relatively mild at the comparably high temperatures that we consider in the present work, and the required computation time is increased by a factor of ten in the most challenging case, which does not constitute a problem.

Let us conclude this section by considering the PIMC estimation of the momentum distribution $n(\mathbf{q})$, which, for a single spin-species $\sigma\in\{\uparrow,\downarrow\}$, is defined as~\cite{Militzer_momentum_HEDP_2019,cep,Dornheim_PRB_nk_2021,Dornheim_PRE_2021}
\begin{eqnarray}\label{eq:momentum_distribution}
n_\sigma(\mathbf{q}) = \frac{(2\pi)^d}{L^3} \left<\sum_{l=1}^{N_\sigma}\delta\left({\mathbf{\hat{q}_l}}-\mathbf{q}\right) \right>\ ,
\end{eqnarray}
with the normalization
\begin{eqnarray}\label{eq:normalization}
\sum_\mathbf{q}n_\sigma(\mathbf{q}) = N_\sigma\ .
\end{eqnarray}
Since the PIMC method is formulated in coordinate space, Eq.~(\ref{eq:momentum_distribution}) constitutes an off-diagonal property and its estimation requires a modification of the configuration space.
Specifically, the PIMC estimator for $n(\mathbf{q})$ is given by
\begin{eqnarray}\label{eq:nk_formula}
n_\sigma(\mathbf{q}) = \frac{1}{L^3} \frac{Z_{\mathbf{r},\mathbf{r'};\sigma}}{Z} \left<
e^{i\mathbf{q}(\mathbf{r}-\mathbf{r}')}
\right>_{\mathbf{r},\mathbf{r'};\sigma}\ ,
\end{eqnarray}
where $\mathbf{r}$ and $\mathbf{r}'$ denote the open ends of a single special trajectory. This is illustrated by the leftmost particle in the right panel of Fig.~\ref{fig:PIMC}, which, in contrast to all other particles in the system, remains open with $\mathbf{r}\neq\mathbf{r}'$. In practice, we employ the extended ensemble approach that was recently introduced in Ref.~\cite{Dornheim_PRB_nk_2021}. This sampling scheme allows us to efficiently switch between closed configurations
contributing to the canonical partition function $Z$, and open trajectories contributing to the off-diagonal partition function $Z_{\mathbf{r},\mathbf{r'};\sigma}$. In addition, this gives us direct access to the ratio of the two functions, which, in turn, means that we can directly evaluate the RHS.~of Eq.~(\ref{eq:nk_formula}) without having to estimate a normalization constant as it was done in previous schemes~\cite{Militzer_momentum_HEDP_2019,cep}.

\subsection{Interaction energy and finite-size effects\label{sec:FSC}}

The interaction energy $V=\braket{\hat{V}}$ is a diagonal property in coordinate space. Therefore, it can be straightforwardly computed in our PIMC simulations as
\begin{eqnarray}\label{eq:VX}
V(\mathbf{X}) = \frac{1}{P} \sum_{\alpha=0}^P \sum_{k=1}^N \sum_{l=1+k}^N W_\textnormal{Ew}(\mathbf{r}_{l,\alpha},\mathbf{r}_{k,\alpha})\ ,
\end{eqnarray}
where $W_\textnormal{Ew}(\mathbf{r}_1,\mathbf{r}_2)$ is the usual Ewald pair potential as it has been introduced e.g.~in Ref.~\cite{Fraser_PRB_1996}, and $\mathbf{r}_{k,\alpha}$ denotes the coordinates of particle $k$ on the imaginary-time slice $\alpha$.
The final Monte Carlo estimate for $V$ is then simply given by averaging Eq.~(\ref{eq:VX}) over the entire Markov chain of configurations,
\begin{eqnarray}
\braket{\hat{V}}_\textnormal{PIMC} = \frac{1}{N_\textnormal{MC}}\sum_{i=1}^{N_\textnormal{MC}} V(\mathbf{X}_i)\ ,
\end{eqnarray}
which becomes exact by increasing the number of samples as $\sim 1/\sqrt{N_\textnormal{MC}}$, see Eq.~(\ref{eq:sqrt}) above.

Naturally, the PIMC method is restricted to a finite number of particles $N$.
In this case, the interaction energy per particle can be expressed as the sum over the discrete reciprocal vectors $\mathbf{G}$ of the simulation cell,
\begin{eqnarray}\label{eq:v_N}
\frac{V^N}{N} = \frac{1}{2V}\sum_{\mathbf{G}\neq\mathbf{0}}\left[S^N(\mathbf{G})-1 \right] \frac{4\pi}{G^2} + \frac{\xi_\textnormal{M}}{2}\ ,
\end{eqnarray}
where $S^N(\mathbf{q})$ denotes the static structure factor of the finite system.
In addition, $\xi_\textnormal{M}$ is the well-known Madelung constant taking into account the interaction of a charge with its own background and periodic array of images~\cite{Fraser_PRB_1996}.
In contrast, the interaction energy per particle in the thermodynamic limit is defined as a continuous integral,
\begin{eqnarray}\label{eq:v_TDL}
v = \frac{1}{\pi} \int_0^\infty \textnormal{d}q\ \left[
S^\textnormal{TDL}(q)-1
\right]\ ,
\end{eqnarray}
with $S^\textnormal{TDL}(q)$ being the static structure factor of the infinite system.

The finite-size error of $V^N/N$ is then given as the difference between Eqs.~(\ref{eq:v_TDL}) and (\ref{eq:v_N}), such that
\begin{eqnarray}
v = \frac{V^N}{N} + \Delta v^N\ .
\end{eqnarray}
It is well-known that there are two potential contributions to $\Delta v^N$~\cite{Chiesa_PRL_2006,Drummond_PRB_2008,Holzmann_FSC_PRB_2016,dornheim_prl,review,Dornheim_JCP_2021}: a) a discretization error $\Delta v^N_\textnormal{d}$ due to the approximation of the continuous integral in Eq.~(\ref{eq:v_TDL}) by a discrete sum; and b) an intrinsic error $\Delta v^N_\textnormal{i}$ due to finite-size effects in the static structure factor itself, i.e., $S^\textnormal{TDL}(\mathbf{q})\neq S^N(\mathbf{q})$. See the comprehensive work by Drummond \emph{et al.}~\cite{Drummond_PRB_2008} for a detailed derivation. Empirically, it has been found both in the ground-state~\cite{Chiesa_PRL_2006,Drummond_PRB_2008,Holzmann_FSC_PRB_2016} and at finite temperature~\cite{dornheim_prl,review,Dornheim_JCP_2021,dornheim_POP} that item a) constitutes the dominant contribution to $\Delta v^N$, as the static structure factor exhibits a very weak dependence on $N$ even for small systems, $N\sim10$.

In practice, we can express the discretization error as 
\begin{eqnarray}\label{eq:FSC_v_d}
\Delta v_\textnormal{d}^N\left[S_\textnormal{trial}(q)\right]  &=& \frac{1}{\pi} \int_0^\infty \textnormal{d}q\ \left[
S_\textnormal{trial}(q)-1
\right] \\ \nonumber & & - \left(
\frac{1}{2V}\sum_{\mathbf{G}\neq\mathbf{0}}\left[S_\textnormal{trial}(\mathbf{G})-1 \right] \frac{4\pi}{G^2} + \frac{\xi_\textnormal{M}}{2}
\right)\ ,
\end{eqnarray}
where $S_\textnormal{trial}(q)$ is a suitable trial function that should qualitatively resemble both $S^\textnormal{TDL}(q)$ and $S^N(q)$. Typical choices are given either by the random phase approximation or the STLS scheme~\cite{stls,stls2,stls_original}, see Sec.~\ref{sec:LRT} below. Moreover, Chiesa \emph{et al.}~\cite{Chiesa_PRL_2006} have pointed out that the first-order contribution to Eq.~(\ref{eq:FSC_v_d}) is the $\mathbf{G}=\mathbf{0}$ term that is missing from the discrete sum. In fact, this contribution can be evaluated analytically, and Brown \emph{et al.}~\cite{Brown_PRL_2013}
have given the finite-$T$ expression
\begin{eqnarray}\label{eq:BCDC}
\Delta v^N_{\textnormal{d,0}} = \frac{\omega_p}{4N}\textnormal{coth}\left( \frac{\beta\omega_p}{2} \right)\ ,
\end{eqnarray}
where $\omega_p=\sqrt{r_s^3/3}$ is the plasma frequency.

For the parameters that are explored in the present work, we find that Eq.~(\ref{eq:BCDC}) is not sufficient, and instead evaluate the complete discretization error Eq.~(\ref{eq:FSC_v_d}) within RPA. In addition, we note that Dornheim and Vorberger~\cite{Dornheim_JCP_2021} have recently presented a new scheme to also reliably estimate the intrinsic error $\Delta v^N_\textnormal{i}$. Yet, our numerical results indicate that this intrinsic error is small and vanishes towards high temperature.

\subsection{Linear response theory and local field correction\label{sec:LRT}}

The density response of an electron gas to an external harmonic perturbation~\cite{Dornheim_PRL_2020} of wave-number $q$ and frequency $\omega$ is---within linear response theory---fully described by the dynamic density response function~\cite{quantum_theory,kugler1}
\begin{eqnarray}\label{eq:chi}
\chi(q,\omega) = \frac{\chi_0(q,\omega)}{1-\frac{4\pi}{q^2}\left[1-G(q,\omega)\right]\chi_0(q,\omega)}\ .
\end{eqnarray}
In particular, the dynamic density response of an ideal Fermi gas $\chi_0(q,\omega)$ can be readily evaluated, and the full wave-vector and frequency-resolved information about exchange--correlation effects is contained in the dynamic local field correction $G(q,\omega)$. Correspondingly, setting $G(q,\omega)=0$ leads to a description of the density response on the mean-field level, which is commonly known as the random phase approximation. Therefore, the LFC is highly important as input for many applications such as the interpretation of X-ray Thomson scattering experiments~\cite{siegfried_review,kraus_xrts,Dornheim_PRL_2020_ESA}, the estimation of ionization-potential depression~\cite{Zan_PRE_2021}, or the construction of electronically screened potentials~\cite{zhandos1,ceperley_potential,zhandos2}. Consequently, many approximate closure relations have been presented in the literature, most notably the schemes by Singwi, Tosi, Land, and Sj\"olander~\cite{stls_original,stls,stls2} (STLS), and by Vashista and Singwi (VS)~\cite{vs_original,stls2,stolzmann}. Additional recent improvements include the hypernetted chain formulation by Tanaka~\cite{tanaka_hnc} and the integral-equation based dielectric scheme by Tolias \emph{et al.}~\cite{Tolias_JCP_2021,castello2021classical}, which both constitute improvements over STLS at strong coupling.

Being formulated in the imaginary-time domain, the PIMC method is by design restricted to the static limit, i.e., 
\begin{eqnarray}
\chi(q) = \lim_{\omega\to0}\chi(q,\omega) \ .
\end{eqnarray}
In this limit, the density response function can be estimated from the imaginary-time version of the fluctuation--dissipation theorem~\cite{bowen2}
\begin{eqnarray}\label{eq:static_chi}
\chi({q}) = -n\int_0^\beta \textnormal{d}\tau\ F({q},\tau) \quad ,
\end{eqnarray}
where $F(q,\tau)$ denotes the usual intermediate scattering function~\cite{siegfried_review}, but evaluated at an imaginary time argument $\tau\in[0,\beta]$,
\begin{eqnarray}\label{eq:F}
F(q,\tau) = \frac{1}{N} \braket{\rho(q,\tau)\rho(-q,0)}\ .
\end{eqnarray}
In particular, the PIMC method has straightforward access to such imaginary-time correlation functions~\cite{Berne_JCP_1983,Dornheim_JCP_ITCF_2021}, and Eq.~(\ref{eq:static_chi}) thus implies that we can estimate the full $q$-dependence of the density response function from a single simulation of the unperturbed system. For completeness, we note that Dornheim \emph{et al.}~\cite{Dornheim_JCP_ITCF_2021} have recently generalized this concept to the nonlinear density response of a system, which, however, is beyond the scope of the present work.
Having obtained the density response function, it is then straightforward to solve Eq.~(\ref{eq:chi}) for the static local field correction 
\begin{eqnarray}
G(q) &=& \lim_{\omega\to0}G(q,\omega) \nonumber\\ &=&
1 - \frac{q^2}{4\pi}\left( 
\frac{1}{\chi_0(q)} - \frac{1}{\chi(q)}
\right)\ .\label{eq:G_static}
\end{eqnarray}
This has recently allowed us to present a neural network representation of $G(q;r_s,\theta)$, that allows to reliably predict the static LFC in the range of $0\leq q \leq 5q_\textnormal{F}$, $0.7\leq r_s \leq 20$, and $0\leq \theta\leq4$, i.e., over the entire parameter space that is relevant for WDM research.

Let us conclude this section by exploring the relation between the LFC and the static structure factor $S(q)$. 
Specifically, the fluctuation--dissipation theorem~\cite{quantum_theory}
\begin{eqnarray}\label{eq:FDT}
S({q},\omega) = - \frac{ \textnormal{Im}\chi({q},\omega)  }{ \pi n (1-e^{-\beta\omega})}
\end{eqnarray}
gives a direct connection between the dynamic structure factor $S(q,\omega)$ and Eq.~(\ref{eq:chi}). The static structure factor is then defined as the normalization of Eq.~(\ref{eq:FDT})
\begin{eqnarray}\label{eq:Sq}
S(q) = \int_{-\infty}^\infty \textnormal{d}\omega\ S(q,\omega)\ ,
\end{eqnarray}
and in this way entails an averaging over the full frequency range. 
In the classical limit, $S(q)$ is directly connected to the static LFC by the relation~\cite{ICHIMARU198791},
\begin{eqnarray}\label{eq:S_classical}
S(q) = \frac{q^2}{q^2+q_\textnormal{D}^2\left[1-G(q)\right]}\ ,
\end{eqnarray}
with the definition of the Debye wave number
\begin{eqnarray}
q_\textnormal{D} =  \left(4\pi e^2 n\beta \right)^{1/2}\ .
\end{eqnarray}
Equivalently, we can solve Eq.~(\ref{eq:S_classical}) for $G(q)$, which gives
\begin{eqnarray}\label{eq:G_classical}
G(q) = 1 + \frac{q^2}{q_\textnormal{D}^2} \left(1-\frac{1}{S(q)}\right)\ .
\end{eqnarray}
Therefore, our new accurate PIMC results for $S(q)$ and $G(q)$ will allow us to assess the accuracy of Eqs.~(\ref{eq:S_classical}) and (\ref{eq:G_classical}) and to study the convergence to the classical limit of these important properties in the high temperature limit.

\section{High-temperature limit of the equation of state and virial expansion benchmarks}\label{analy}
\subsection{Thermodynamic relations and virial expansions}

Analytical expression for the thermodynamic properties of the OCP can be obtained from many-particle theory~\cite{kremp_book}. Within a perturbation theory, for instance using the method of thermodynamic Green's functions, results are obtained which become exact in the limits of high density, or low density, or high temperature. We are interested in the low-density, high-temperature limit of the equations of state.

A well-known example is the virial expansion for the pressure 
$p^{\rm sr}(T,n)$ of a system with short-range interaction (sr). 
At {\it fixed} $T$, the dependence on the density $n=N/\Omega$ is expanded as
\begin{equation}
 p^{\rm sr}(T,n)=b^{\rm sr}_1(T)n+b^{\rm sr}_2(T)n^2+b^{\rm sr}_3(T)n^3+\dots.
 \label{shortr}
\end{equation}
The first coefficient is $b^{\rm sr}_1(T) = k_BT$. For classical systems, expressions for the higher virial coefficients are known from text books on statistical physics, using the Mayer cluster expansion. In particular, we have for the second virial coefficient of a classical system  
\begin{equation}
b^{\rm sr}_2(T)= k_BT \int d^3r \left(e^{-V(r)/k_BT} -1\right).
\end{equation}
For quantum systems, the second virial coefficient is determined according to the Beth-Uhlenbeck equation \cite{bethu1936} by the contribution of bound states 
and the continuum contribution which is expressed in terms of the scattering phase shifts.

Other equations of state are of interest in this work such as the caloric one which expresses the internal energy as function of $T, N$ and the volume $\Omega$.
In general, all equations of state are derived consistently if a thermodynamic potential is known, 
in our case the free energy $F(T,\Omega,N)$. For instance, the pressure follows as
\begin{equation}
 p(T,n)=\left(\frac{\partial}{\partial \Omega}F(T,\Omega,N)\right)\Big|_{T,N}
 \label{ptherm}
\end{equation}
and the internal energy as
\begin{eqnarray}
 \label{Utherm}
 U(T,\Omega,N)&=&F+TS\\&=& F(T,\Omega,N)-T\left(\frac{\partial}{\partial T}F(T,\Omega,N)\right)\Big|_{\Omega,N}.\nonumber
\end{eqnarray}

As is well known, for the long-range Coulomb interaction, the virial expansion
(\ref{shortr}) is not converging. The classical expression for the second virial coefficient given above diverges, and the quantum Beth-Uhlenbeck \cite{bethu1936} formula contains scattering phase shifts which are not defined as usual for the Coulomb interaction.
Convergent results have been obtained after partial summation of so-called ring diagrams so that the Debye-H{\"u}ckel limiting law is obtained as a result. 
The analytical form of the low-density/high-temperature behavior of the thermodynamic quantities is thus changed. 
For the free energy of the uniform electron gas (UEG), we have according to Friedman (1962) \cite{friedman1962}  (see Sec. 2.3 of \cite{kraeft1986quantum}), $\Lambda^2=2 \pi \hbar^2/(m k_BT)$,
\begin{eqnarray}
\label{Fvir}
F(T,\Omega,N)&=&\Omega k_BT \left\{n \ln n + [\ln(\Lambda^3)-1] n
\right. 
\nonumber \\ &&\left.-A_0(T)n^{3/2}-A_1(T)n^2 \ln n-A_2(T) n^2\right. 
\nonumber \\ &&\left. -A_3(T)n^{5/2} \ln n-A_4(T) n^{5/2}+{\cal O}(n^3\ln n)\right\}.
\end{eqnarray}

From this virial expansion of the free energy, we obtain, with (\ref{ptherm}) and (\ref{Utherm}), the virial expansions for other equations of state
\begin{eqnarray}
\label{Uvir1}
U(T,\Omega,N)&=&\Omega k_BT^2 \left\{\frac{3}{2T}n+\frac{\partial}{\partial T}A_0(T)n^{3/2}
\right. 
\nonumber \\ &&\left.+\frac{\partial}{\partial T}A_1(T)n^2 \ln n+\frac{\partial}{\partial T}A_2(T) n^2 +{\cal O}(n^{5/2}\ln n)\right\},
\end{eqnarray}
\begin{eqnarray}
\label{pvir1}
p(T,n)&=& k_BT \left\{n+\frac{1}{2}A_0(T)n^{3/2}
\right. 
\nonumber \\ &&\left.+A_1(T)n^2 \ln n+[A_2(T)+A_1(T)] n^2+{\cal O}(n^{5/2}\ln n)\right\}.
\end{eqnarray}
We consider here the caloric equation of state (\ref{Uvir1}). The virial coefficients  $A_i(T)$ may be calculated within a Green's function approach if
the partial summation of the corresponding class of Feynman diagrams is performed. For instance, $A_0(T)$ is determined by the sum of ring diagrams, whereas $A_2(T)$ follows from the sum of ladder diagrams. The derivations and final expressions are found, for example, 
in \cite{kraeft1986quantum,kremp_book}. For the linear direct term, see Sec.~\ref{sec:xi/6}.

\subsection{Virial coefficients and benchmarks}

The expressions for the lowest order virial coefficients are
\begin{eqnarray}
 A_0(T)&=&\kappa^3/(12 \pi n^{3/2}),\\
 A_1(T)&=&\frac{\pi}{6}\left(\frac{e^2}{4 \pi \epsilon_0 k_BT}\right)^3,\\
 A_2(T)&=&2 \pi \lambda^3 K(\xi)+\frac{\pi}{3} \left(\frac{e^2}{4 \pi \epsilon_0 k_BT}\right)^3 \ln (\kappa \lambda/n^{1/2})\ .
\end{eqnarray}
Here, we used the notations $\kappa^2=ne^2/(\epsilon_0k_BT)$ for the screening parameter, $\lambda^2=\hbar^2/(mk_BT)$ for the thermal wavelength, and $\xi=-e^2/(4 \pi \epsilon_0 k_BT \lambda)$ for the Born parameter.

The expression for the virial coefficient  $A_2(T)$ is given in \cite{kremp_book}. It contains the quantum virial function 
\begin{equation}
K(\xi)=-D(\xi)-\frac{1}{2} E(\xi)
\end{equation}
with the direct
\begin{equation}\label{D}
D(\xi)=-\frac{\sqrt 
\pi}{8}\xi^2-\frac{1}{6}\left(\frac{1}{2}C+ \ln 
3-\frac{1}{2}\right)\xi^3
+\sqrt 
{\pi}\sum_{m=4}^{\infty}\frac{\zeta(m-2)}{\Gamma\left(\frac{1}{2}m+1\right)}\left(\frac{\xi}{2}
\right)^m\,, 
\end{equation}
and exchange term
\begin{eqnarray}\label{E}
E(\xi)&=&\frac{1}{4}\sqrt {\pi}+\frac{1}{2}\xi+\frac{1}{4}\sqrt{\pi}(\ln 
2)\xi^2+\frac{1}{72}\pi^2\xi^3\nonumber\\
&&+\sqrt{\pi}\sum_{m=4}^{\infty}
(1-2^{2-m})\frac{\zeta(m-1)}{\Gamma\left(\frac{1}{2}m+1\right)}\left(\frac{\xi}{2}\right)^m\,.
\end{eqnarray}
Here, we used the Riemann $\zeta$ function, and $C=0.57721\ldots$ is Euler's constant.
The derivation of 
Eqs. (\ref{D}) and (\ref{E}) is given in \cite{kremp_book}, sects. 6.5.3 and 6.5.4. 

For the virial coefficient $A_2(T)$ another expression was also discussed,
\begin{equation} 
\label{wrong}
 B_2(T)=A_2(T)- \frac{\pi}{3} \lambda^3 \xi;
\end{equation}
see \cite{kraeft1986quantum}. This additional $\xi/6$-term appears from an approximate treatment of the Coulomb interaction, see \cite{WDK2005,Kraeft_2015}, and \cite{EHK1967,EKK1970} and also \ref{apa}. This additional term which appears also in \cite{kraeft1986quantum} when defining the function $D(\xi)$ as $-\xi/6$ has been corrected recently \cite{Kraeft_2015}. We mention that the exchange function E (\ref{E}) contains a linear term $\xi/2$ in contrast to the $D$ function (\ref{D}) which does not.  

Altogether, from the relations given above, we can derive an expression for the mean potential energy $V$ which reads up to the second virial coefficient, i.e. neglecting terms of the order ${\cal O}(n^{5/2}\ln n)$
\begin{eqnarray}
\frac{V}{N\,k_BT}&=&-\frac{\kappa^3}{8\pi n}-2\pi n \lambda^3\left[\frac{\partial K(\xi)}{\partial (e^2)}-\frac{\xi^3}{2}\ln\kappa\lambda-\frac{\xi^3}{12}\right]\ .
\label{full_virial}
\end{eqnarray}
The derivative of the virial function is the inverse of the charging procedure, i.e., a derivative with respect to the coupling parameter. 

For small values of $\xi$ one might expand the expression $K(\xi)$ given in Eqs. (\ref{D}) and (\ref{E}). If we  restrict to orders up to $\xi^3 \propto e^6$, this leads to~\cite{KraeftKremp68}
\begin{eqnarray}
\frac{V}{Nk_BT}&=&-\frac{\kappa^3}{8\pi n}-2\pi n \lambda^3\left[-\frac{\xi}{4}-\frac{\sqrt{\pi}}{4}\xi^2(1+\ln 2)\right.\nonumber\\
 &&\left.-\frac{\xi^3}{2}\left(\ln\kappa\lambda+\frac{C}{2}+\ln 3 -\frac{1}{3}-\frac{\pi^2}{24}\right)\right].
\label{(ne2)2}
 \end{eqnarray}
The linear term in Eq.~(\ref{(ne2)2}) is an exchange contribution.

For the comparison with PIMC results, we use atomic units, $\hbar = m = a_B = e^2/4 \pi \epsilon_0 =1$. Instead of $T$ we can use $T_{\rm Ha}= k_BT/E_{\rm Ha}$ with the 
Hartree energy $E_{\rm Ha} = 27.211386$~eV. We have $\lambda = -\xi=1/\sqrt{T_{\rm Ha}}$. Instead of the density $n$, the Brueckner parameter $r_s$ is given by $n= (3/4 \pi) (r_s a_B)^{-3}$. The parameter $\theta$ defined above is related to these parameters as $\theta=2 (9 \pi/4)^{-2/3} T_{\rm Ha} r_s^2$.
For the screening parameter we have $\kappa^2=6 (4/9 \pi)^{2/3}/(\theta r_s)$.


We give the internal energy per particle in units of $E_{\rm Ha}$ and introduce corresponding virial coefficients,
\begin{eqnarray}
\frac{U_{\rm Ha}}{N}(T_{\rm Ha},r_s)&=&\frac{3}{2}T_{\rm Ha}+U_0(T_{\rm Ha})r_s^{-3/2} 
+U_1(T_{\rm Ha})r_s^{-3}  \ln (r_s^{-3})+U_2(T_{\rm Ha}) r_s^{-3}  
\nonumber \\ 
&&+{\cal O}(n^{3/2} \ln n). 
\end{eqnarray}
Expressions for the coefficients are
\begin{equation}
U_0(T_{\rm Ha})=-\frac{3^{1/2}}{2} T_{\rm Ha}^{-1/2},
\end{equation}
\begin{equation}
U_1(T_{\rm Ha})=-\frac{3}{8} T_{\rm Ha}^{-2},
\end{equation}
\begin{equation}
U_2(T_{\rm Ha})=\frac{3}{2} T_{\rm Ha}^{-3/2}  \frac{d}{d T_{\rm Ha}} K\left(-T_{\rm Ha}^{-1/2}\right)-\frac{3}{8} T_{\rm Ha}^{-2} \ln\left(3 T_{\rm Ha}^{-2}\right).
\end{equation}
(Note that the last term can be added to $U_1$ as a factor below the logarithmic term.) These expressions may be considered as exact results and may serve as benchmarks for any approaches to evaluate the low-density limit of 
the thermodynamic quantities. Expressions for higher order virial coefficients are given in Ref.~\cite{kraeft1986quantum} and will not be considered here. 
Note that the approximate value $ B_2(T)$, Eq. (\ref{wrong}), will give the expression
\begin{equation} 
\label{wrong1}
 W_2(T_{\rm Ha})=U_2(T_{\rm Ha})- \frac{1}{2 T_{\rm Ha}}.
\end{equation}

The virial coefficients $U_0$ to $U_2$ are benchmarks and should be reproduced  by any exact treatment. It is of interest to see whether 
these are also seen by the PIMC calculations.

\section{Results\label{sec:results}}

\subsection{Fermion sign problem\label{sec:FSP}}

\begin{figure}\centering
\hspace*{-0.06\textwidth}\includegraphics[width=0.55\textwidth]{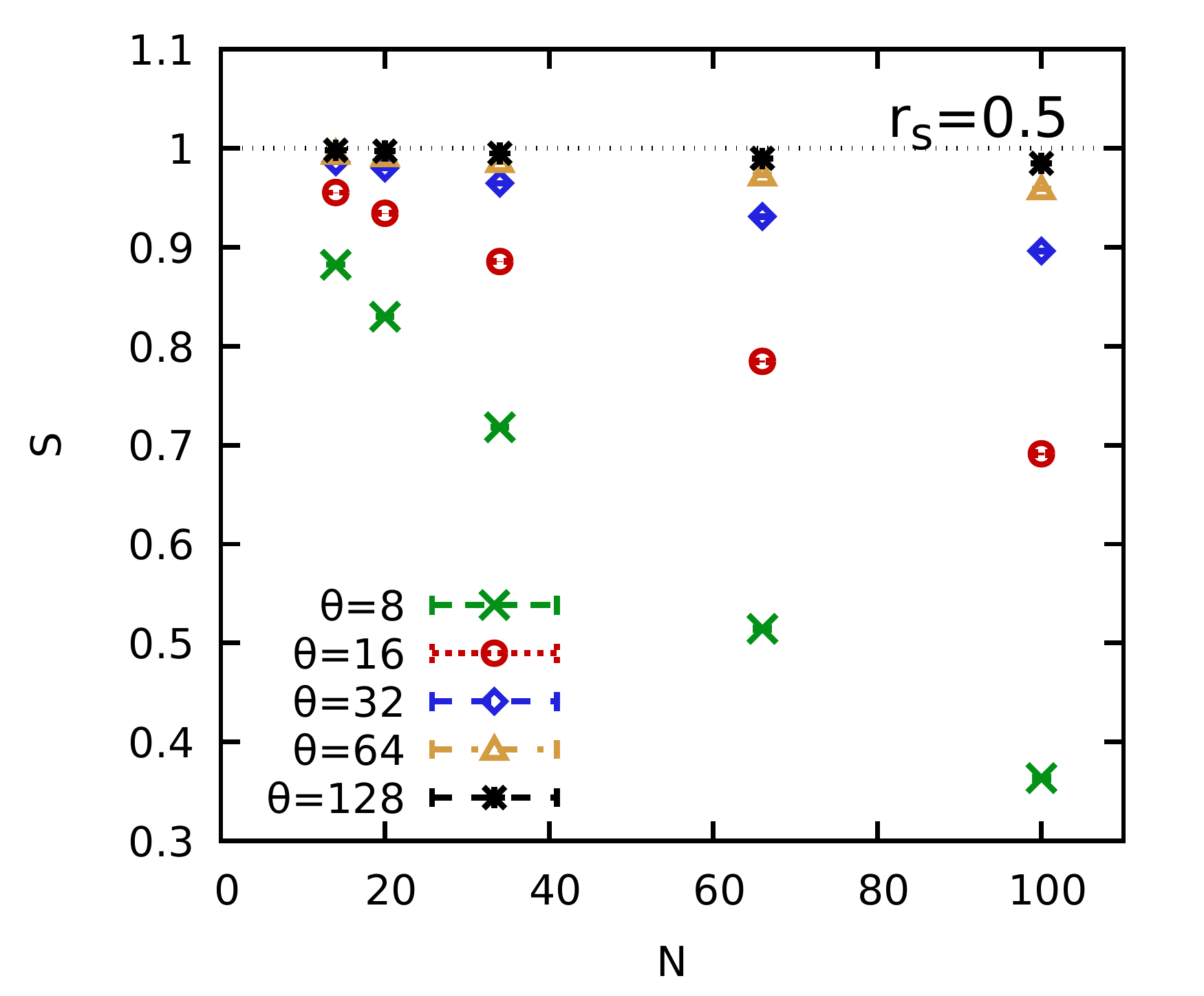}\includegraphics[width=0.55\textwidth]{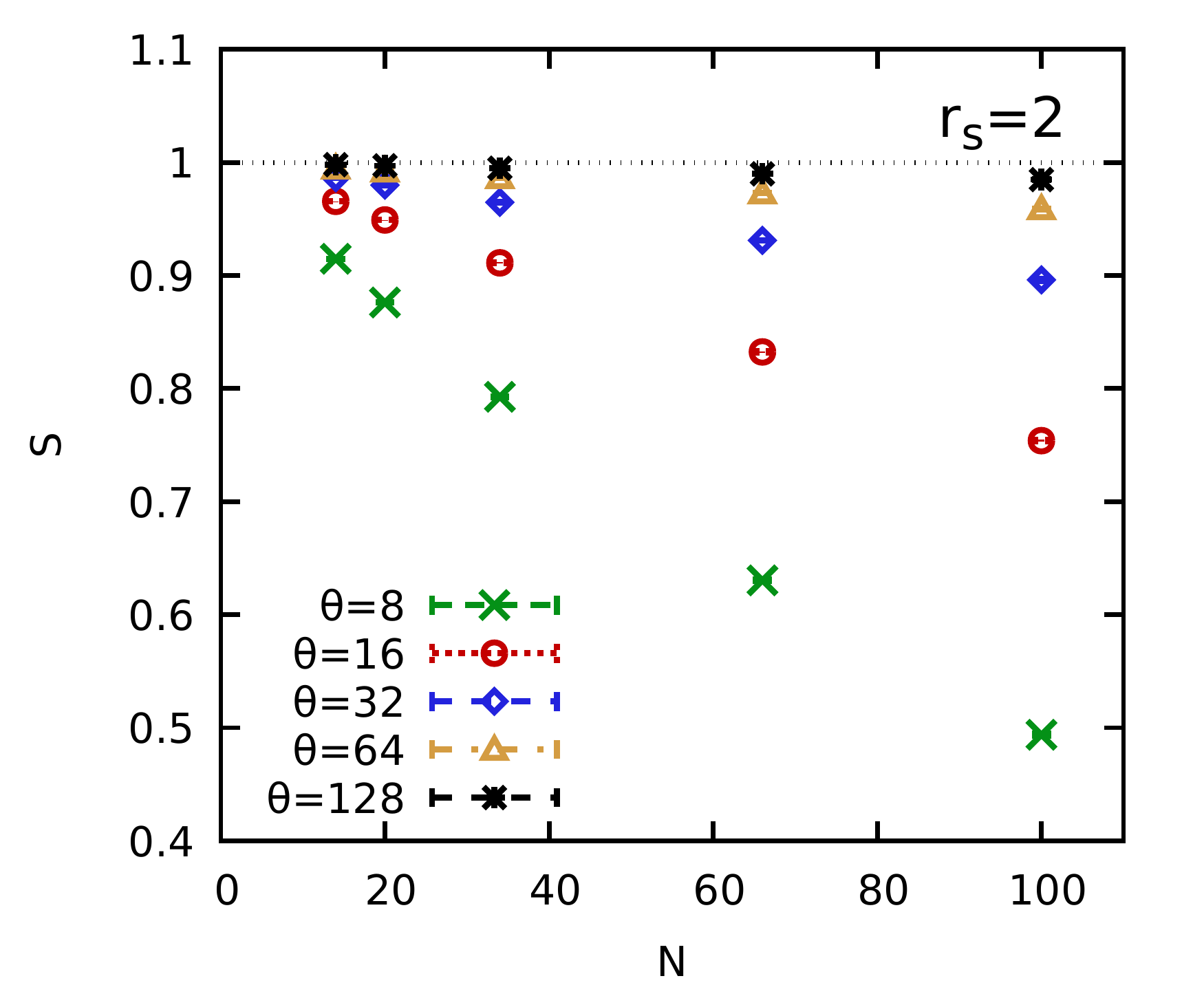}\\\includegraphics[width=0.55\textwidth]{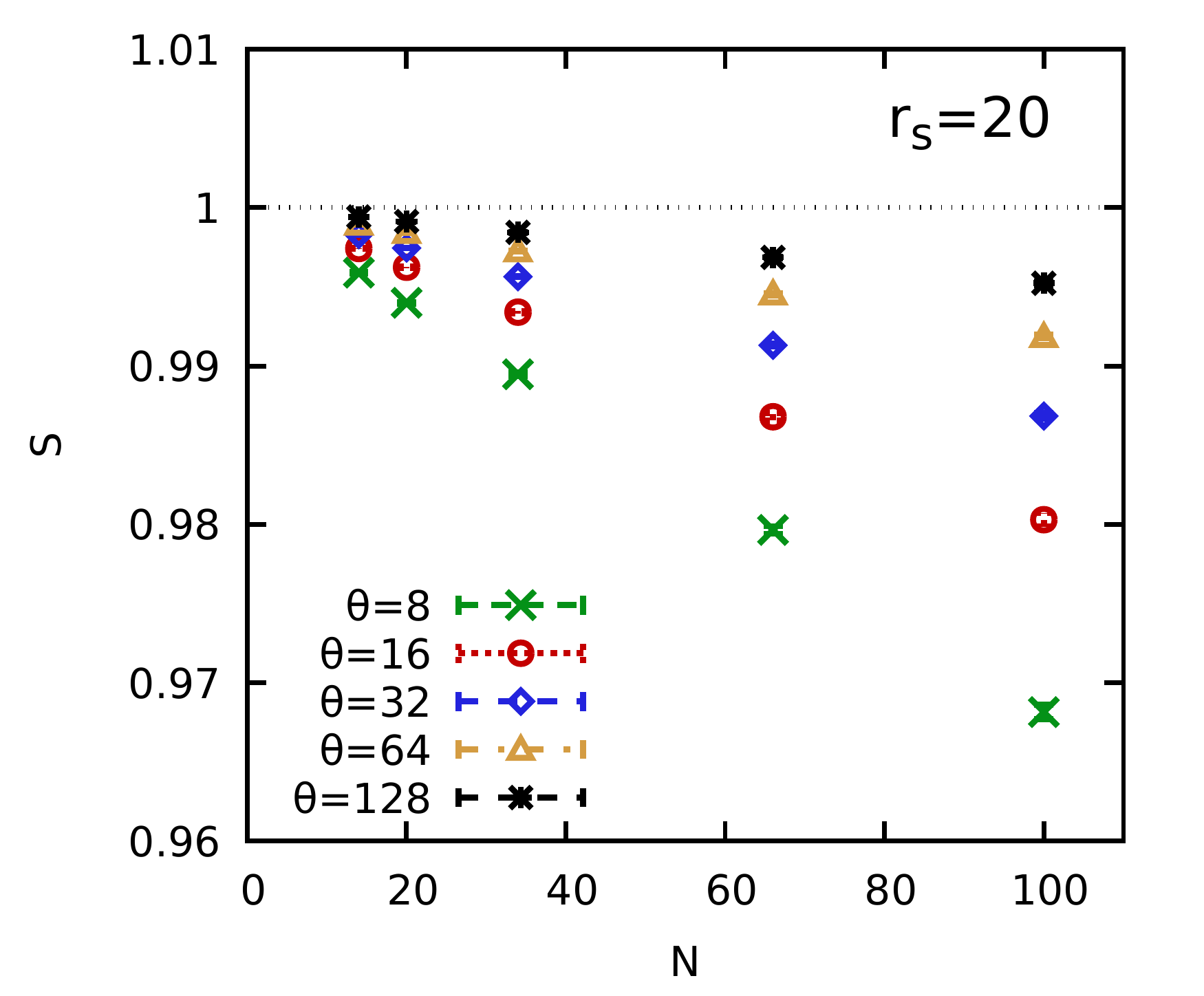}
\caption{\label{fig:Sign_rs2_N}
PIMC results for the dependence of the average sign $S$ on the system-size $N$ for different values of the density parameter $r_s$. The green crosses, red circles, blue diamonds, yellow triangles, and black stars show results for $\theta=8$, $\theta=16$, $\theta=32$, $\theta=64$, and $\theta=128$, respectively.
}
\end{figure}

Let us start our PIMC investigation of the UEG in the high-temperature regime by briefly touching upon the fermion sign problem~\cite{dornheim_sign_problem}. To this end, we show our PIMC results for the average sign $S$ in Fig.~\ref{fig:Sign_rs2_N} for different values of $r_s$ and $\theta$. In particular, the top row corresponds to the high-density regime ($r_s=0.5$, left) and a metallic density ($r_s=2$, right) and we find fairly similar results for $S$. We note that the data sets are strictly ordered with respect to $\theta$, which can be understood as follows: for lower temperatures, the extension of the paths of individual particles, which is proportional to the thermal wavelength $\lambda_\beta\sim\sqrt{\beta}$, increases. Consequently, the overlap of such paths becomes more likely, which, in turn, can lead to the formation of permutation-cycles within the PIMC simulation; see Ref.~\cite{Dornheim_permutation_cycles} for an extensive discussion of this point. These permutation-cycles are the source of the sign changes of the configuration weights $W(\mathbf{X})$, which immediately results in lower values of $S$ for lower temperatures. 
In addition, we see a monotonic decrease of the average sign with the system size $N$, as it is expected~\cite{dornheim_sign_problem,Loh_sign_problem}. In fact, it is well-known that $S$ exponentially decays with $N$ for a uniform system such as the UEG~\cite{dornheim_sign_problem}.
As a consequence of these two trends, the lowest value of $S$ that we have encountered in our simulations appears at $r_s=0.5$, $\theta=8$ and $N=100$. This particular value of $S\gtrsim0.3$ leads to an increase in the required computation time by a factor of approximately ten, which does not constitute a serious obstacle in practice.

The bottom panel of Fig.~\ref{fig:Sign_rs2_N} shows the same information, but at a substantially lower density, $r_s=20$. This constitutes the boundary to the strongly coupled electron liquid regime~\cite{dornheim_electron_liquid,dornheim_dynamic,dynamic_folgepaper}, and the sign problem is substantially less severe. More specifically, the strong Coulomb repulsion between individual electrons drastically reduces the likelihood of permutation-cycles within the PIMC simulation. Therefore, sign changes of $W(\mathbf{X})$ are less frequent, and the average value of $S$ remains large.
At the same time, we observe the formation of exchange-cycles with nonzero probability even for $N=14$, $r_s=20$, and $\theta=128$, which is a strong indication for an ergodic exploration of the full configuration space of our sampling scheme. This is achieved due to the local update structure in the worm algorithm by Boninsegni \emph{et al.}~\cite{boninsegni1,boninsegni2}, which is used to sample different permutations in our extended ensemble approach~\cite{Dornheim_PRB_nk_2021}.

\subsection{Static structure factor and density response\label{sec:static}}

\begin{figure}\centering
\hspace*{-0.06\textwidth}\includegraphics[width=0.55\textwidth]{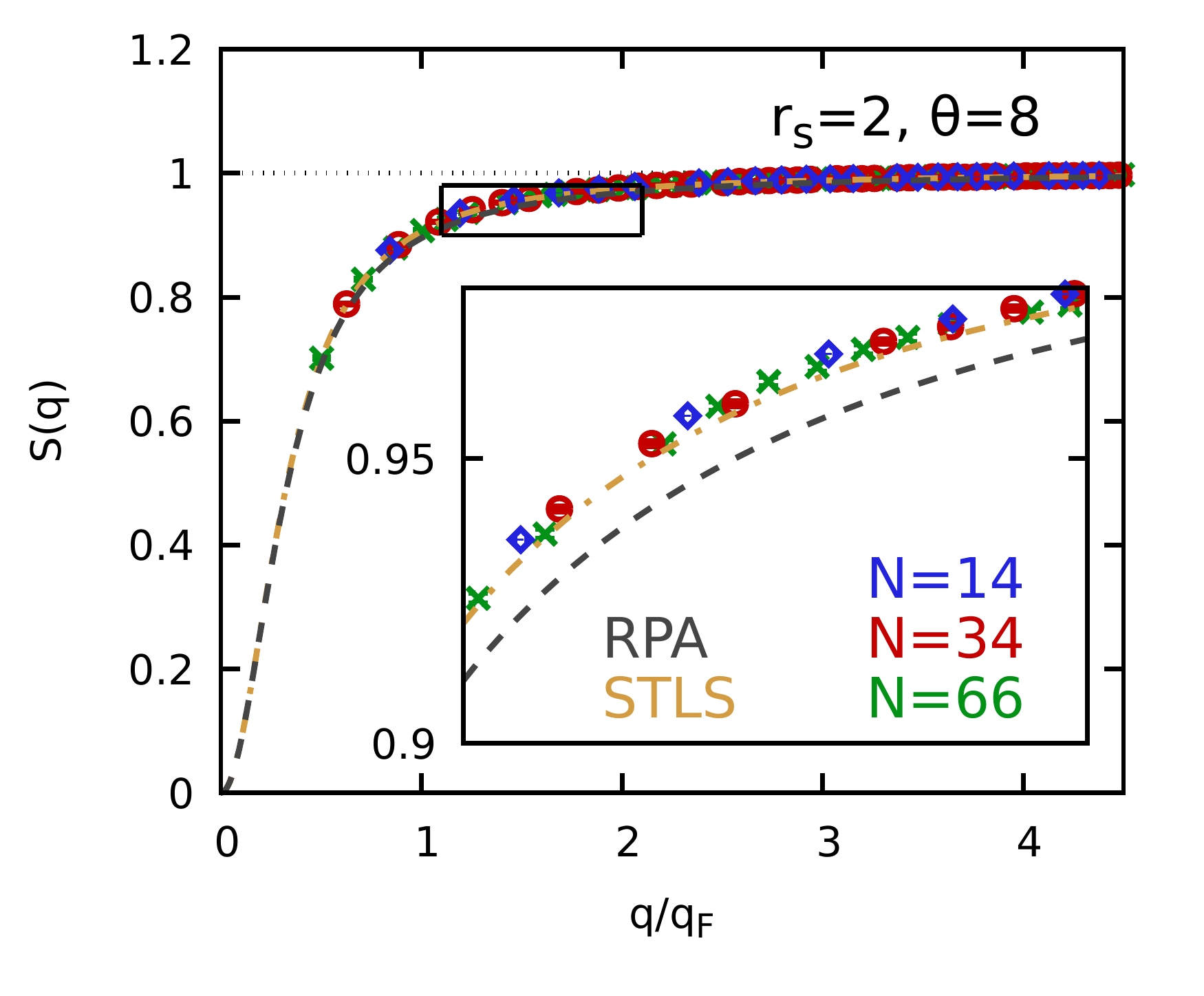}\includegraphics[width=0.55\textwidth]{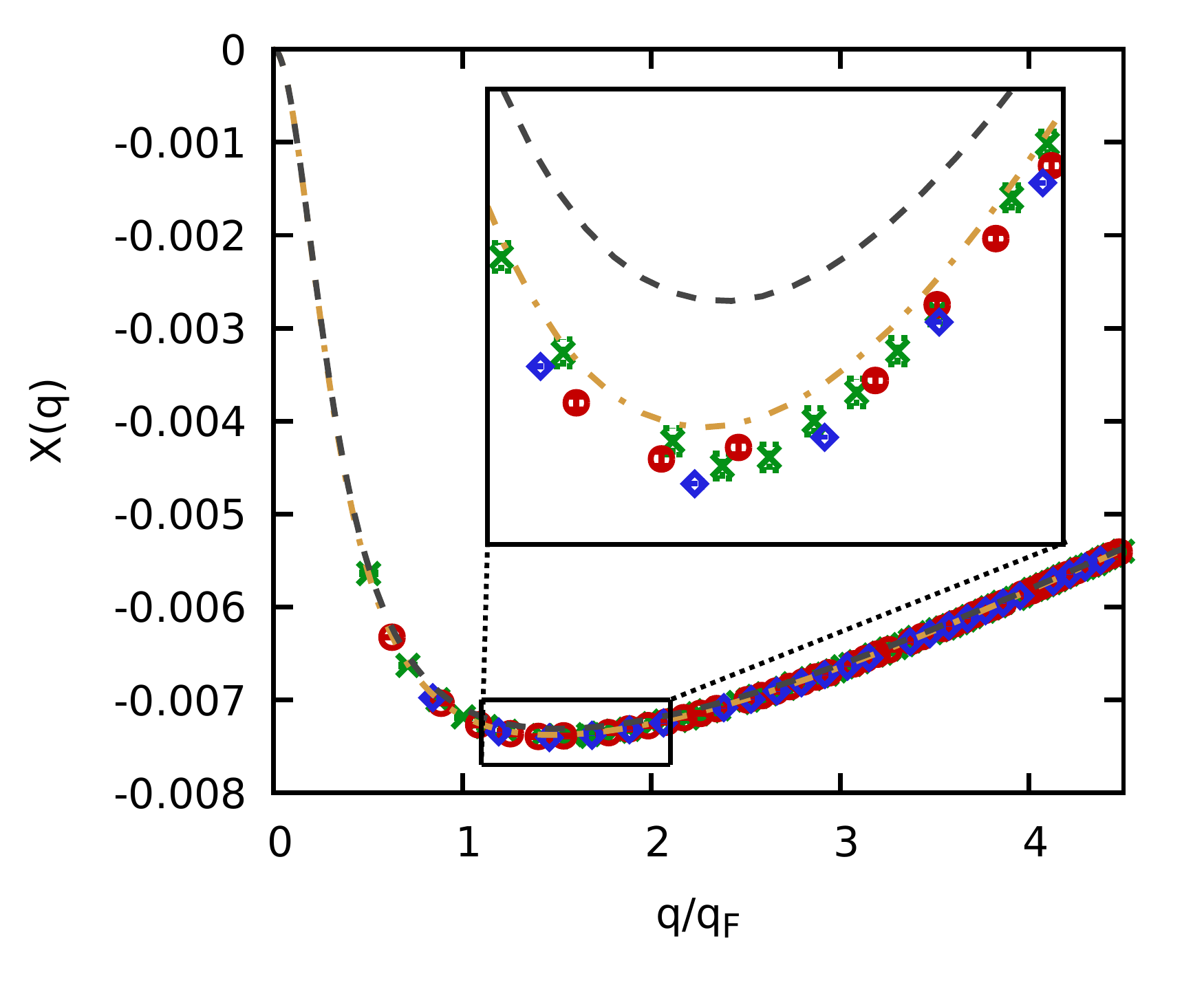}\\\vspace*{-1cm}\hspace*{-0.06\textwidth}\includegraphics[width=0.55\textwidth]{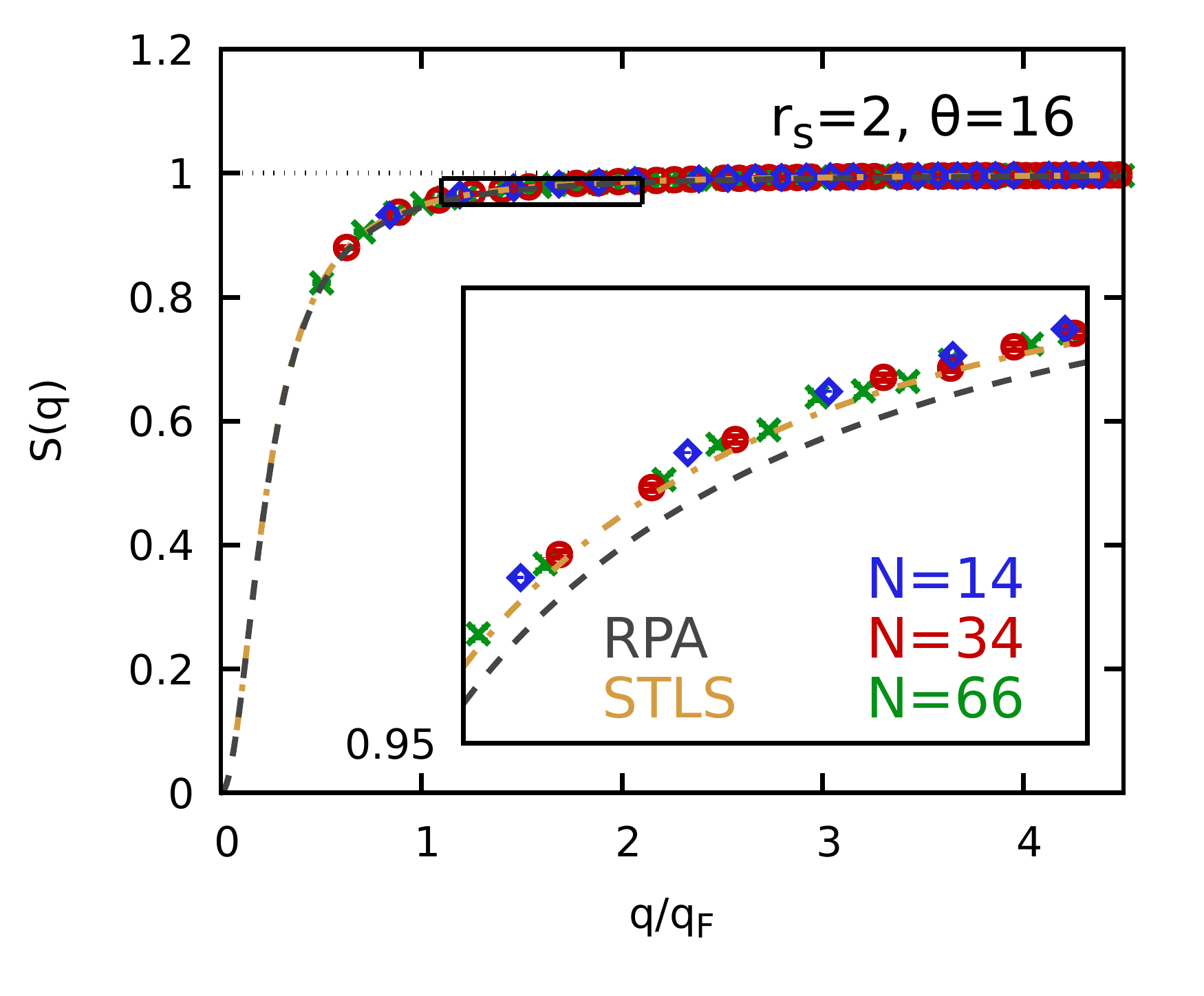}\includegraphics[width=0.55\textwidth]{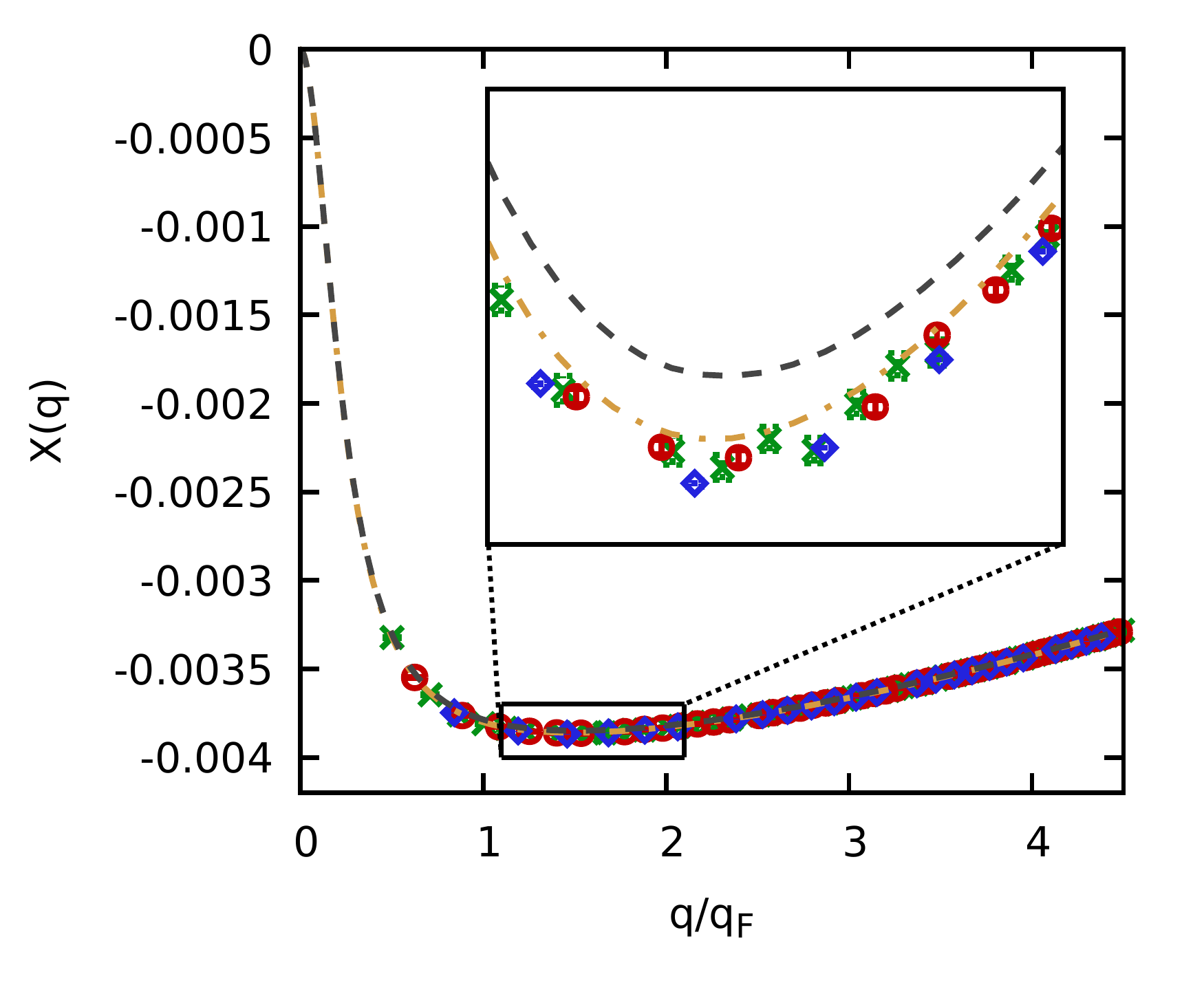}\\\vspace*{-1cm}\hspace*{-0.06\textwidth}\includegraphics[width=0.55\textwidth]{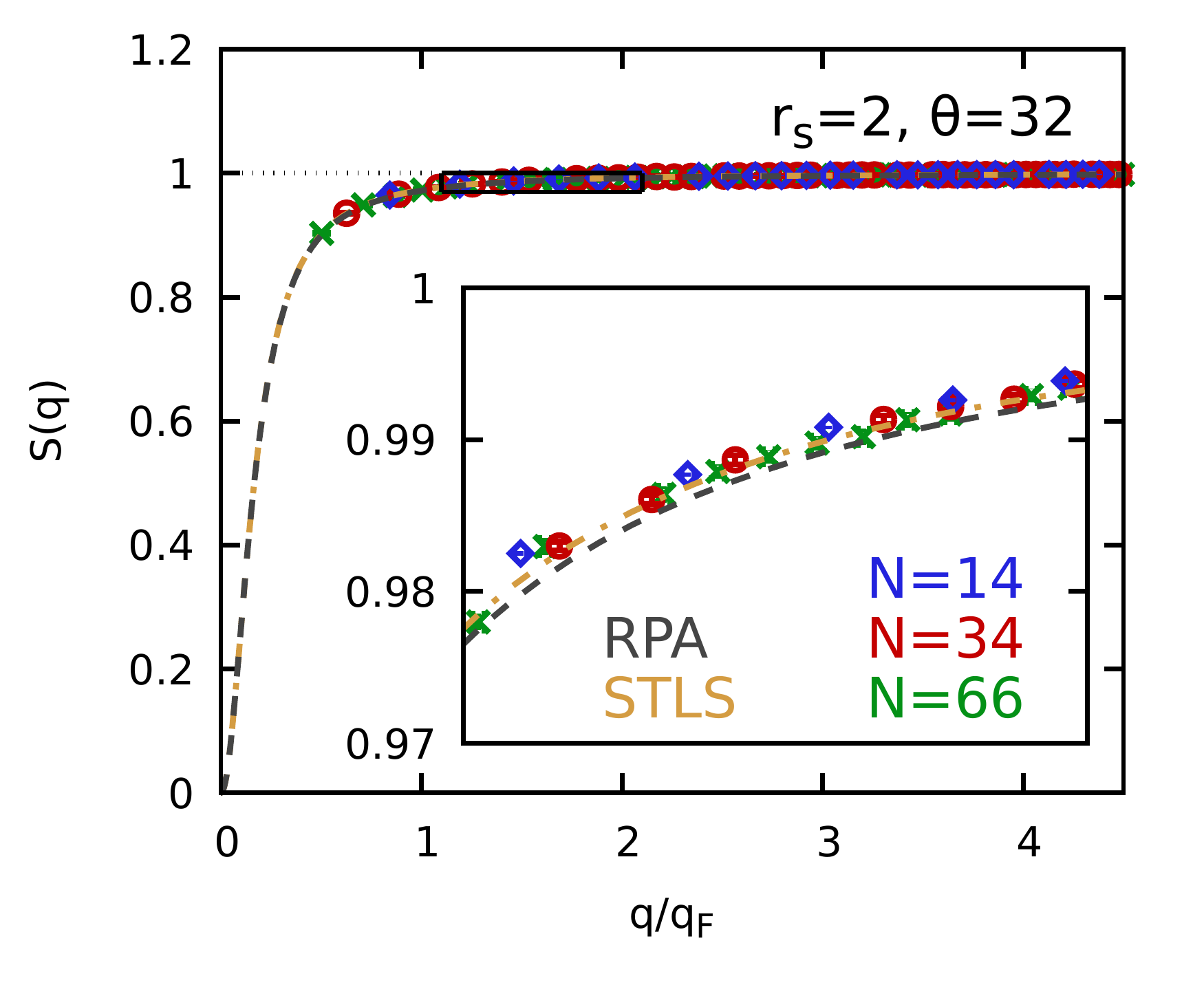}\includegraphics[width=0.55\textwidth]{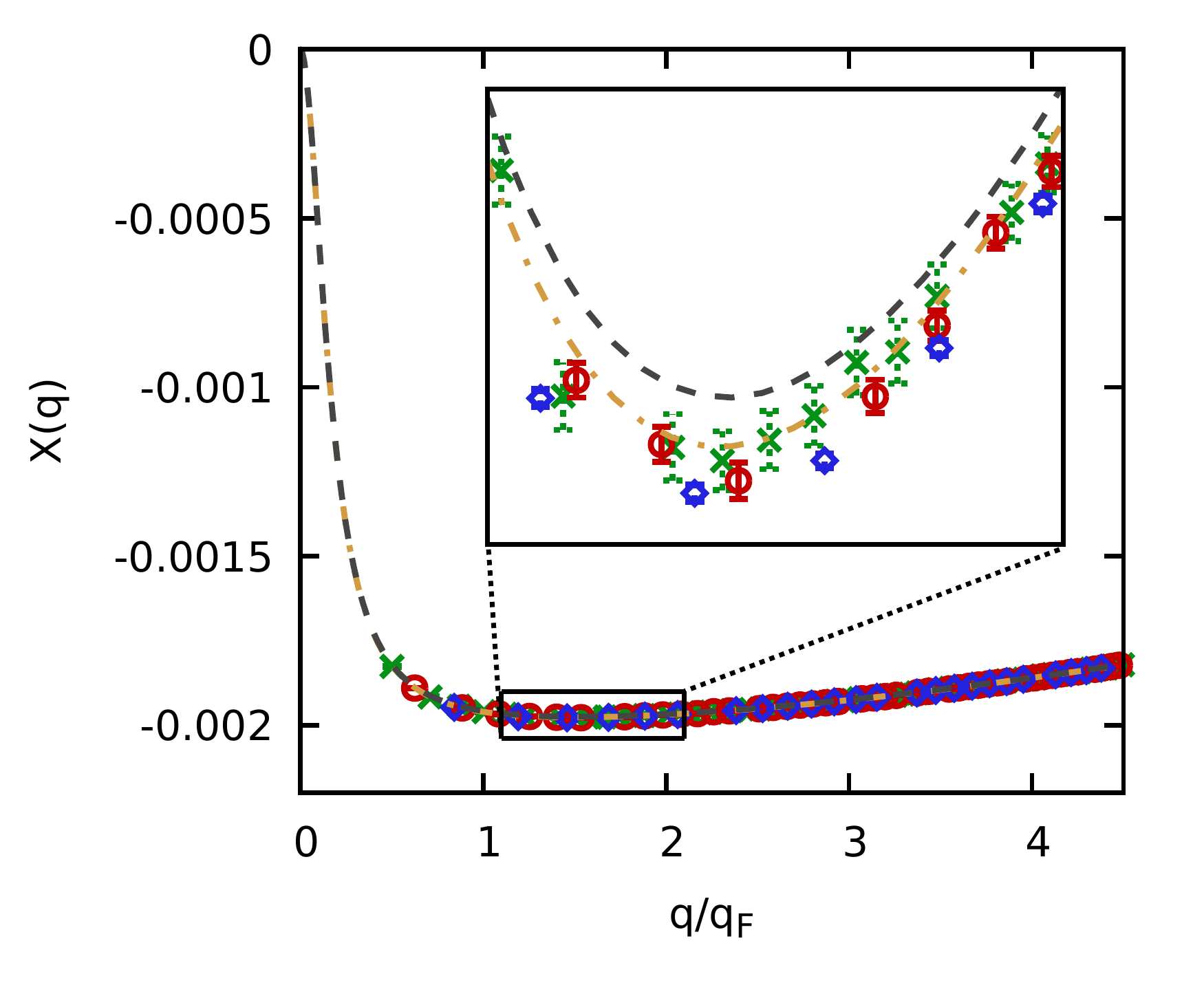}
\caption{\label{fig:rs2_theta}
PIMC results for the static structure factor $S(q)$ [left] and static density response function $\chi(q)$ [right] at $r_s=2$ and $\theta=8$ (top), $\theta=16$ (center), and $\theta=32$ (bottom) for $N=14$ (blue diamonds), $N=34$ (red circles), and $N=66$ (green crosses) unpolarized electrons. The dashed black and dash-dotted yellow curves have been computed within RPA and finite-$T$ STLS~\cite{stls2,stls}.
}
\end{figure}



Let us proceed our investigation of the UEG in the high-temperature regime by considering its structural properties. In Fig.~\ref{fig:rs2_theta}, we show simulation results for $r_s=2$ at $\theta=8$ (top row), $\theta=16$ (center row), and $\theta=32$ (bottom row). Let us first examine the left column showing the static structure factor $S(q)$. In particular, the blue diamonds, red circles, and green crosses show our PIMC data for $N=14$, $N=34$, and $N=66$ unpolarized electrons. In addition, we have also included RPA and STLS results as the dashed grey and dash-dotted yellow curves, which can be used as a guide to the eye. The inset shows a magnified segment, and even here hardly any significant intrinsic system-size dependence can be resolved in our PIMC results for $S(q)$ within the given error bars. This is a strong indication that the dominant finite-size error in the interaction energy is given by the discretization error due to the momentum quantization in the finite simulation cell~\cite{dornheim_cpp}, as it will be discussed in detail in Sec.~\ref{sec:potential} below. Naturally, PIMC data are only available above a minimum $q$-value of $q_\textnormal{min}=2\pi/L$. At the same time, we note that both RPA and STLS become exact in the limit of $q\to0$, which can be used to construct an accurate representation of $S(q)$ over the entire $q$-range~\cite{dornheim_cpp}. In addition, both RPA and STLS give the correct qualitative behaviour of $S(q)$ for all three depicted values of $\theta$, although the approximate STLS scheme is considerably more accurate for $\theta=8$ and $\theta=16$; for $\theta=32$ the impact of exchange--correlation effects as they are encoded in the LFC is very small, and RPA and STLS can hardly be distinguished with the naked eye.

The right column of Fig.~\ref{fig:rs2_theta} gives the same information for the static density response function $\chi(q)$. Here, too, finite-size effects are hardly significant within the given level of statistical uncertainty. Furthermore, the comparison of our PIMC results to RPA and STLS gives the same trends. Finally, we note that $\chi(q)$ decreases in magnitude when $\theta$ is increased, as the system becomes less correlated. 
This trend is well-known from other Fermi systems, and has recently been reported from a PIMC simulation of normal liquid $^3$He~\cite{dornheim2021pathhe}.

\begin{figure}\centering
\hspace*{-0.06\textwidth}\includegraphics[width=0.55\textwidth]{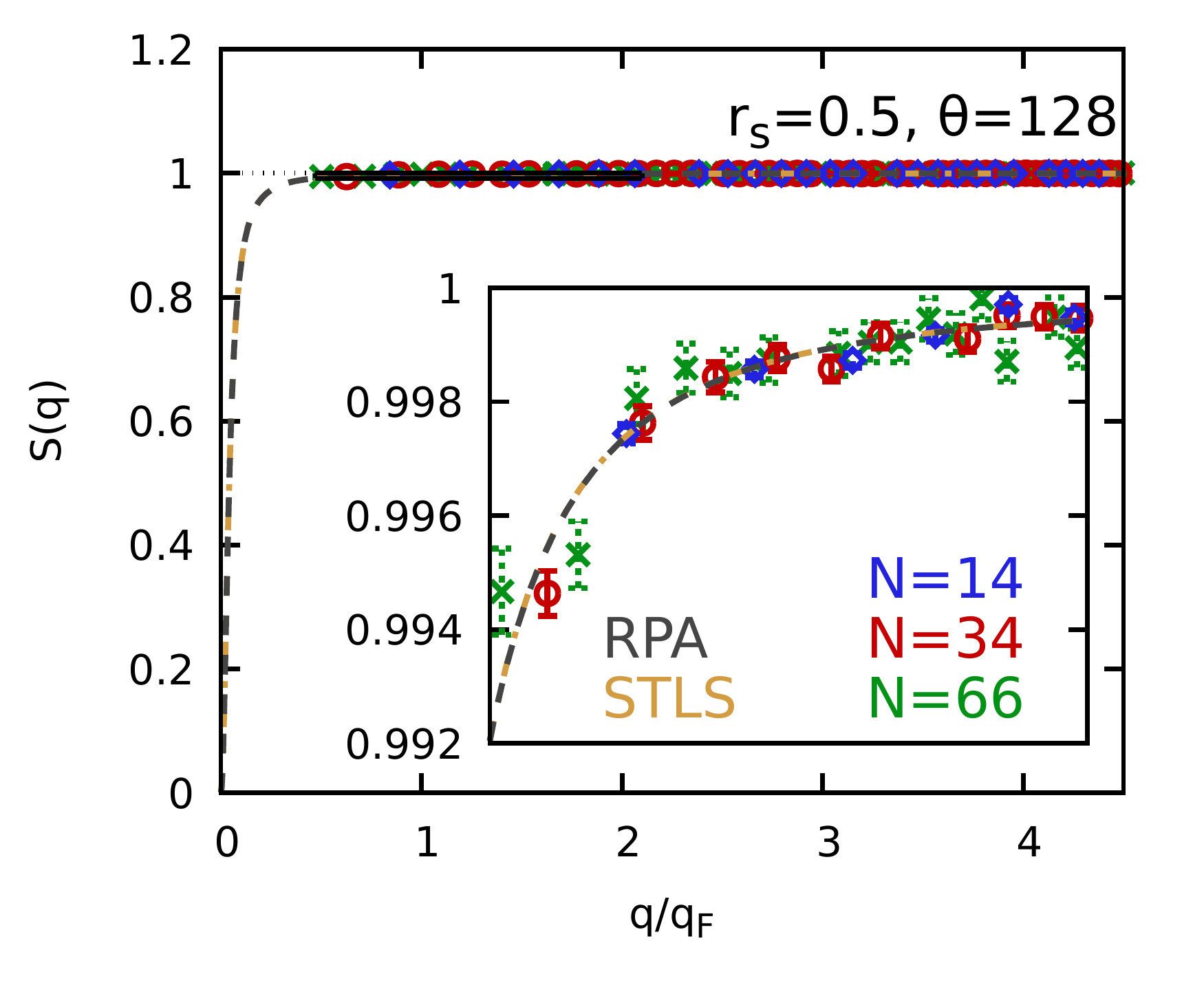}\includegraphics[width=0.55\textwidth]{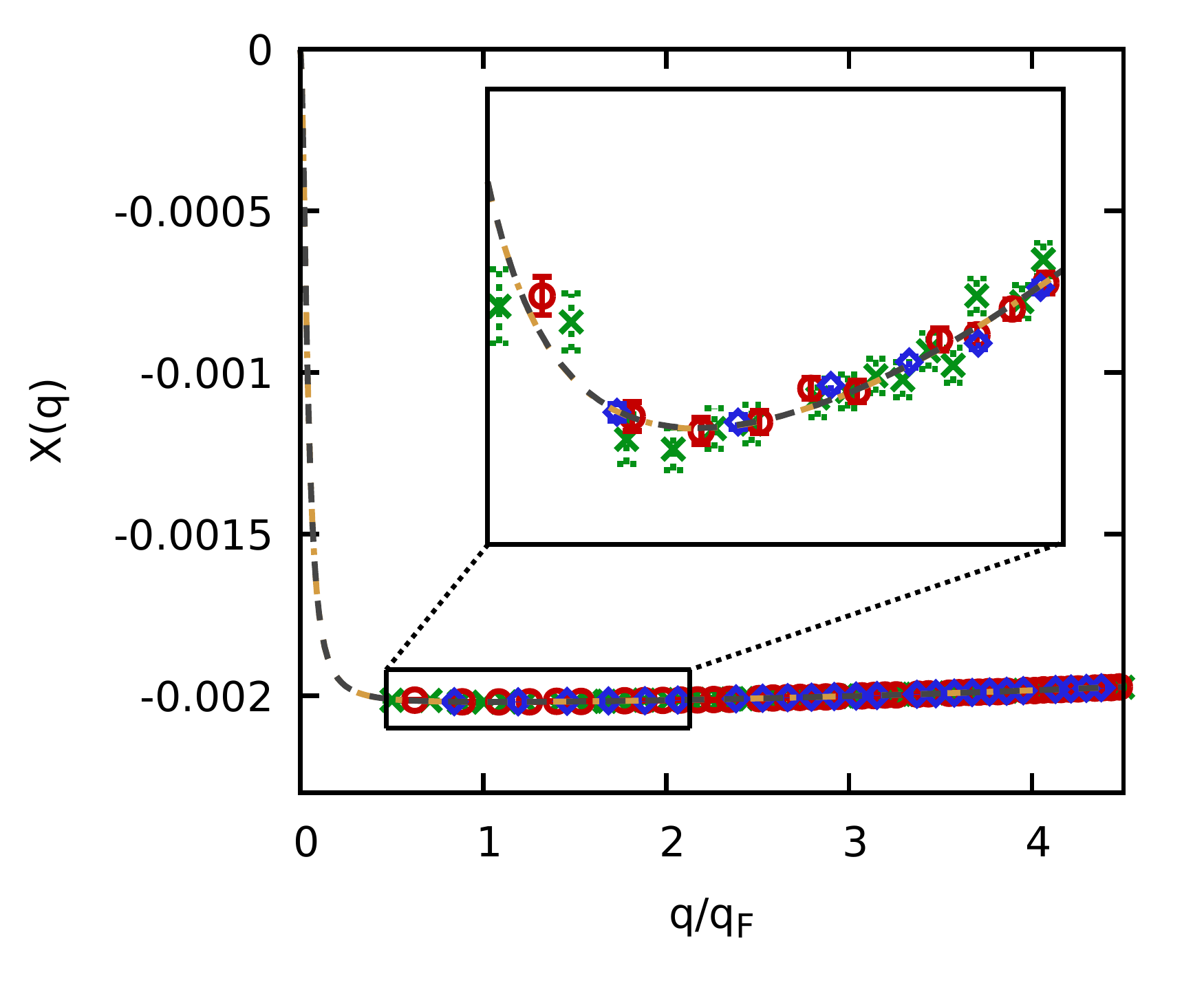}\\\vspace*{-1cm}\hspace*{-0.06\textwidth}\includegraphics[width=0.55\textwidth]{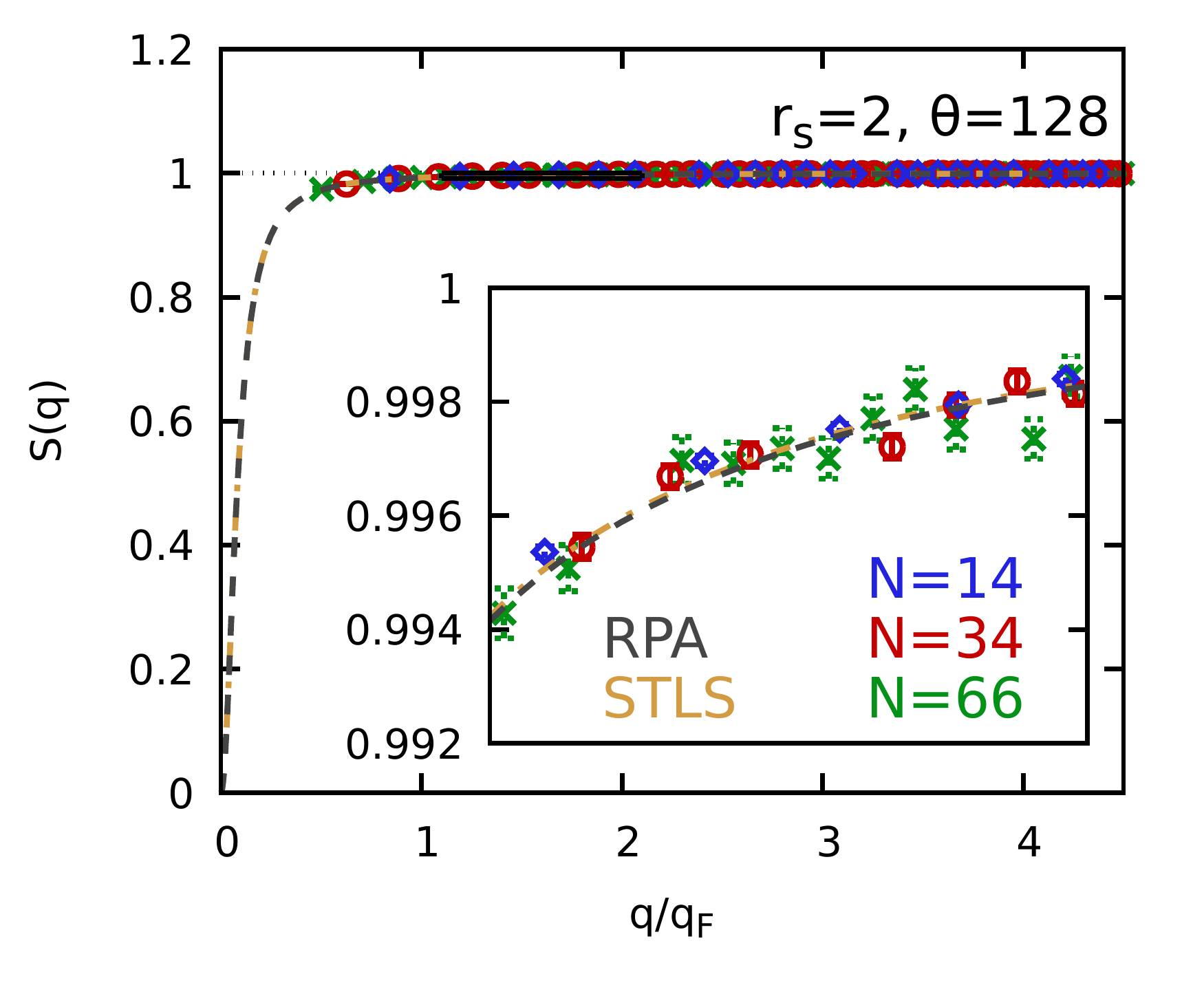}\includegraphics[width=0.55\textwidth]{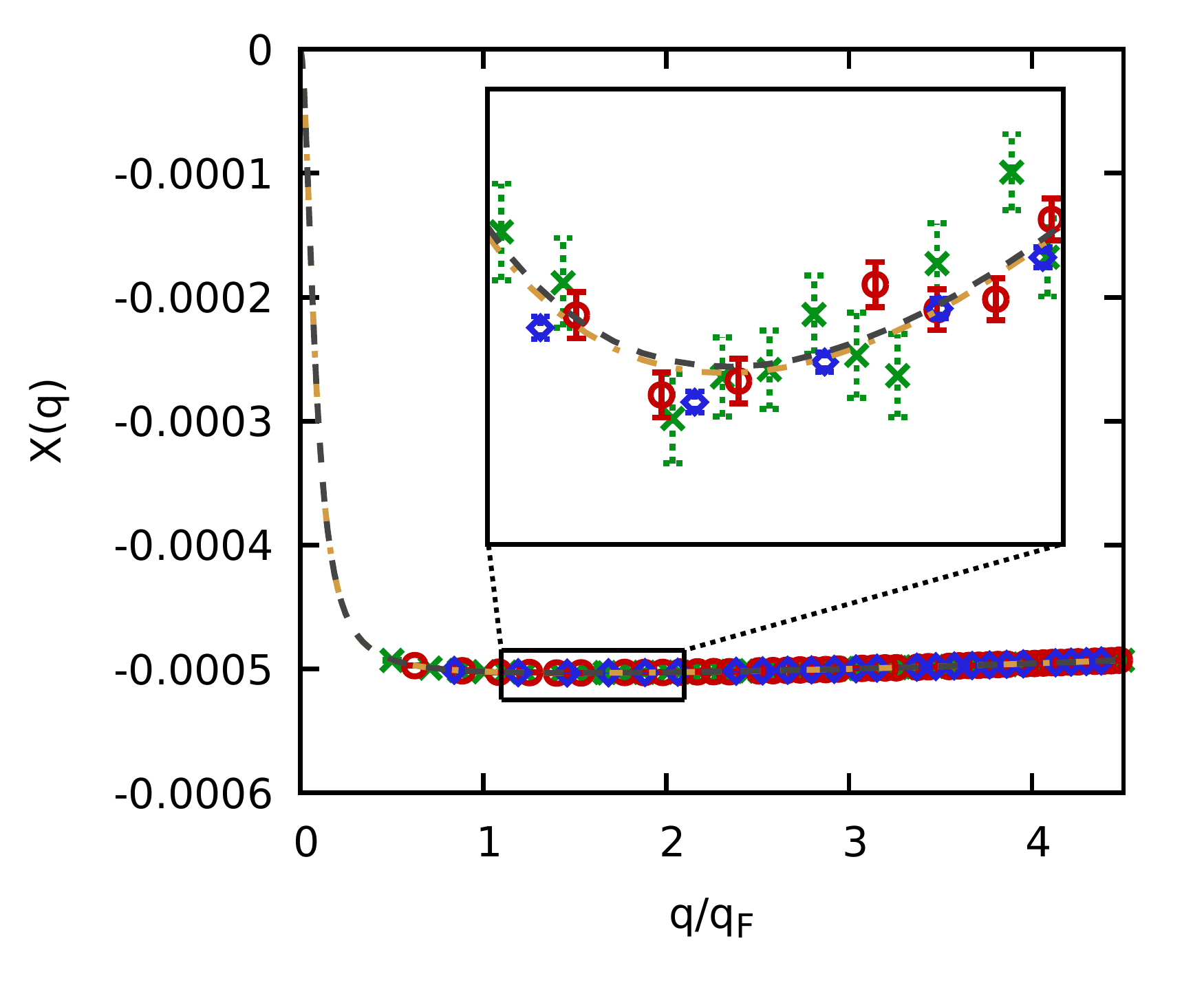}\\\vspace*{-1cm}\hspace*{-0.06\textwidth}\includegraphics[width=0.55\textwidth]{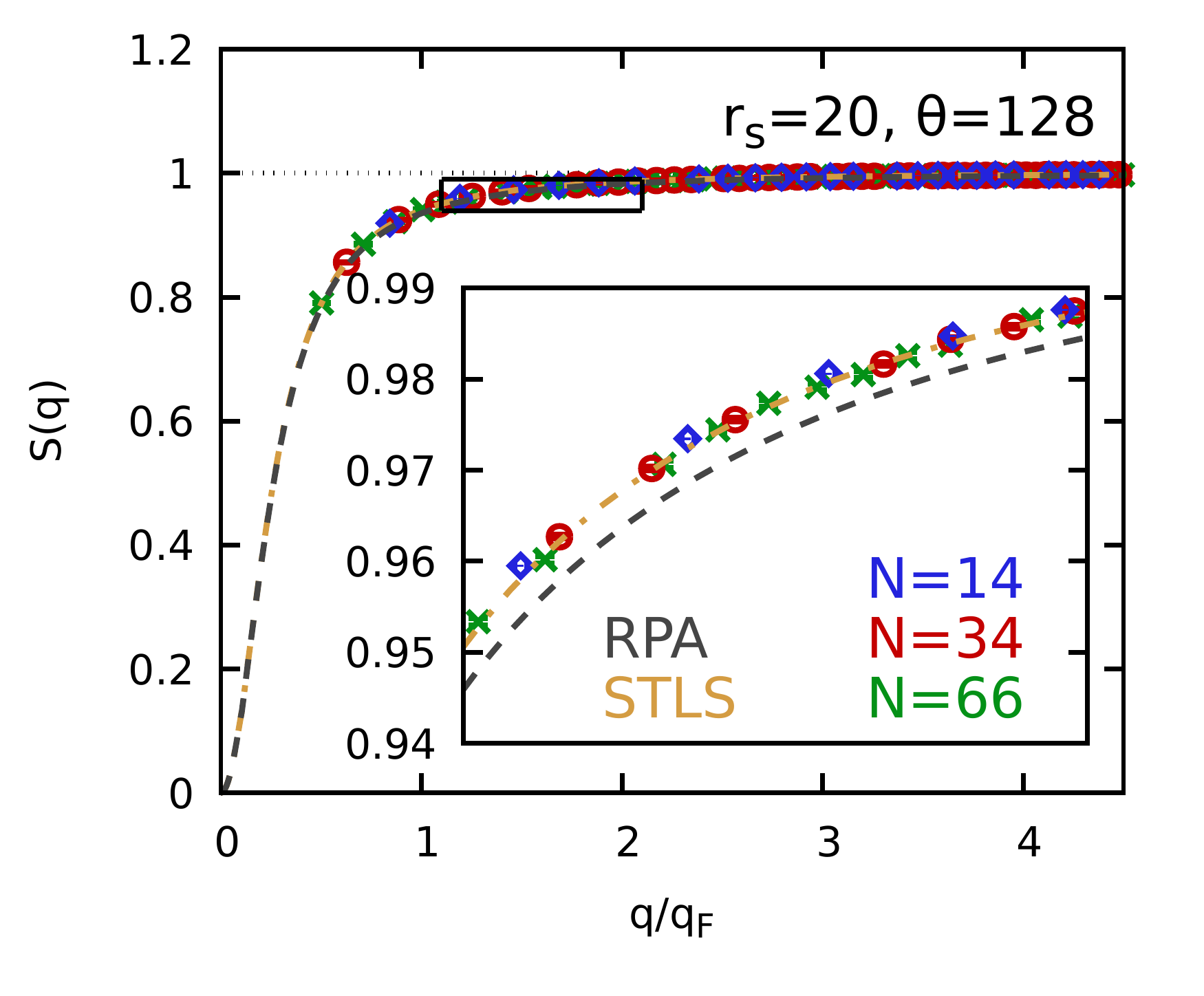}\includegraphics[width=0.55\textwidth]{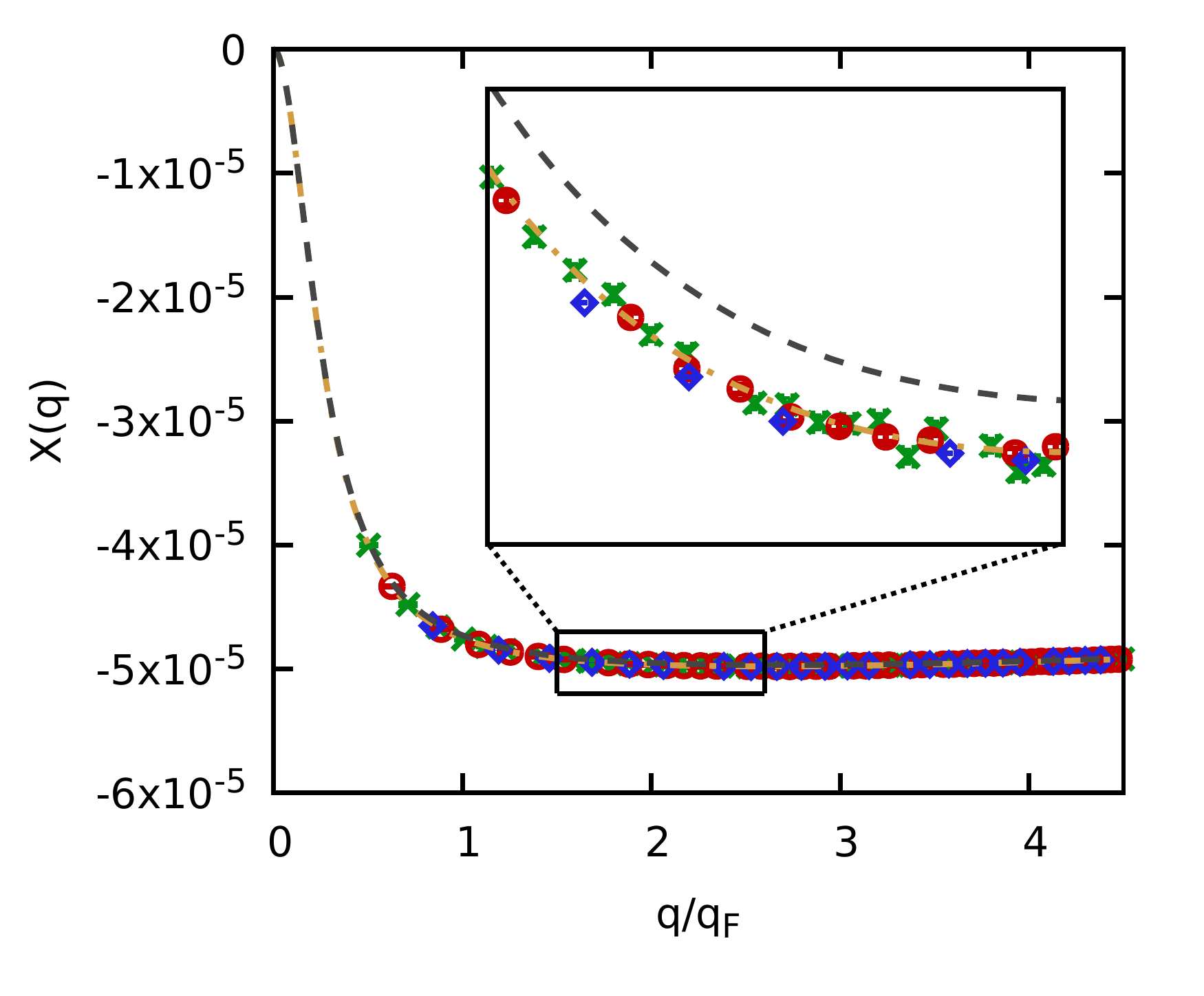}
\caption{\label{fig:rs2_theta128}
PIMC results for the static structure factor $S(q)$ [left] and the static density response function $\chi(q)$ [right] at $\theta=128$ and $r_s=0.5$ (top), $r_s=2$ (center), and $r_s=20$ (bottom) for $N=14$ (blue diamonds), $N=34$ (red circles), and $N=66$ (green crosses) unpolarized electrons. The dashed black and dash-dotted yellow curves have been computed within RPA and finite-$T$ STLS~\cite{stls2,stls}.
}
\end{figure}

Let us next proceed to Fig.~\ref{fig:rs2_theta128}, where we explore the static properties of the UEG at an extreme temperature, $\theta=128$. The top row corresponds to a high density, $r_s=0.5$. In this case, the static structure factor rapidly converges to unity with increasing wave number $q$, and only deviates in the small-$q$ limit due to the perfect screening in the UEG~\cite{kugler_bounds}. Further, RPA, STLS, and PIMC results are virtually indistinguishable for both $S(q)$ and the density response function $\chi(q)$ over the entire $q$-range.
In the center row, we show results for $r_s=2$, and observe the same qualitative trends. Here, too, electronic exchange--correlation effects as they are encoded in the static LFC have negligible impact. This changes only for $r_s=20$ (bottom row), where the stronger Coulomb repulsion leads to a noticeable deviation between RPA and STLS, with the latter being in excellent agreement to the PIMC data.

\begin{figure}\centering
\hspace*{-0.06\textwidth}\includegraphics[width=0.55\textwidth]{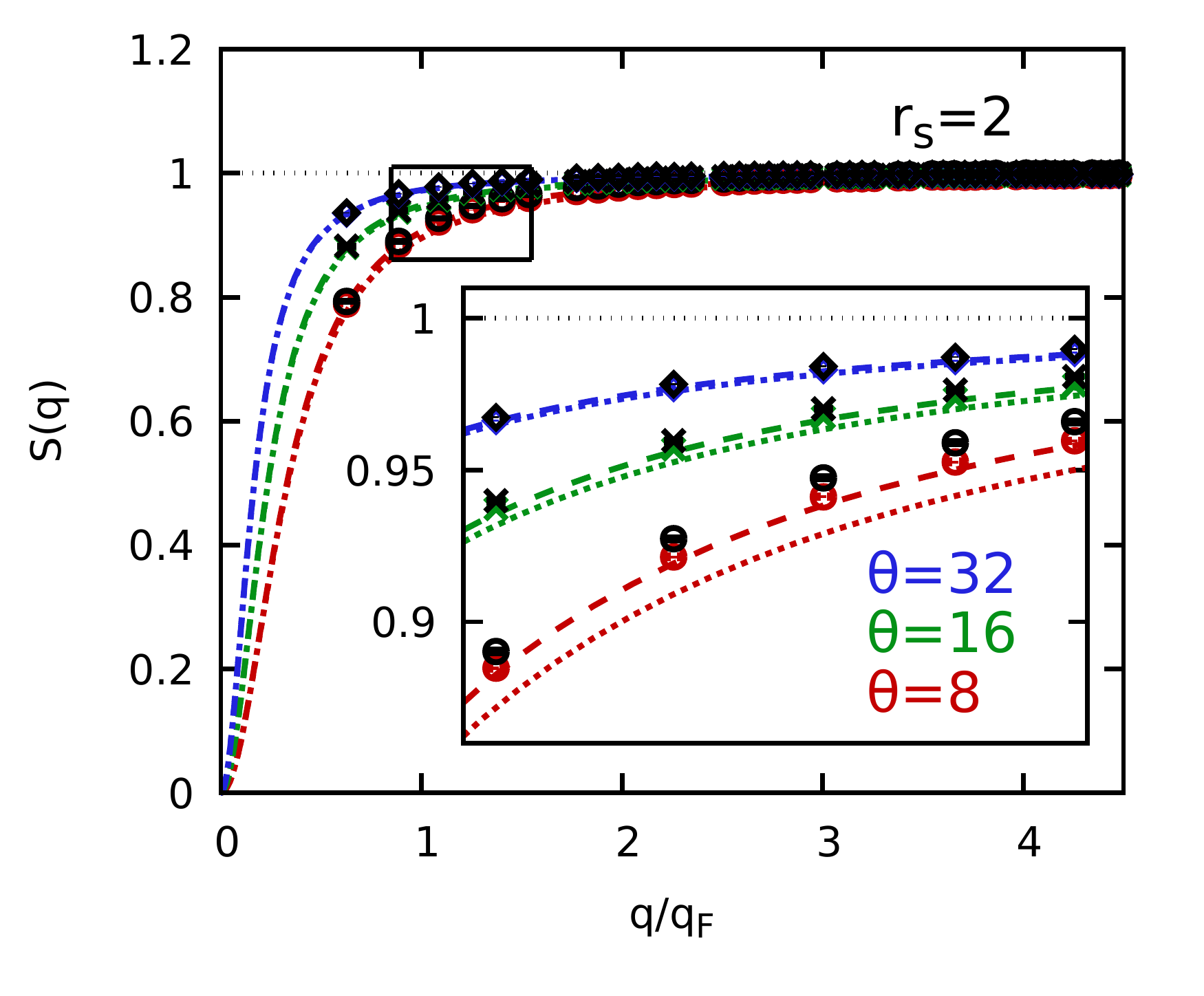}\includegraphics[width=0.55\textwidth]{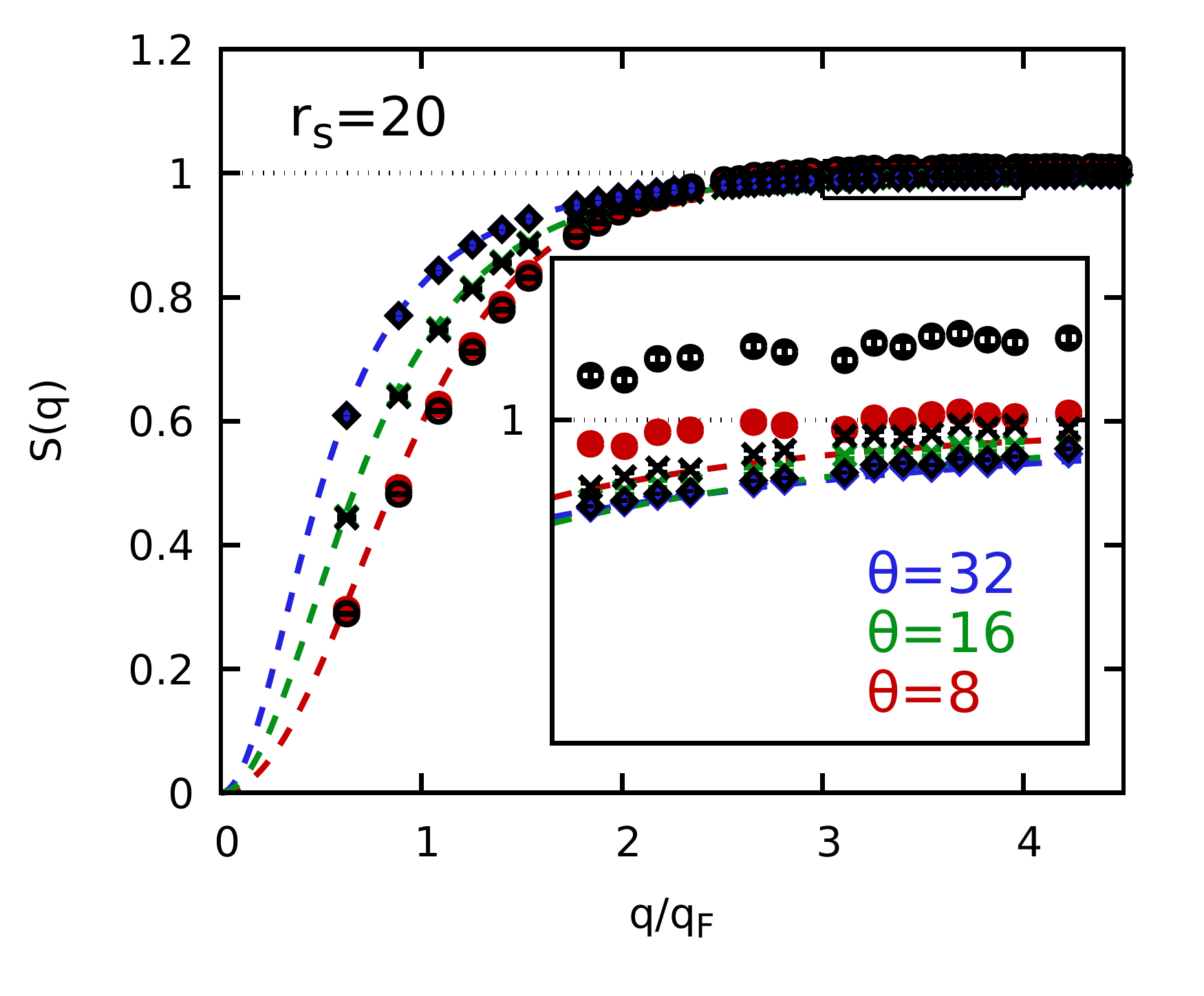}
\caption{\label{fig:rs2_classical}
Analysis of the classical relation between $S(q)$ and the static LFC $G(q)$, cf.~Eq.~(\ref{eq:S_classical})~\cite{ICHIMARU198791} for $r_s=2$ (left) and $r_s=20$ (right). The red, green, and blue curves distinguish results for $\theta=8$, $\theta=16$, and $\theta=32$, respectively, and the dotted and dashed lines depict results within RPA and STLS. The coloured [black] data points show direct PIMC results [have been obtained from $G(q)$ by evaluating Eq.~(\ref{eq:S_classical})].
}
\end{figure}

Let us next investigate the convergence of the structural properties of the UEG towards the classical limit, which can be done by analyzing the classical relation between $S(q)$ and $G(q)$ given in Eq.~(\ref{eq:S_classical}). This is done in Fig.~\ref{fig:rs2_classical}, where the left panel corresponds to $r_s=2$, and the red circles, green crosses, and blue diamonds show our PIMC results for $\theta=8$, $\theta=16$, and $\theta=32$. The dashed (dotted) curves of the same color show the corresponding results within STLS (RPA) and have been included as a reference. Finally, the dark symbols show our results for Eq.~(\ref{eq:S_classical}) using as input the LFC that has been obtained from our PIMC simulations by evaluating Eq.~(\ref{eq:G_static}). Evidently, the classical relation is rather accurate, and we find good quantitative agreement to the exact PIMC data, which becomes better when the temperature is increased.

The right panel of Fig.~\ref{fig:rs2_classical} shows the same information for $r_s=20$, and we find the same qualitative trend. Still, the relative deviations between PIMC and Eq.~(\ref{eq:S_classical}) are larger. This can be seen particularly well for $\theta=8$ in the inset showing a magnified segment.

\begin{figure}\centering
\hspace*{-0.06\textwidth}\includegraphics[width=0.55\textwidth]{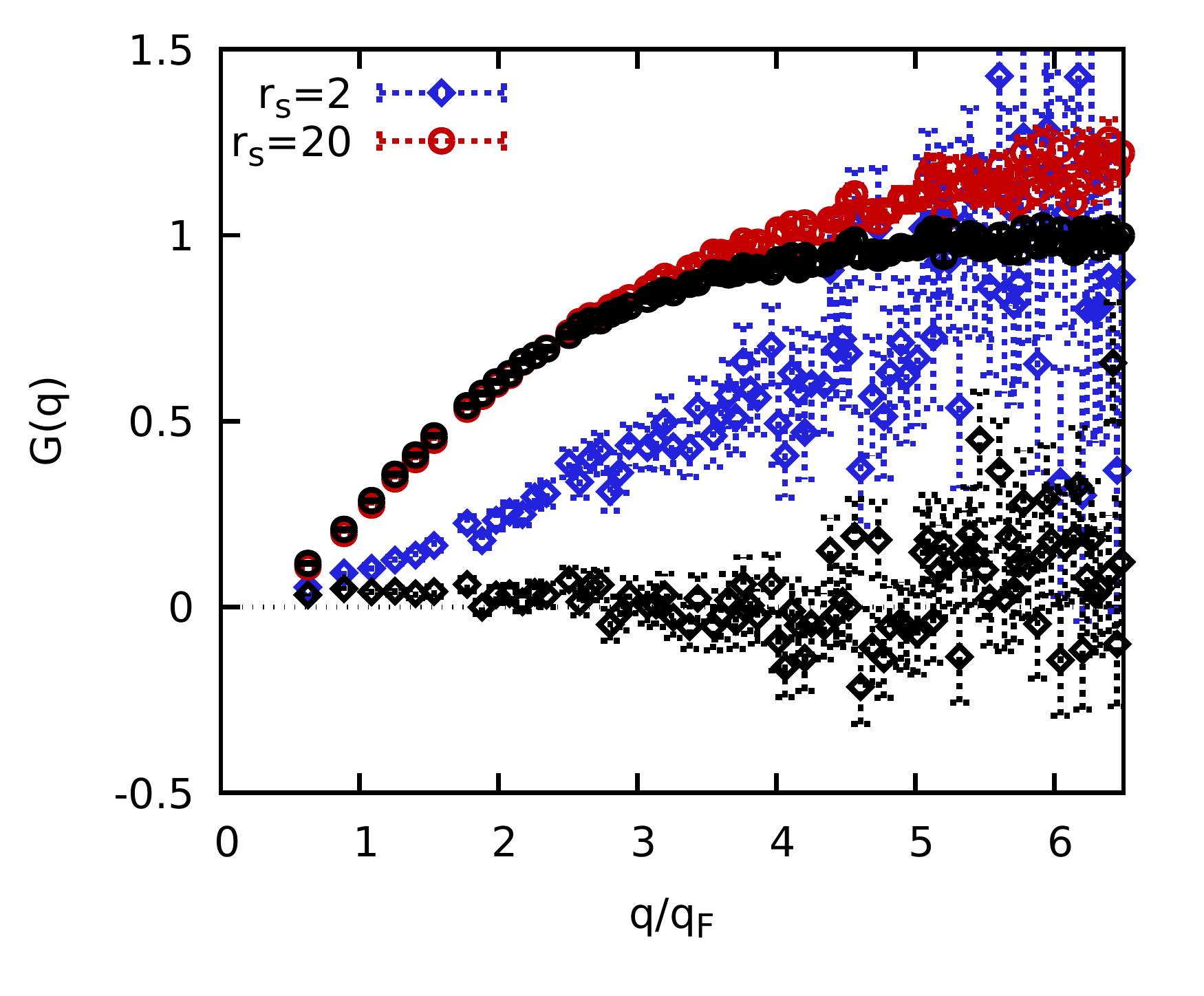}\includegraphics[width=0.55\textwidth]{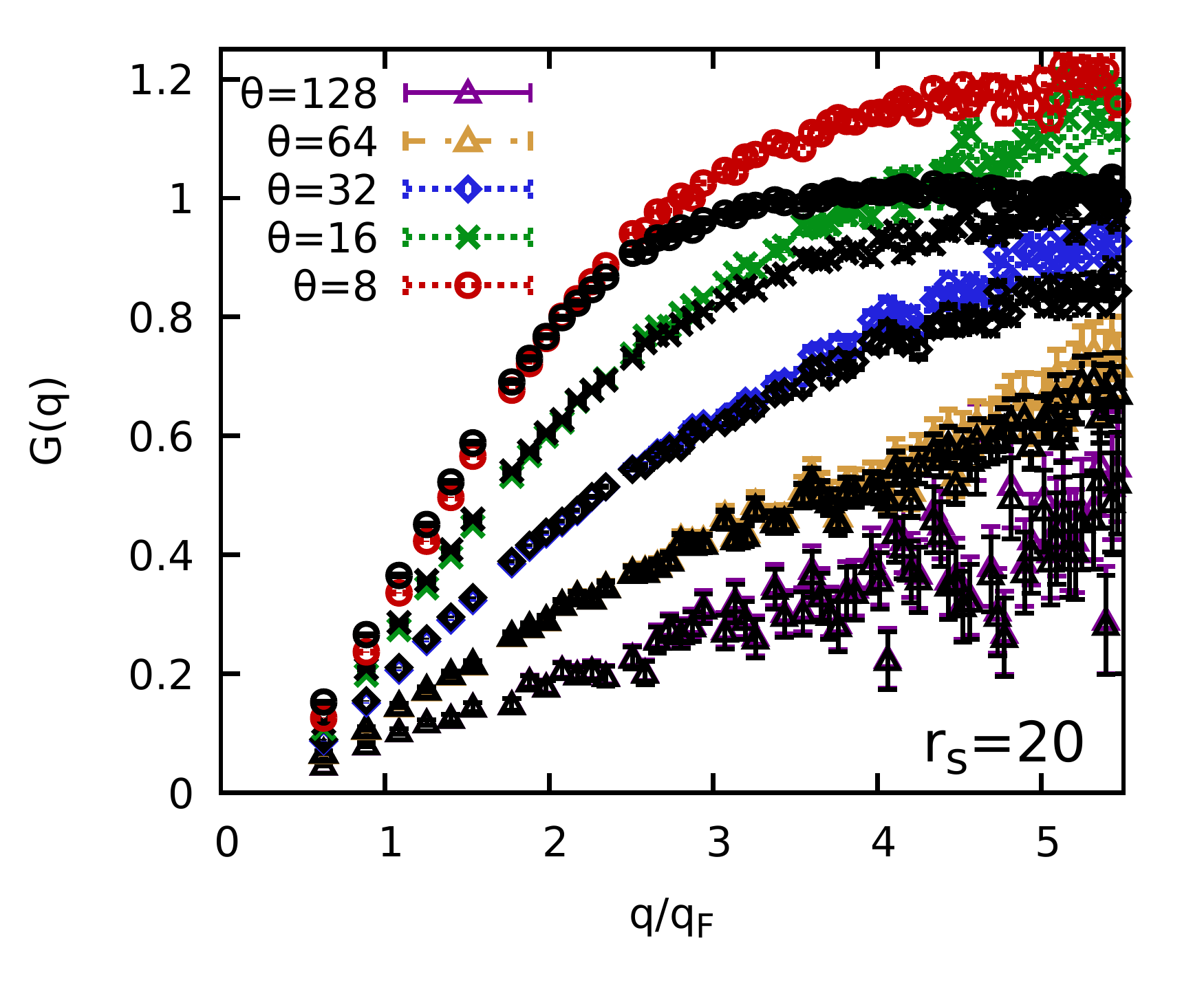}
\caption{\label{fig:LFC_rs2_classical}
Analysis of the classical relation~\cite{ICHIMARU198791} between $S(q)$ and the static LFC $G(q)$. Left: Static LFC for $\theta=16$, with the blue diamonds and red circles depicting direct PIMC results for $r_s=2$ and $r_s=20$; the corresponding black symbols have been obtained from $S(q)$ by evaluating Eq.~(\ref{eq:G_classical}). Right: Static LFC $G(q)$ at $r_s=20$ for different values of the degeneracy temperature $\theta$.
}
\end{figure}

At a first glance, these findings might seem to indicate that the static LFC $G(q)$ is indeed sufficient to describe the static structure factor at high temperatures. However, the fact of the matter is somewhat more subtle which can be seen in Fig.~\ref{fig:LFC_rs2_classical}, where we investigate the corresponding behaviour of the LFC itself. 
In particular, the left panel shows results for $\theta=16$ and the blue diamonds and red circles are our direct PIMC results [Eq.~(\ref{eq:G_static})] for $r_s=2$ and $r_s=20$, respectively. In addition, the corresponding dark symbols have been obtained by evaluating the classical expression for the LFC, Eq.~(\ref{eq:G_classical}), using as input the exact data for the static structure factor from PIMC. Evidently, we find good qualitative agreement in the case of $r_s=20$, as one might expect given the previously observed good agreement in $S(q)$, cf.~Fig.~\ref{fig:rs2_classical}. As a side note, we mention that the agreement between the two data sets deteriorates with increasing $q$, which can be understood as follows: large wave numbers directly correspond to small distances in coordinate space. Since every system exhibits quantum mechanical behaviour on a sufficiently small scale when the density and temperature are being kept fixed, it is intuitive that Eq.~(\ref{eq:G_classical}) will eventually fail in the large-$q$ limit where single-particle effects dominate. This can also be nicely seen in the right panel of Fig.~\ref{fig:LFC_rs2_classical}, where we show the LFC at $r_s=20$ for five different values of $\theta$. Here it becomes also clear that Eq.~(\ref{eq:G_classical}) becomes more accurate towards larger $T$, and that the aforementioned single-particle quantum behaviour starts only at larger $q$ in that case.

\begin{figure}\centering
\hspace*{-0.06\textwidth}\includegraphics[width=0.7\textwidth]{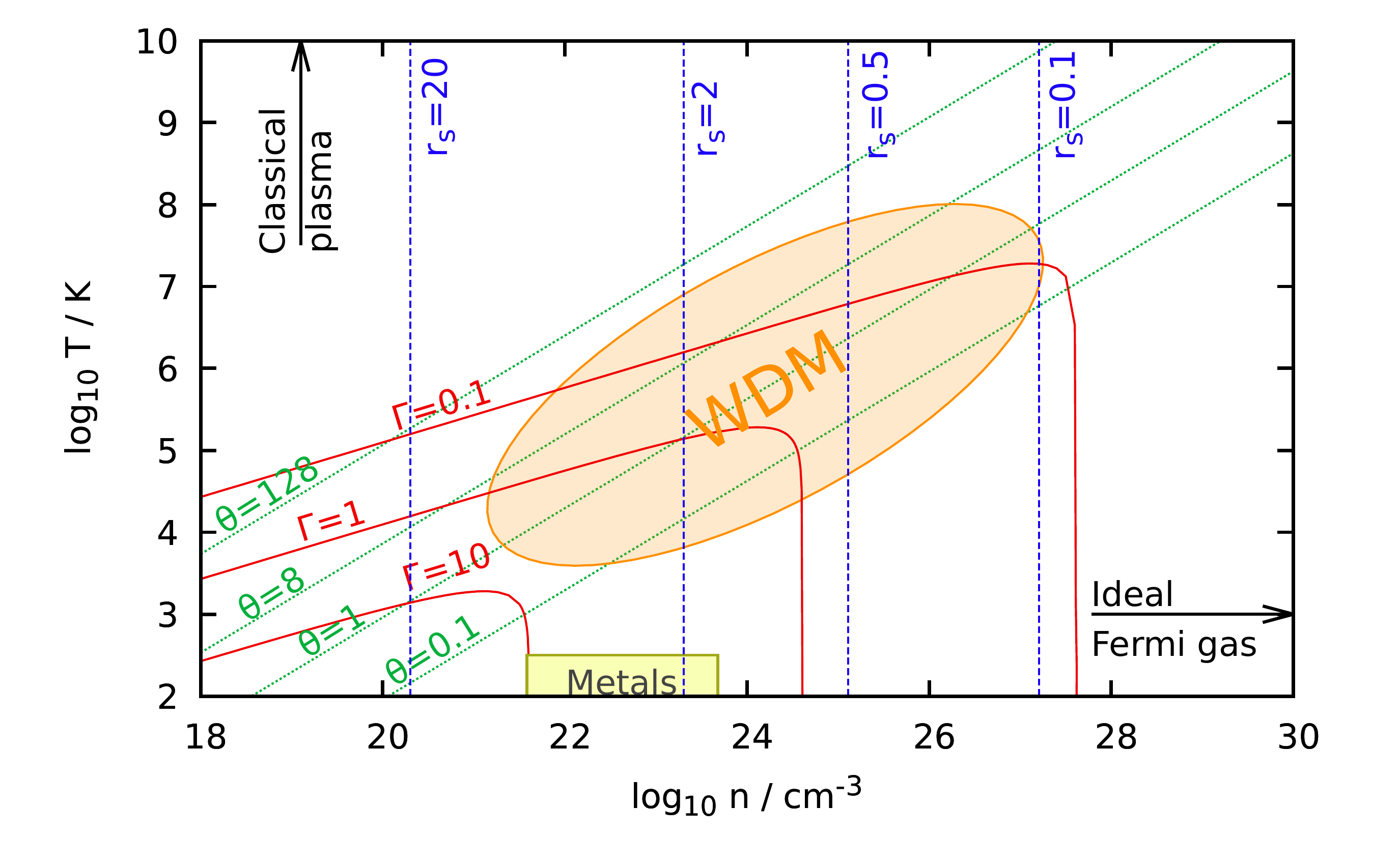}
\caption{\label{fig:over}
Density-temperature plain around the warm dense matter (WDM) regime (orange bubble). Adapted from Ref.~\cite{review}.
}
\end{figure}

Remarkably, this picture completely changes at $r_s=2$, where the two data sets for the static LFC starkly disagree for all $q$, both qualitatively and quantitatively. To understand this seemingly unintuitive behaviour, we might consider the different scaling of the characteristic parameters in the density-temperature plain shown in Fig.~\ref{fig:over}. Specifically, the dashed blue and dotted green line correspond to different fixed values of the density parameter $r_s$ and the degeneracy temperature $\theta$, respectively. In addition, we have also included the coupling parameter 
as solid red lines. Further, the orange shaded area around the center depicts the commonly assumed boundaries of the WDM regime~\cite{new_POP,wdm_book,review}.
Let us begin our analysis by considering $r_s=20$. Evidently, the system is strongly coupled at the Fermi temperature, where we find $\Gamma\approx10$. Upon increasing $T$ and $\theta$, the system reaches $\Gamma=1$ around $\theta=10$ and $\Gamma=0.1$ for $\theta\gtrsim100$. In other words, both coupling and degeneracy effects vanish towards large $T$, but the system becomes first non-degenerate and then ideal. Therefore, the classical relation Eq.~(\ref{eq:G_classical}) for the static local field correction gives meaningful results as electronic exchange--correlation effects substantially shape the physical behaviour of the system even at $\theta=128$.
We also note that when electrons are weakly degenerate and strongly correlated at the same time ( i.e., at $\theta>1$ and $\Gamma>1$),  
the repulsive Coulomb pair interaction does not allow electrons to approach each other close enough, preventing the manifestation of Pauli blocking (exchange interaction).

On the other hand, we might consider the case of high densities, $r_s=0.1$. In this regime, the system is very weakly coupled ($\Gamma<0.1$) even at $\theta=1$, where quantum effects are still highly important. Thus, the system will first reach the ideal quantum gas limit upon increasing $T$, whereas quantum effects only start to become negligible at even higher temperatures. In that case, the classical relation Eq.~(\ref{eq:S_classical}) will eventually become meaningful. Yet, this has nothing to do with the static LFC $G(q)$, which has a negligible impact in this regime. As a consequence, the classical relation for the LFC, Eq.~(\ref{eq:G_classical}), does not give access to the (absent) electronic exchange--correlation effects, but approximately vanishes within the given level of uncertainty.

The regime of metallic densities, which includes the example of $r_s=2$ shown above, falls between these two extremes, but is physically closer to the high-density case. Therefore, Eq.~(\ref{eq:G_classical}) breaks down (cf.~Fig.~\ref{fig:LFC_rs2_classical}), whereas the static structure factor is still accurate.

\subsection{Potential energy\label{sec:potential}}

\begin{figure}\centering
\hspace*{-0.06\textwidth}\includegraphics[width=0.55\textwidth]{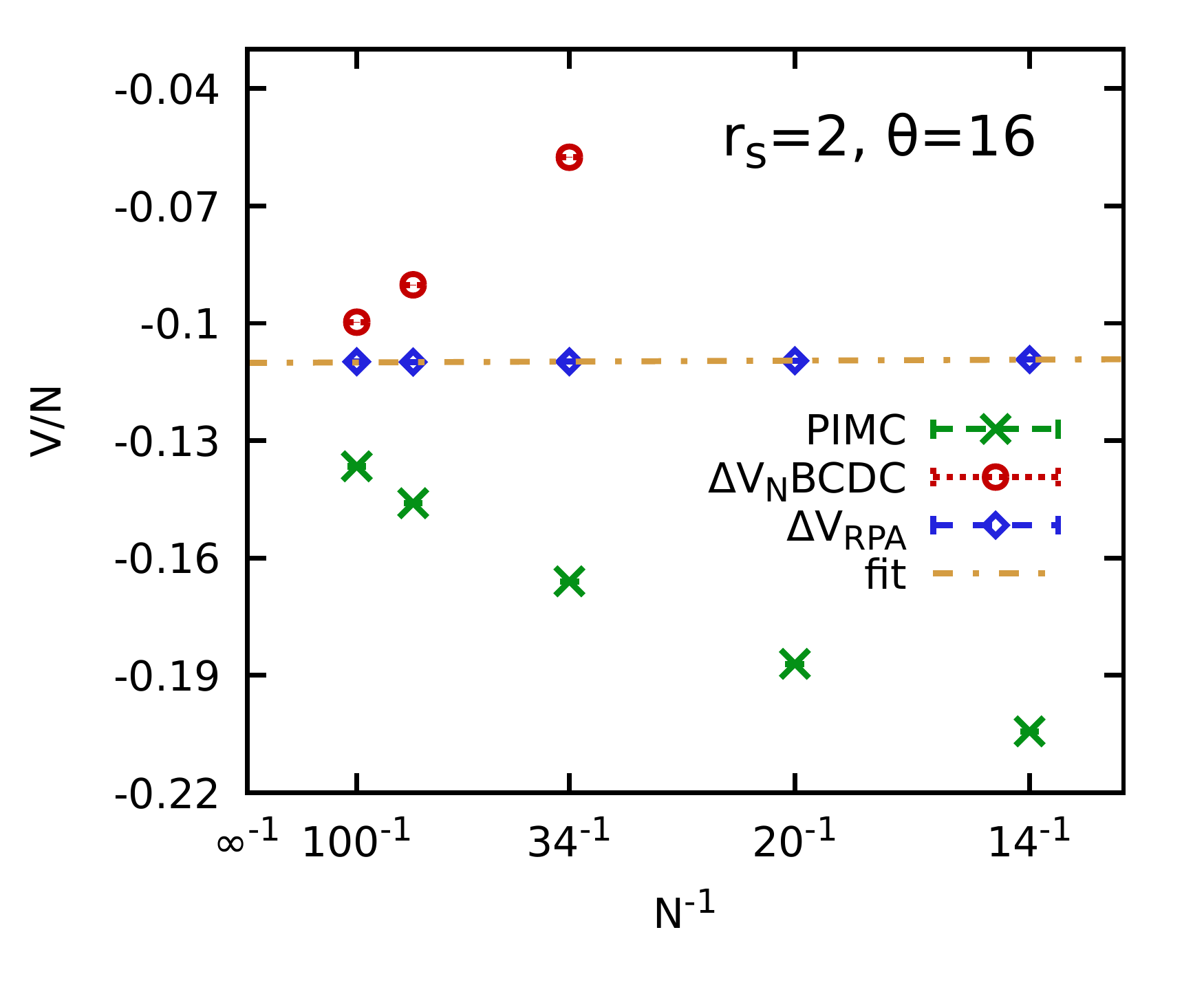}\includegraphics[width=0.55\textwidth]{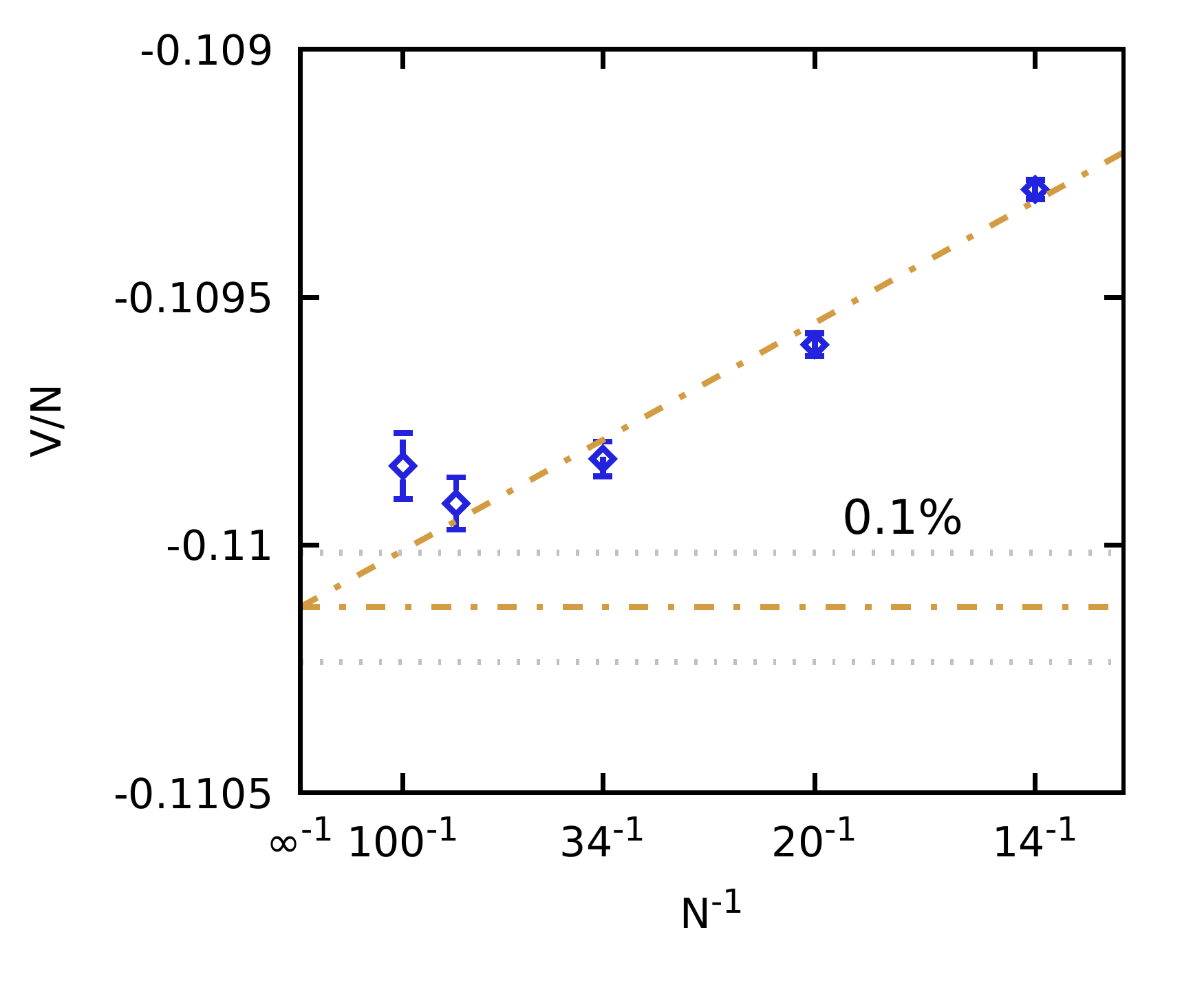}
\\\vspace*{-0.82cm}\hspace*{-0.06\textwidth}\includegraphics[width=0.55\textwidth]{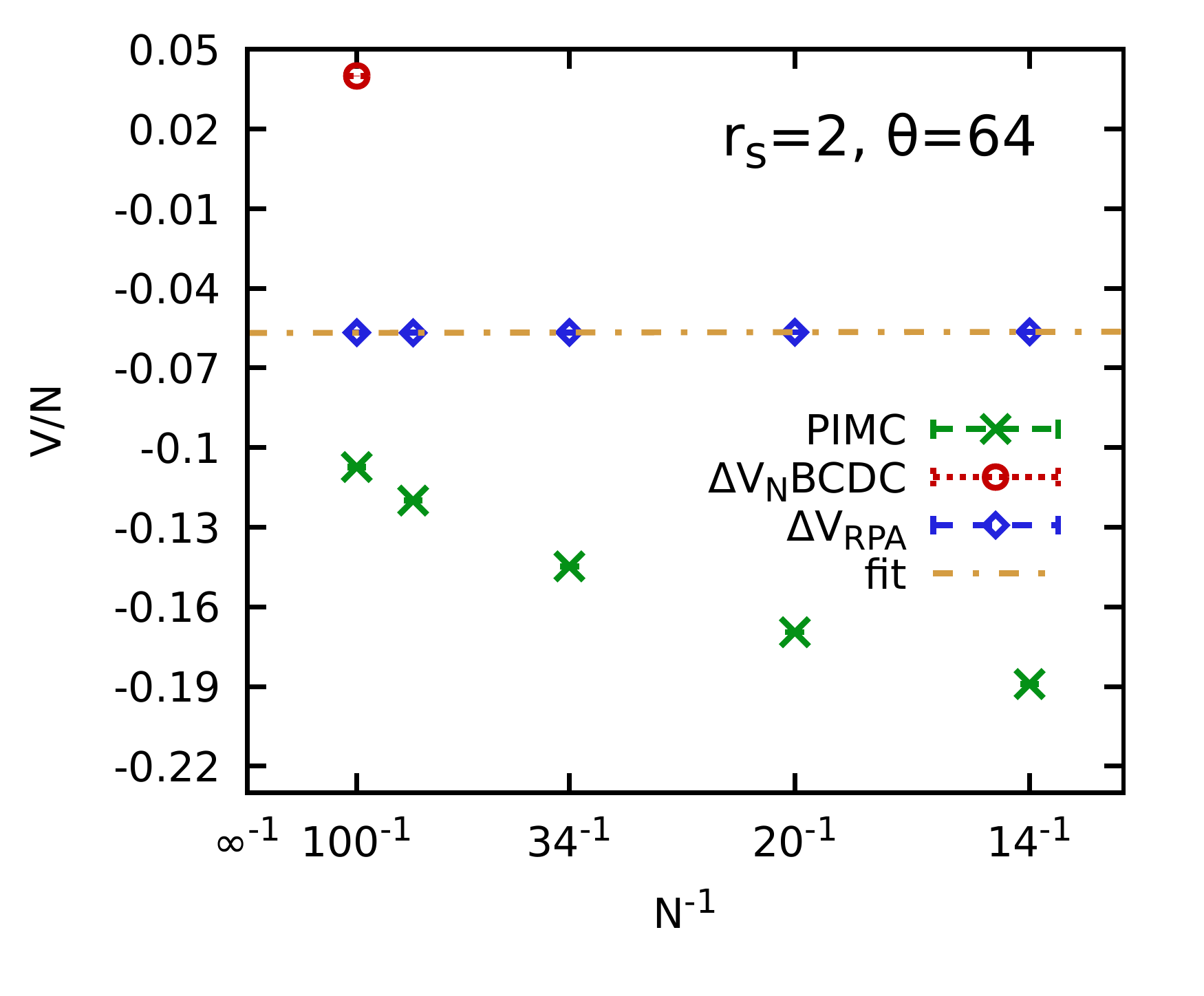}\includegraphics[width=0.55\textwidth]{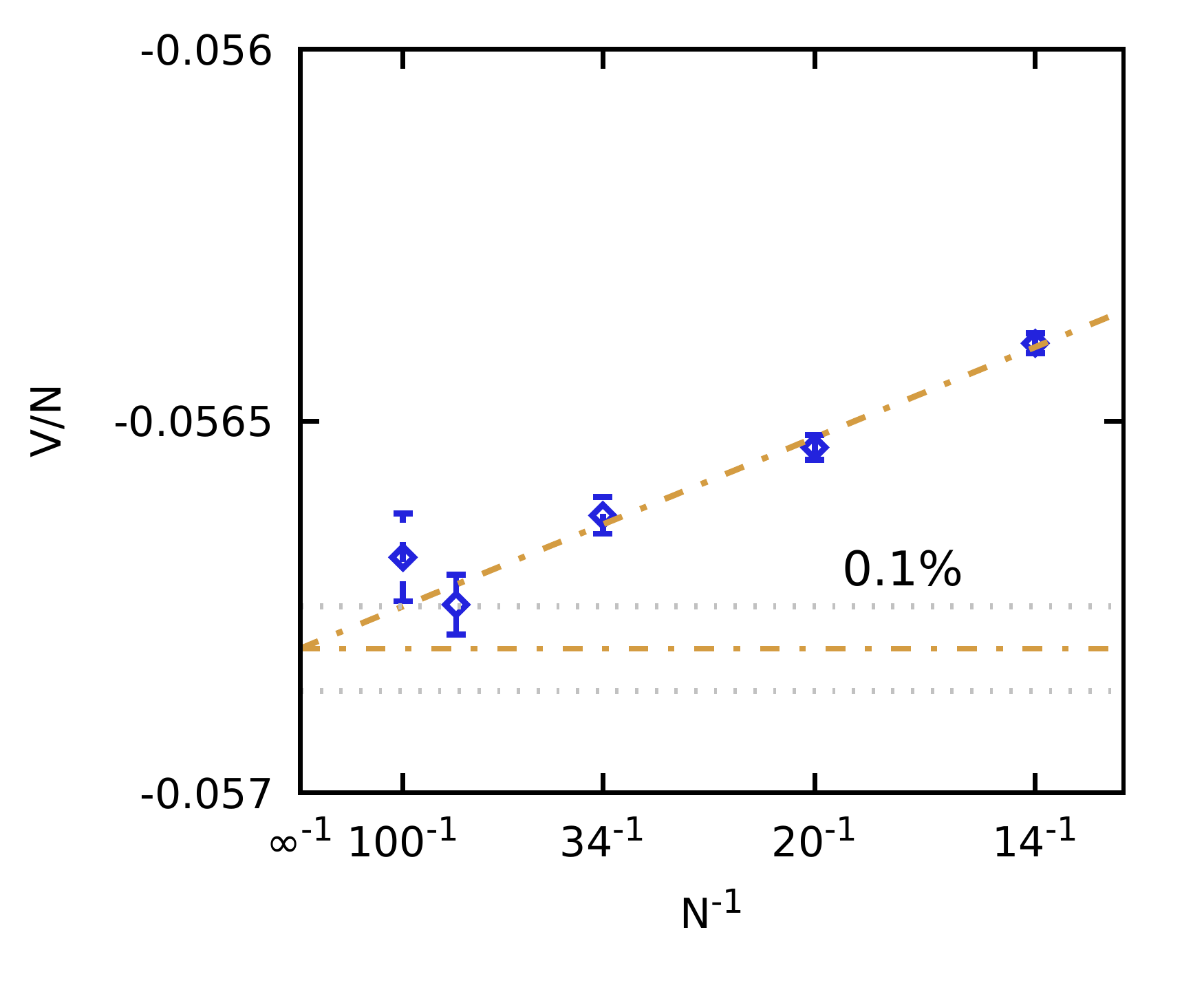}\\\vspace*{-0.82cm}\hspace*{-0.06\textwidth}\includegraphics[width=0.55\textwidth]{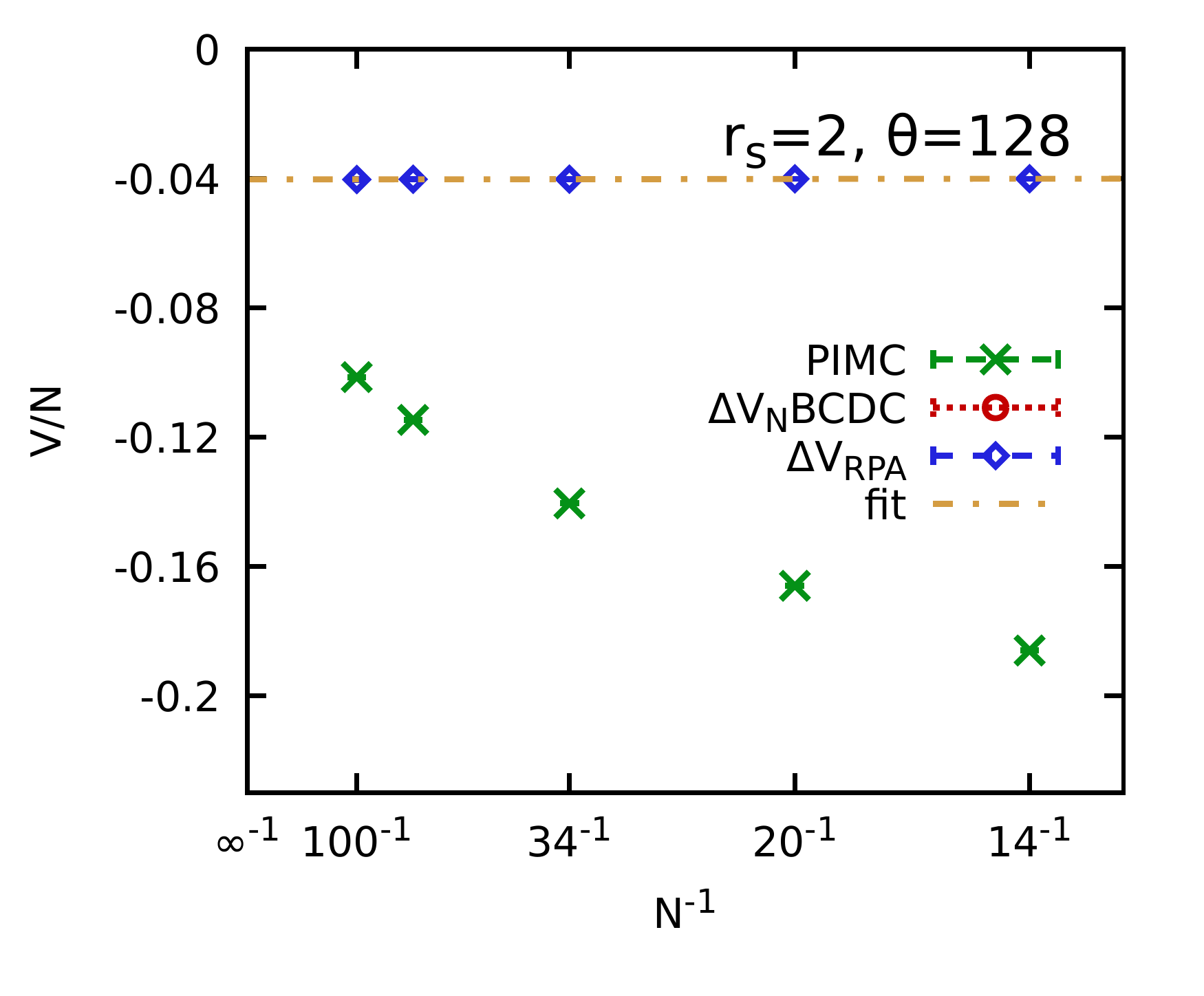}\includegraphics[width=0.55\textwidth]{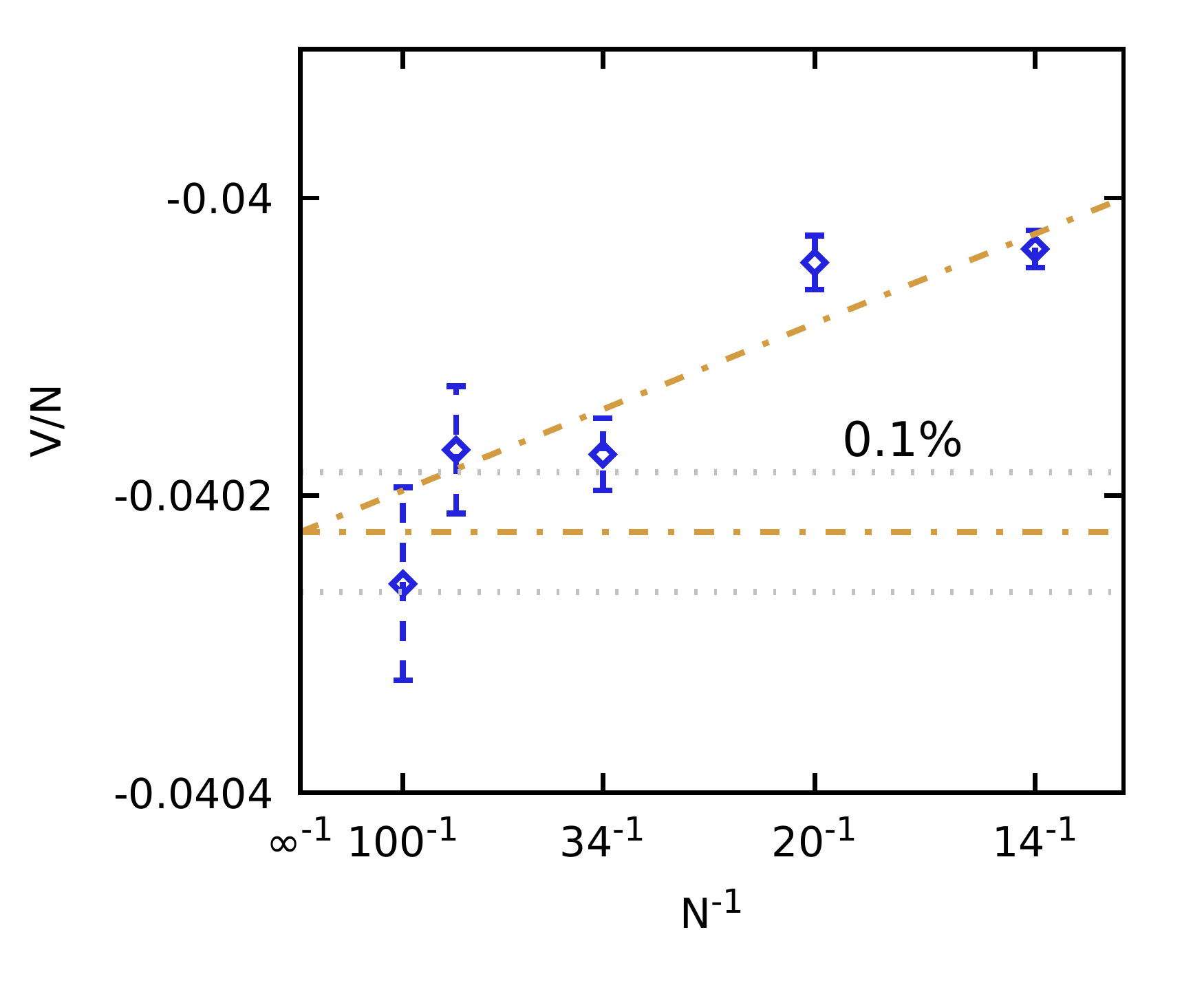}
\caption{\label{fig:Pot_rs2_N}
PIMC results for the dependence of the interaction energy per particle $V/N$ on the system-size $N$ for $r_s=2$ and $\theta=16$ (top), $\theta=64$ (center), and $\theta=128$ (bottom). The green crosses show our direct PIMC results for finite $N$ and the red circles have been obtained by adding onto those data the first-order correction by Brown \emph{et al.}~\cite{Brown_PRL_2013} [BCDC, Eq.~(\ref{eq:BCDC})]. The blue circles correspond to the full estimation of the discretization error, Eq.~(\ref{eq:FSC_v_d}), within RPA, and the dash-dotted yellow curves to a simple linear fit.
}
\end{figure}

A further interesting research topic regarding the UEG in the high-temperature regime is the behaviour of the potential energy $v$.
In particular, the adiabatic connection formula~\cite{review,groth_prl,ksdt}
\begin{eqnarray}\label{eq:adiabatic}
f_\textnormal{xc}(r_s,\theta) = \frac{1}{r_s^2} \int_0^{r_s} \textnormal{d}\overline{r}_s\ v(\overline{r}_s,\theta)\overline{r}_s\ ,
\end{eqnarray}
relates $v$ to the exchange--correlation free energy $f_\textnormal{xc}$ of the system, from which, in principle all thermodynamic properties can be derived. Moreover, as PIMC simulations do not give direct access to $f_\textnormal{xc}$ itself, Eq.~(\ref{eq:adiabatic}) constitutes the basis for the construction of representations thereof with respect to density and temperature, $f_\textnormal{xc}(r_s,\theta)$~\cite{stls2,tanaka_hnc,groth_prl,review,ksdt}. 
On the other hand, any representation of $f_\textnormal{xc}(r_s,\theta)$ can be used to estimate the potential energy via the relation
\begin{eqnarray}\label{eq:v_from_fxc}
v(r_s,\theta) = 2f_\textnormal{xc}(r_s,\theta) + r_s \left. \frac{\partial f_\textnormal{xc}(r_s,\theta)}{\partial r_s}\right|_\theta\ .
\end{eqnarray}
Of particular interest is the comparison to the representation by Groth \emph{et al.}~\cite{groth_prl}, which is based on highly accurate Configuration PIMC (CPIMC)~\cite{groth_prb_2016,Schoof_PRL_2015} and Permutation blocking PIMC (PB-PIMC)~\cite{Dornheim_NJP_2015,dornheim_jcp,Dornheim_CPP_2019} data in the temperature range of $0\lesssim\theta\leq8$. As a first step, we have to perform an extrapolation to the thermodynamic limit, which is presented in Fig.~\ref{fig:Pot_rs2_N} for $r_s=2$ and three different values of the degeneracy temperature $\theta$.
More specifically, the top row shows results for $\theta=16$, and the green crosses depict our raw PIMC data without any finite-size correction. Evidently, the data points exhibit a pronounced dependence on the system size, and the application of a suitable finite-size correction is mandatory. The red circles have been obtained by adding to the PIMC data the first-order correction from Eq.~(\ref{eq:BCDC}) by Brown \emph{et al.}~\cite{Brown_PRL_2013} (BCDC). Yet, this correction is not sufficient to remove the dependence on $N$, and still no direct extrapolation to $N\to\infty$ is possible. To overcome this obstacle, we numerically evaluate the full discretization error, Eq.~(\ref{eq:FSC_v_d}), using as a trial function the static structure factor within RPA. We note that this is a reasonable choice, as the RPA closely resembles the exact PIMC data for $S(q)$ for all parameters considered in this work, see Sec.~\ref{sec:static} above.
The results are shown as the blue diamonds in Fig.~\ref{fig:Pot_rs2_N}. Evidently, the system-size dependence has been reduced by at least two orders of magnitude, and no dependence can be resolved on this scale with the naked eye.
The right panel shows a magnified segment around the thus finite-size corrected points, and the dash-dotted yellow curve has been obtained from a simple linear fit over all corrected data points. We find that the residual error $\Delta v^N_\textnormal{i}$ is of the order of $\Delta v/v\sim0.1\%$ and can be reliably removed following the linear extrapolation. 

The center and bottom rows of Fig.~\ref{fig:Pot_rs2_N} show the same information for higher temperatures, $\theta=64$ and $\theta=128$, and we find the same qualitative behaviour. The two main differences to the case of $\theta=16$ are 1) the relative increase in the discretization error $\Delta v_\textnormal{d}^N$ and 2) the decrease of the intrinsic error $\Delta v^N_\textnormal{i}$.
We note that similar types of analysis have been performed for $r_s=0.5$ and $r_s=20$; all extrapolated data are freely available online~\cite{repo}.



\begin{figure}\centering
\hspace*{-0.06\textwidth}\includegraphics[width=0.55\textwidth]{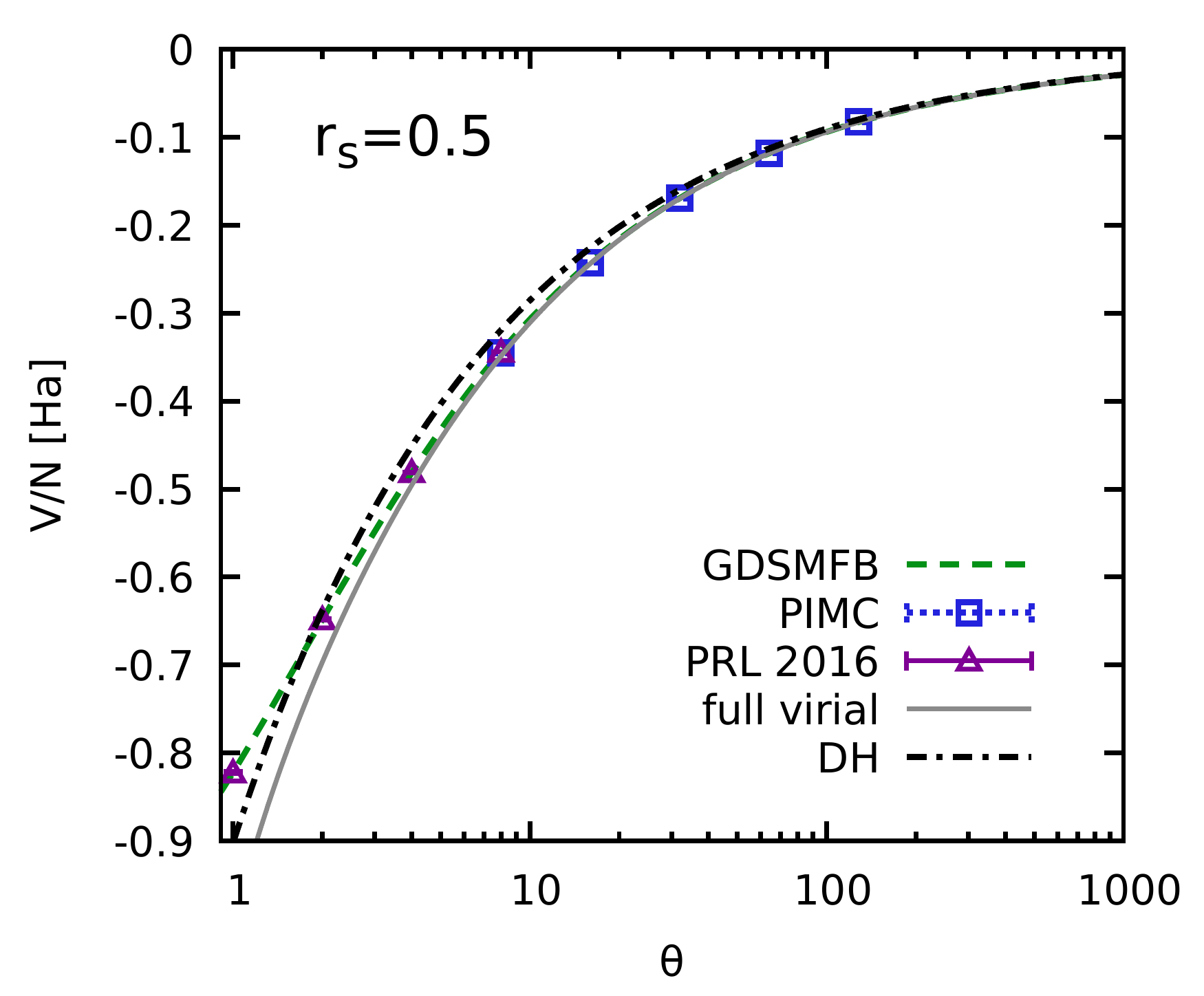}\includegraphics[width=0.55\textwidth]{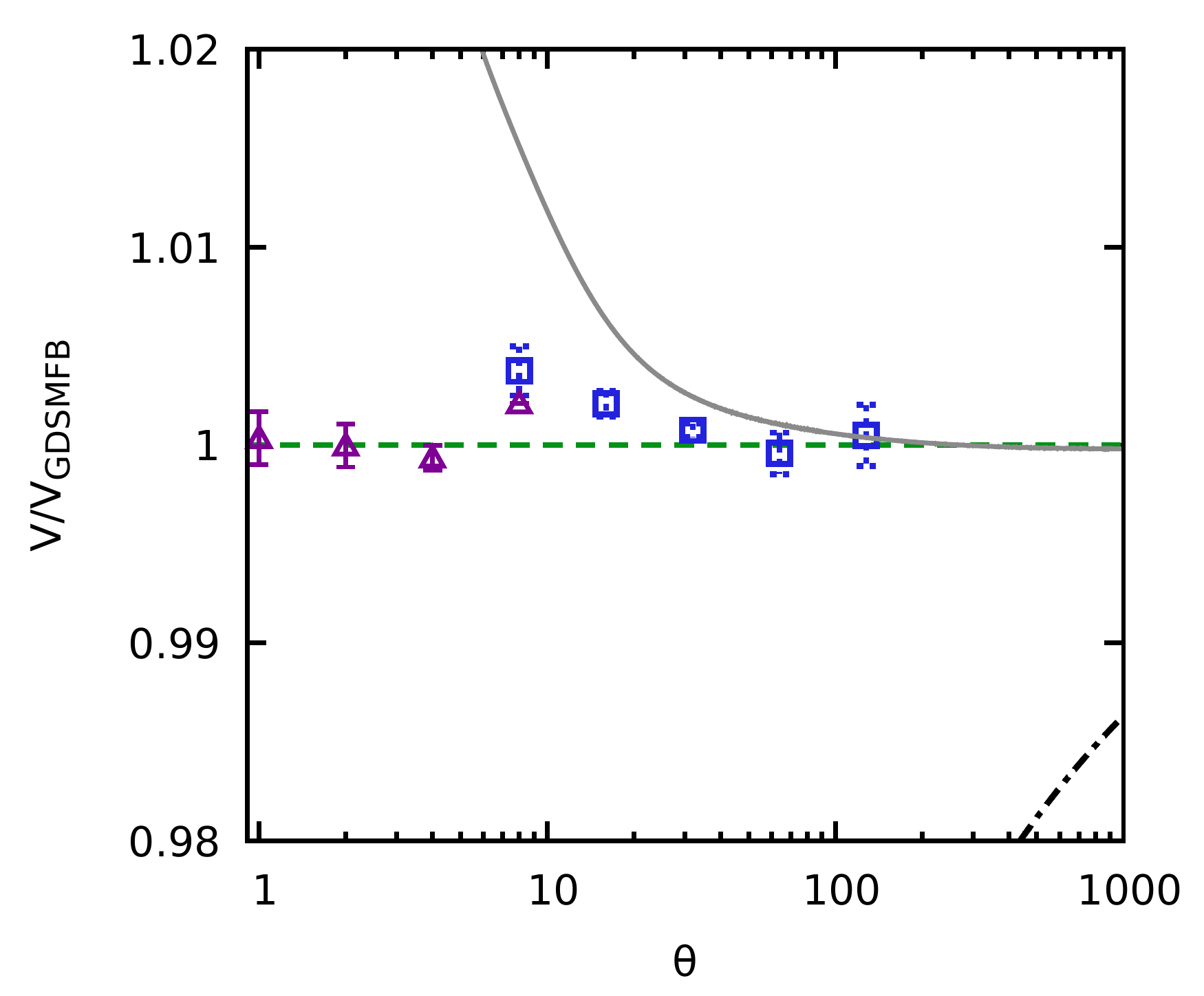}\\\hspace*{-0.06\textwidth}\includegraphics[width=0.55\textwidth]{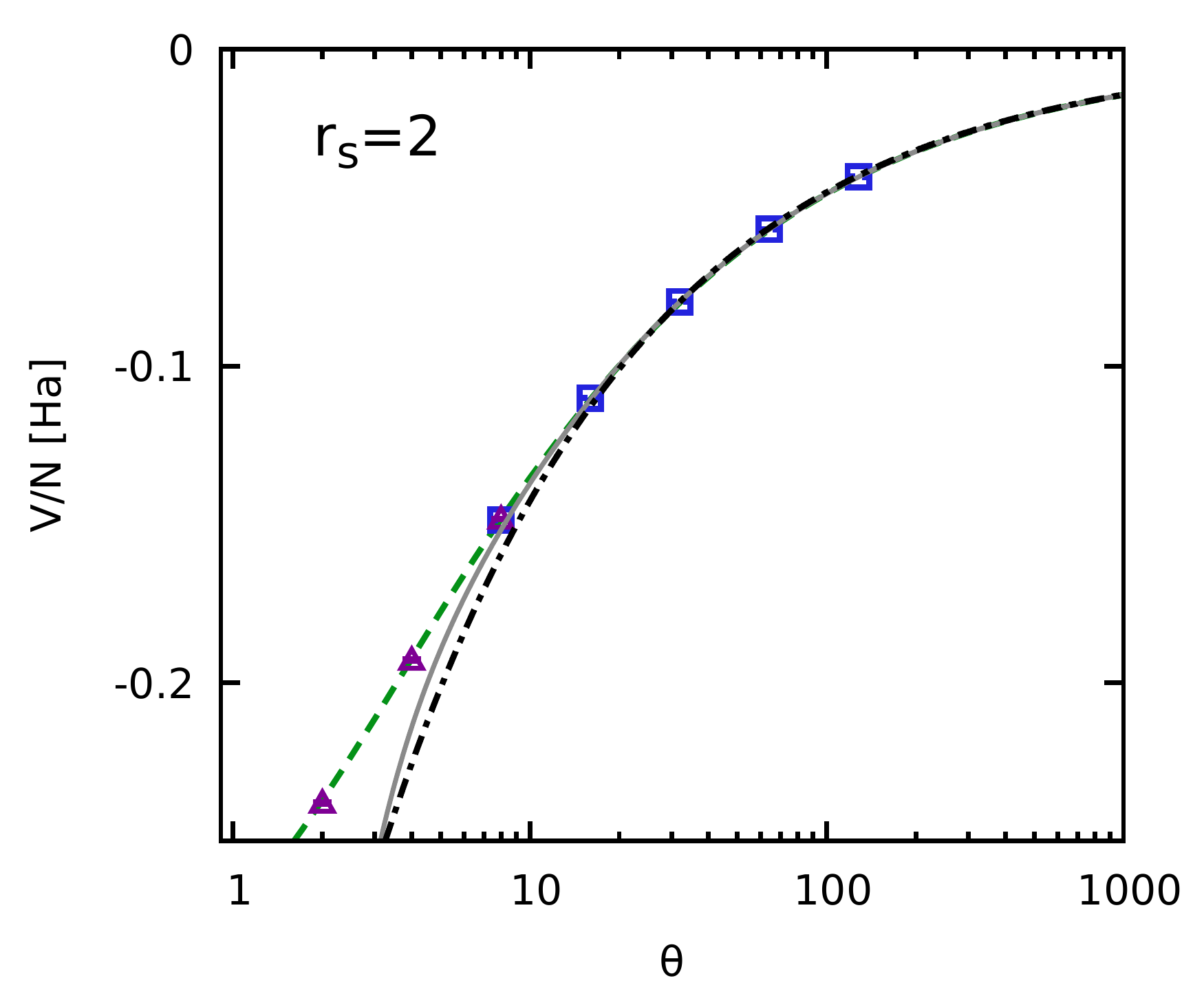}\includegraphics[width=0.55\textwidth]{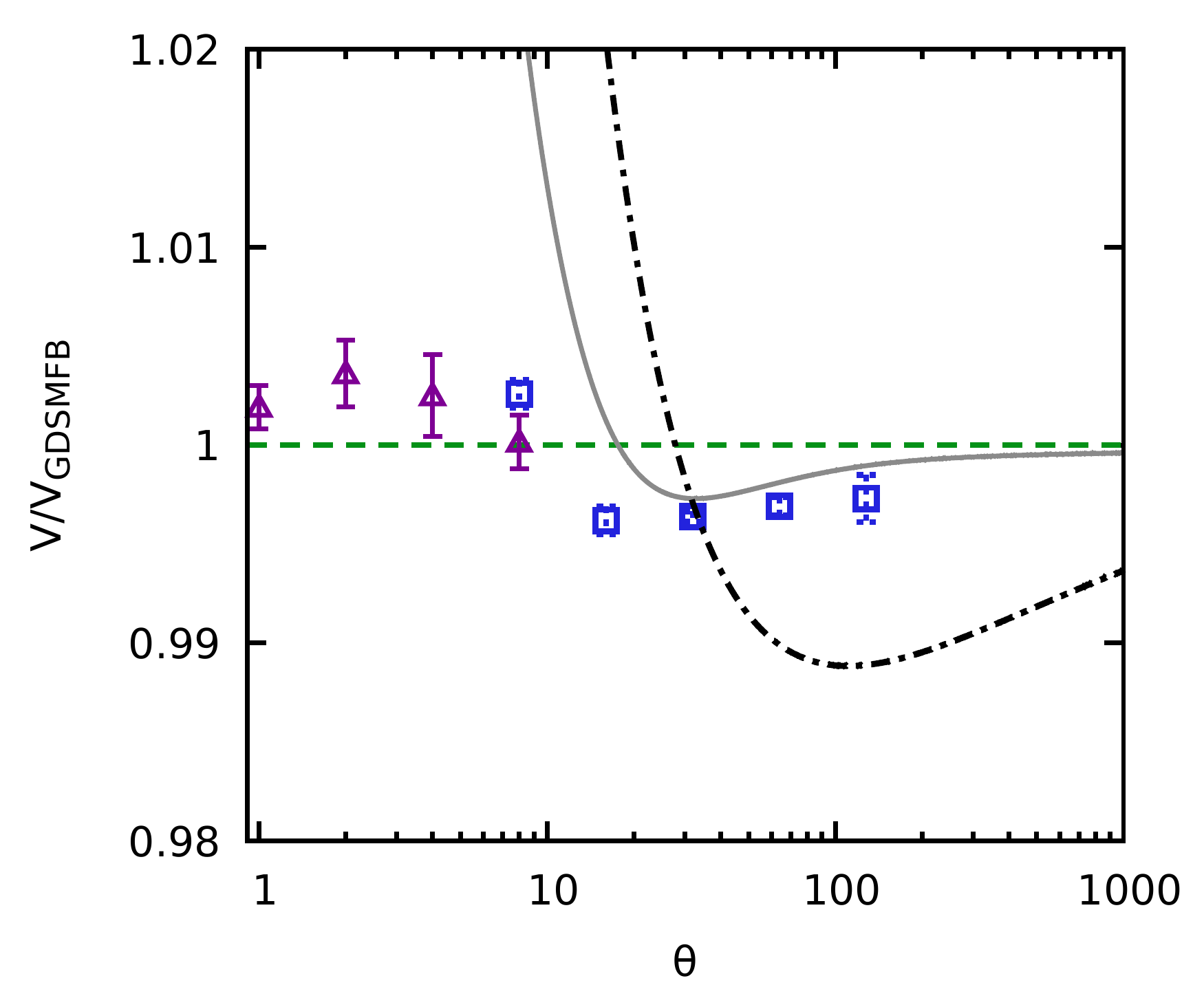}\\\hspace*{-0.06\textwidth}\includegraphics[width=0.55\textwidth]{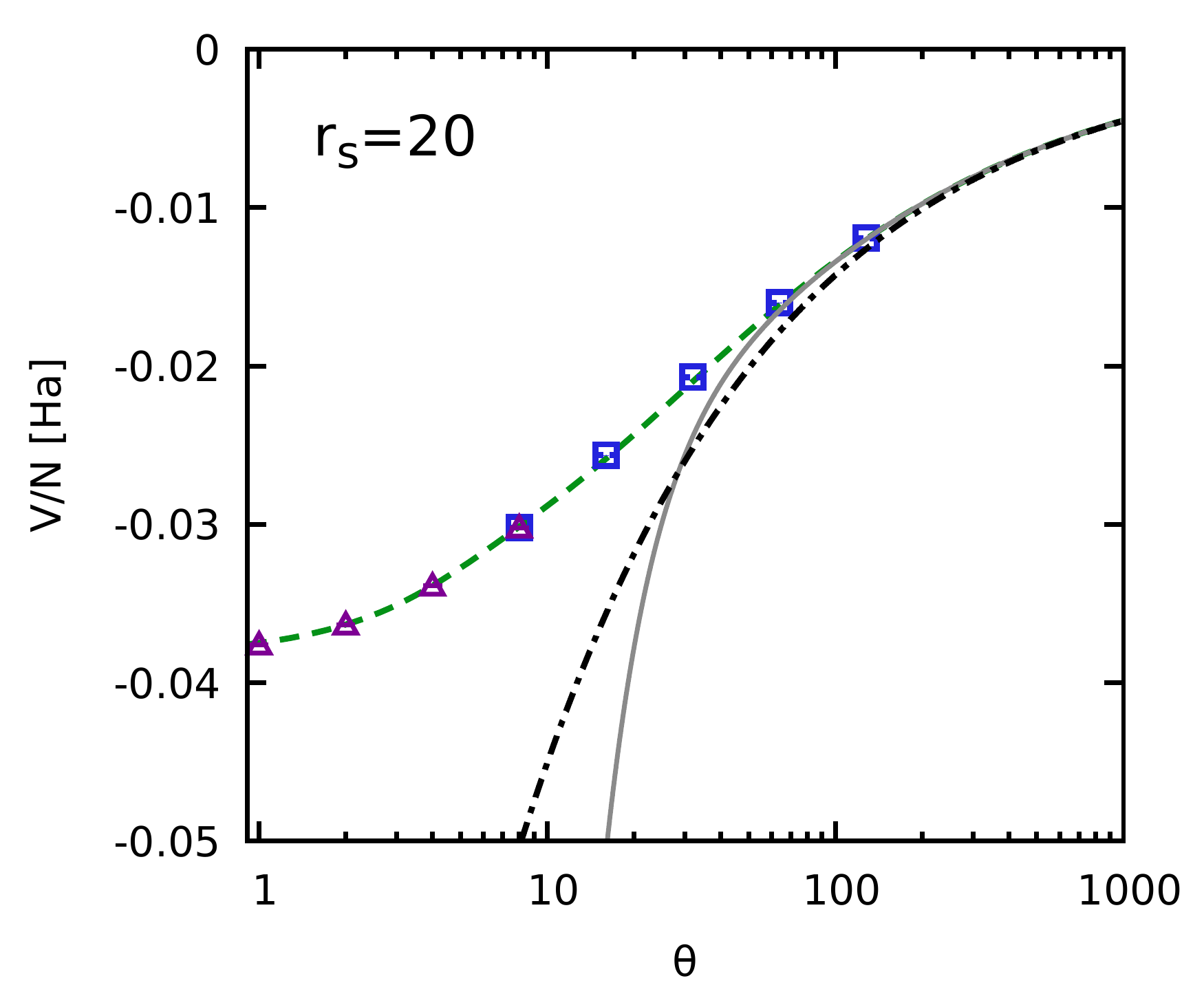}\includegraphics[width=0.55\textwidth]{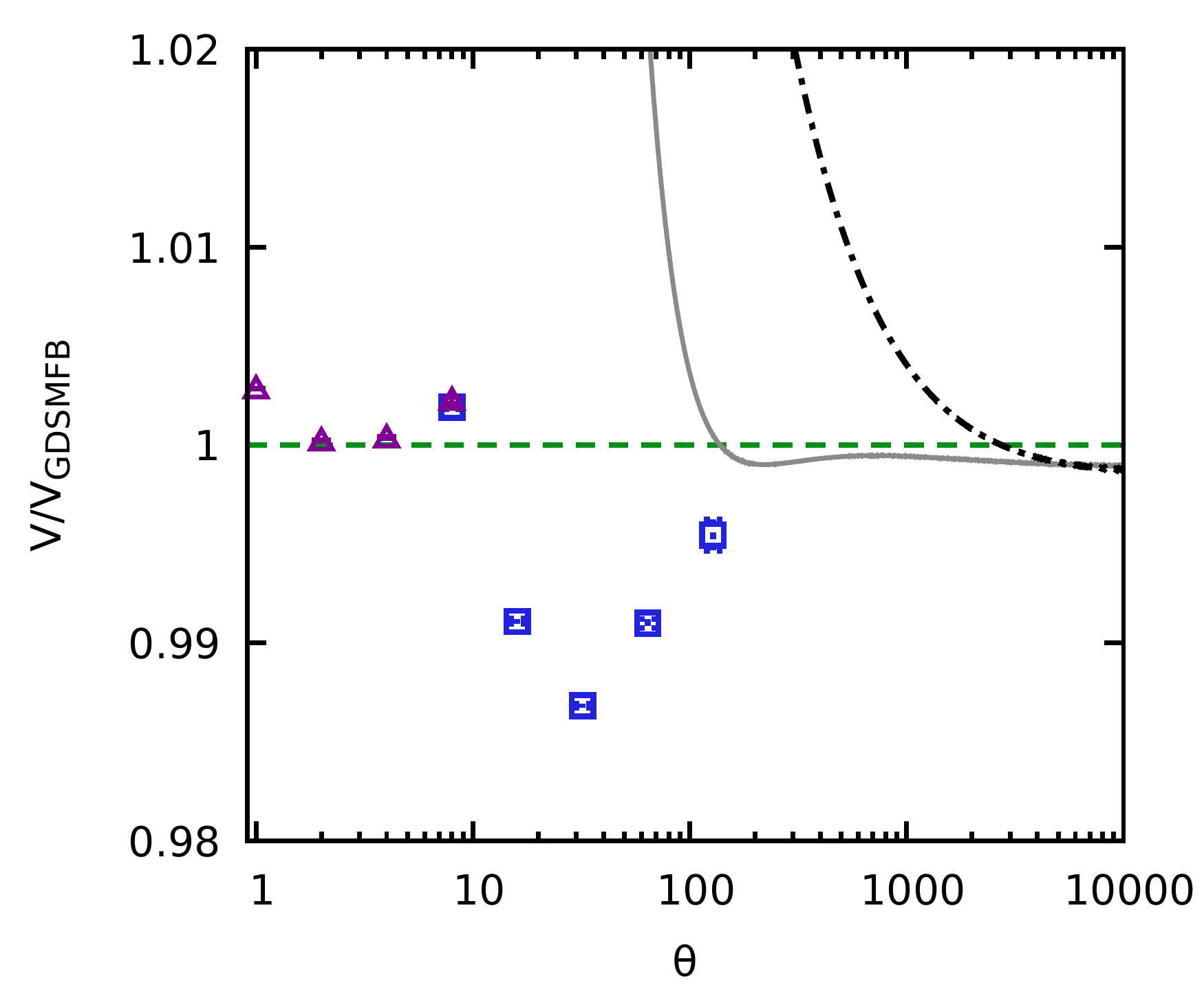}
\caption{\label{fig:V_theta}
Temperature dependence of the interaction energy per particle $V/N$ (left), and $v/v_\textnormal{GDSMFB}$ (right). Blue squares: extrapolated PIMC data from this work; purple triangles: extrapolated PB-PIMC and CPIMC data from Refs.~\cite{dornheim_prl,groth_prl}; dashed green: parametrization by Groth \emph{et al.}~\cite{groth_prl}; dash-dotted grey: Debye-H\"uckel (DH);  solid grey line: expansion including the full virial function Eq.~(\ref{full_virial}).
}
\end{figure}

Having obtained accurate results for the interaction energy of the UEG in the high-temperature regime allows us to gauge the accuracy of existing parametrizations and analytical theories. This is shown in Fig.~\ref{fig:V_theta} for $r_s=0.5$ (top row), $r_s=2$ (center row), and $r_s=20$ (bottom row). Specifically, the left column shows the $\theta$-dependence of the interaction energy per particle, with the blue squares depicting our new PIMC results, and the purple triangles showing previous CPIMC and PB-PIMC results from Refs.~\cite{dornheim_prl,groth_prl}. In addition, the dashed green curve has been obtained by evaluating Eq.~(\ref{eq:adiabatic}) using as input for $f_\textnormal{xc}$ the parametrization by Groth \emph{et al.}~\cite{groth_prl} (GDSMFB), which, by definition, attains the Debye-H\"uckel limit (dash-dotted grey) for high temperatures. Moreover, we have evaluated the full low-density expansion Eq.~(\ref{full_virial}).
First and foremost, we find that all data sets converge towards the same analytically known Debye-H\"uckel limit for large temperatures, as it is expected. In addition, our new PIMC results are fully consistent to earlier investigations~\cite{groth_prl,dornheim_prl}, and extend them towards higher temperature. Regarding the analytical curves, we find that the GDSMFB curve exhibits the by far best agreement to the PIMC data, as it interpolates between numerical results in the WDM regime and the exact high-$T$ asymptotic behaviour. The low-density/high-temperature expansion Eq.~(\ref{full_virial}) perfectly bridges the gap between Debye-Hückel and the first (few) PIMC data points. In this temperature range, it is much better then DH and about as good or slightly better than the GDSMFB-fit. For lower temperatures, Eq.~(\ref{full_virial}) then becomes increasingly inaccurate with decreasing $\theta$, that is, with increasing coupling strength $\Gamma$. 

A very detailed comparison is available in the right column of Fig.~\ref{fig:V_theta}, where we show the ratio of the different data sets to the interaction energy evaluated from GDSMFB. At $r_s=0.5$, the PIMC data are in good agreement to the latter over the entire $\theta$-range, and we find a maximum of $\Delta v/v\sim0.3\%$ at $\theta=8$. The full virial Eq.~(\ref{full_virial}) meets the first PIMC point and then starts deviating due to strong coupling. 
Let us next consider the metallic density $r_s=2$ shown in the center row, which exhibits a qualitatively similar behaviour. Here, too, the new PIMC data are in very good agreement to the GDSMFB parametrization, and we find a maximum deviation of $\Delta v/v\lesssim0.5\%$ around $\theta=16$. Remarkably, the curve of the full virial expansion~(\ref{full_virial}) exhibits an even better agreement to the PIMC data for $\theta\gtrsim20$, but becomes increasingly inaccurate for $\theta\lesssim10$. 
Finally, the bottom row shows results for $r_s=20$, which is most interesting for a number of reasons. Firstly, the Debye-H\"uckel limit only becomes accurate for a larger value of $\theta$ compared to $r_s=0.5$ and $r_s=2$, which means that existing PIMC-informed parametrizations of $f_\textnormal{xc}$~\cite{groth_prl,ksdt} interpolate over a larger effective temperature-range. Consequently, we find the largest discrepancies between the new PIMC results and GDSMFB in this case, with a maximum of $\Delta v/v\sim1.5\%$ around $\theta=32$. However, we can show here with the high-temperature expansion (\ref{full_virial}) that the GDSMFB parametrization meets the correct limiting law way before Debye-H\"uckel. At the same time, we note that both the interaction energy $v$ and the XC free energy $f_\textnormal{xc}$ are only a small fraction of the total (free) energy at such a high temperature~\cite{status}. Therefore, it is highly unlikely that the observed inaccuracies have any impact on practical applications in this regime. 
It is important to note that at such a relatively low density of $r_s=20$, strong coupling appears at larger $\theta\sim 10$ than for $r_s=2$ or $r_s=0.5$ and thus the full virial expansion is needed but will still fail for $\theta<100$ due to missing higher order virial coefficients.


\subsection{In depth comparison with virial expansion\label{sec:xi/6}}
\begin{table}[th]
\begin{center}
 \begin{tabular}{|c|c|c|c|c|c|}
\hline
$\theta$ & $T$ [K] &$-\xi$& $v_\textnormal{PIMC}$ [Ha] &   $v_{\rm vir}$ [Ha] & $w$ [Ha] \\
\hline
128&2.97742e8 & 0.0325664   &-0.0826214  &-0.082597     &-0.084745   \\
64&1.48871e8 &0.0460558       &-0.1180456 &-0.118221      &-0.122539 \\
32&7.44354e7 &0.0651327       &-0.1692720 &-0.169595      &-0.178304 \\
16&3.72177e7 &0.0921116     &-0.2423993 &-0.243352     &-0.260926   \\
8&1.86089e7 &0.130265     &-0.3447641   &-0.348638   &-0.383984     \\
\hline
 \end{tabular}
\caption{Mean potential energy per perticle of the UEG at $r_s=0.5$:  results from PIMC are compared to  calculations with the full second virial coefficient, Eq. (\ref{full_virial}), $v_{\rm vir}$. The calculation with the low-$\xi$ expansion (\ref{(ne2)2}) up to the order $n^2e^4$ plus the questionable term $\xi/6$ is denoted by $w$.}
\label{Tab:4}
\end{center}
\end{table}

We follow a method to extract the virial coefficients already used for the virial expansion of the electrical conductivity~\cite{R21}. In particular, we would like to discuss in more detail the influence of the direct $\xi/6$ term in the mean potential energy and its relation to the PIMC data. To determine the correct value of the linear term in $\xi$ from the PIMC simulations we should subtract the trivial terms $v_1=V_1/N$ so that
\begin{equation}
 \frac{V}{N\,k_BT}=\frac{V_1}{N\,k_BT}+\frac{V_2}{N\,k_BT}
\end{equation}
with
\begin{equation}
 \frac{V_1}{N\,k_BT}=-\frac{\kappa^3}{8 \pi n}+\pi n \lambda^3 \xi^3 \ln(\kappa \lambda).
\end{equation}
These terms (Debye-H\"uckel and logarithmic $\xi^3$-term) are exact results that are not debated.

For the remaining part, we are interested in the high-temperature limit of the second virial coefficient according to Eq. (\ref{(ne2)2}), where contributions of the order $n^2$ are considered. 
To identify the linear term from the PIMC simulations, we consider the reduced potential energy
\begin{equation}
v^{\rm red}=\frac{\Delta v}{\pi n \lambda^3 \xi\,k_BT} =\left[\frac{V}{N\,k_BT}-\frac{V_1}{N\,k_BT}\right] \frac{1}{\pi n \lambda^3 \xi}.
\label{vred}
\end{equation}
From the second virial coefficient (\ref{(ne2)2}), we expect the value $0.5$ in the limit $\xi \to 0$. The incorrect expression $B_2$, Eq. (\ref{wrong}), for the second virial coefficient containing the additional term $\xi/6$ would give the limit 5/6. 

In atomic units, using dimensionless parameters $r_s, \theta$, we have
\begin{eqnarray}
 \frac{v_1}{E_{\rm Ha}}&=&-\left(\frac{3}{2}\right)^{1/2} \left(\frac{4}{9 \pi}\right)^{1/3}\frac{1}{(r_s \theta)^{1/2}} \\
 &&-\frac{3}{2}\left(\frac{4}{9 \pi}\right)^{4/3} \frac{r_s}{\theta^2}\ln\left[12\left(\frac{4}{9 \pi}\right)^{4/3} \frac{r_s}{\theta^2}\right].\nonumber
 \end{eqnarray}
According Eq.~(\ref{(ne2)2}), up to the order $n^2e^4$ we have from the virial expansion the approximation for $v_2$
\begin{equation}
 \frac{v_{n^2e^4}}{E_{\rm Ha}}\approx -\frac{3}{2}\frac{1}{r_s^3}\left[\frac{1}{2}
\left(\frac{4}{9 \pi}\right)^{2/3} \frac{r_s^2}{\theta}-\left(\frac{\pi}{2}\right)^{1/2} (1+\ln 2)\frac{4}{9 \pi} \frac{r_s^3}{\theta^{3/2}}\right].
\end{equation}

\begin{figure}
\centering
\includegraphics[width=0.55\textwidth]{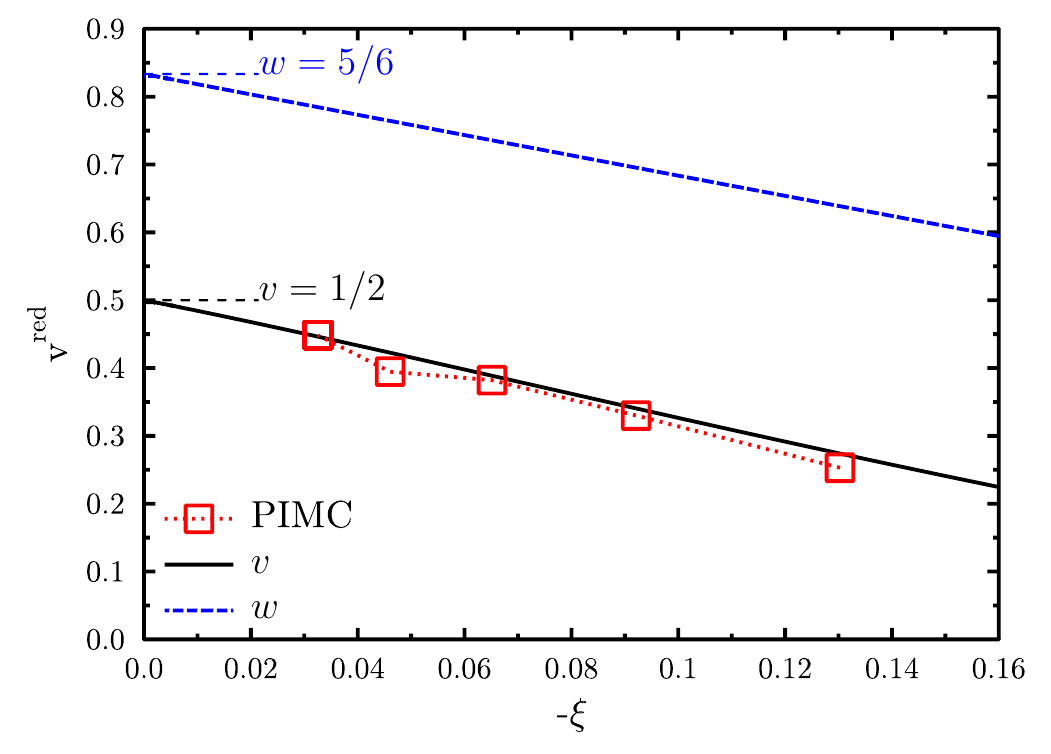}
\caption{\label{fig:V_xi}
Dependence of the reduced interaction energy per particle $v^\textnormal{red}$ (\ref{vred}) on the Born parameter $\xi$ for the density of $r_s=0.5$. Red squares: PIMC data from this work, solid black line labelled $v$: virial expansion Eq.~(\ref{full_virial}), dashed blue line: including the debated $\xi/6$ term (denoted by $w$).
}
\end{figure}

The values for 
the reduced potential energy at $r_s=0.5$ are plotted in Fig.~\ref{fig:V_xi} and can be easily obtained from Table \ref{Tab:4} by using Eq.~(\ref{vred}). The trend is a linear decrease and thus the expansion in terms of powers of the Born parameter $\xi$ seems reasonable in this parameter range. The values for $v^{\rm red}$ calculated with the numerical results $v_\textnormal{PIMC}$ of the PIMC simulations are in excellent agreement with the results of the virial expansions (\ref{full_virial}) \& (\ref{(ne2)2}), which are hardly distinguishable for the parameters of the plot (demonstrating the good convergence of the expansion). The test case containing the questionable term $\xi/6$ in the second virial coefficient $B_2$ (\ref{wrong}) shows however a limiting behavior not in agreement with PIMC data.

\subsection{Momentum distribution\label{sec:momentum}}

\begin{figure}\centering
\hspace*{-0.06\textwidth}\includegraphics[width=0.55\textwidth]{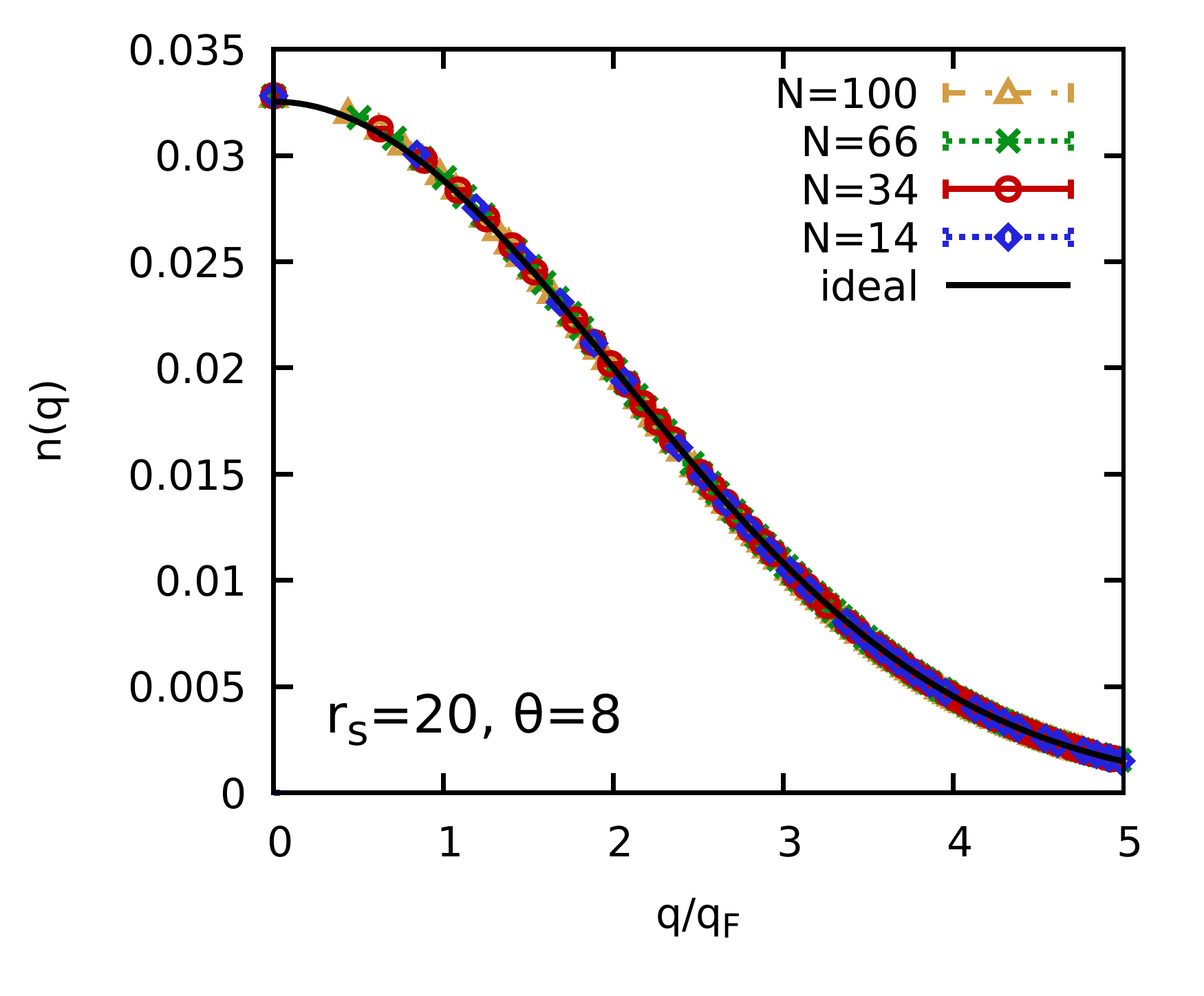}\includegraphics[width=0.55\textwidth]{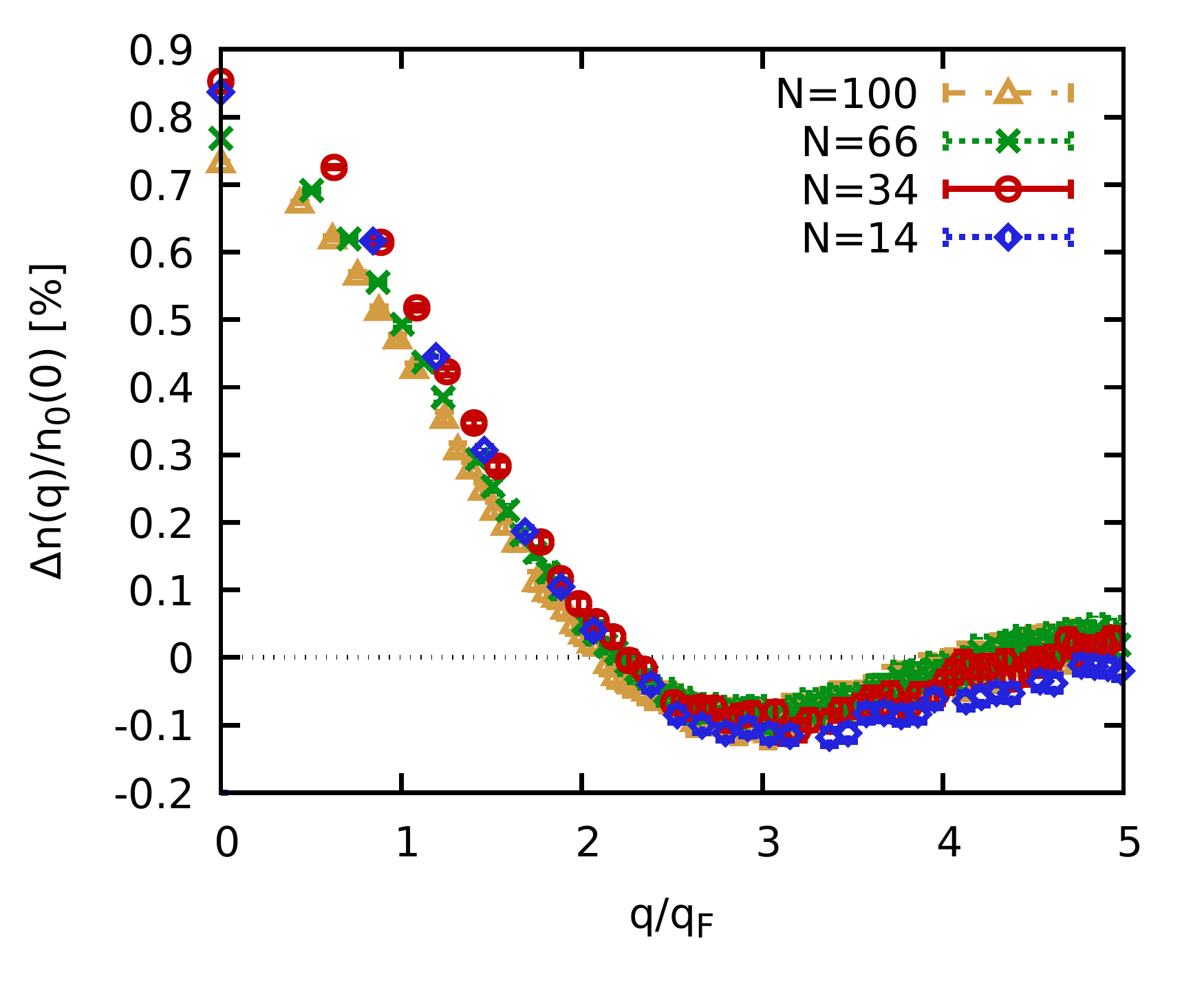}\\\vspace*{-1cm}\hspace*{-0.06\textwidth}\includegraphics[width=0.55\textwidth]{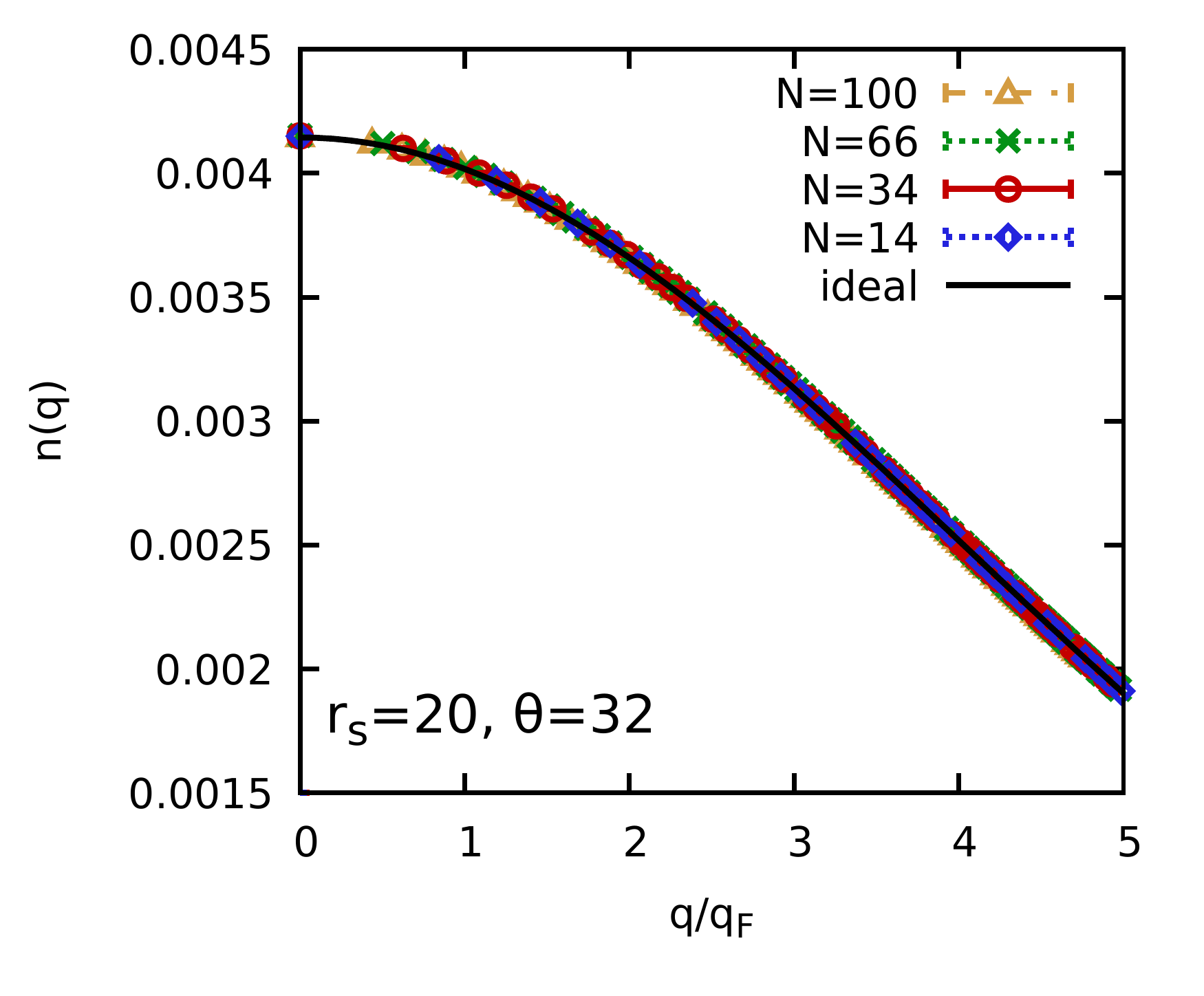}\includegraphics[width=0.55\textwidth]{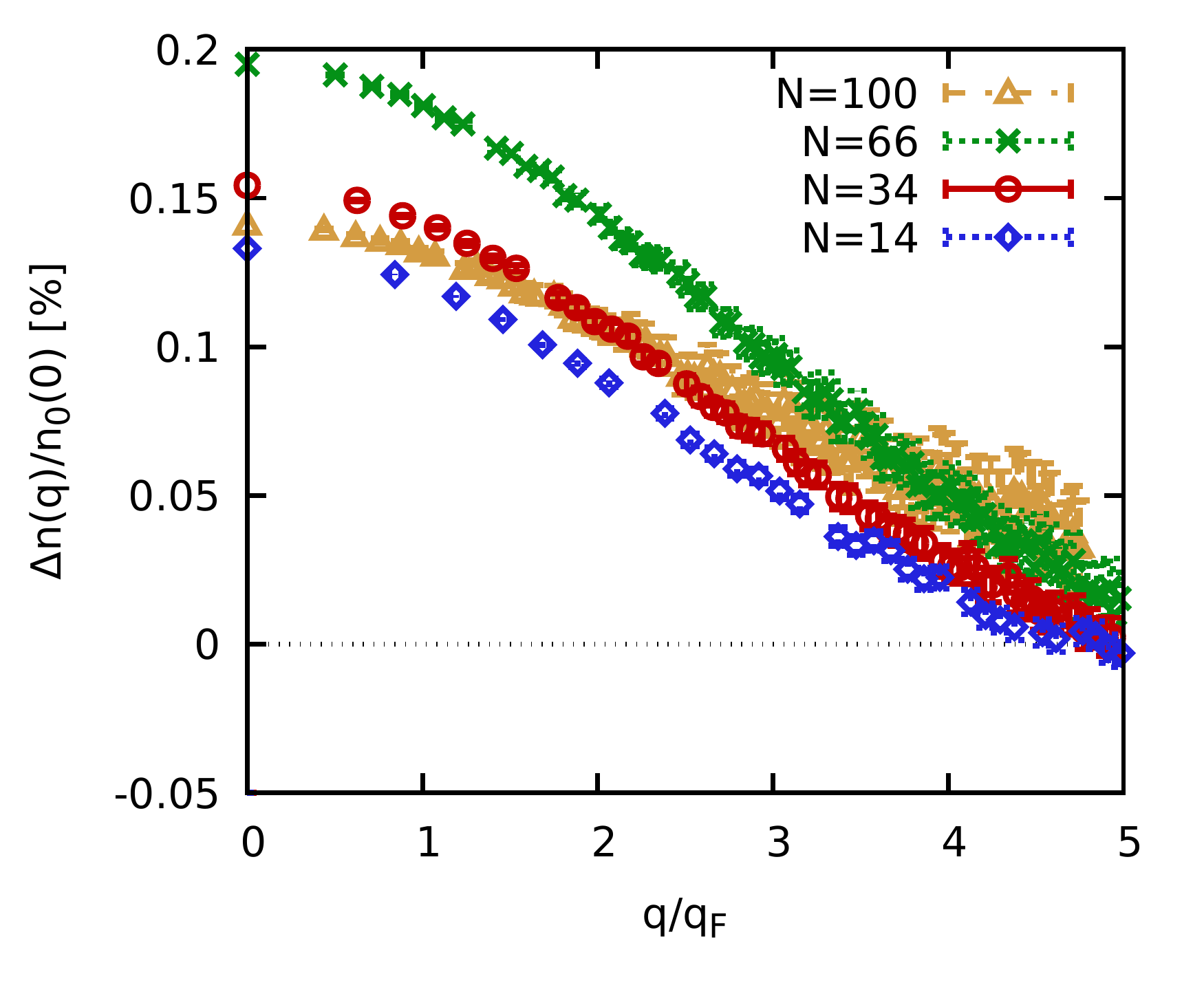}
\caption{\label{fig:nk_rs20}
Left: Momentum distribution $n(q)$ at $r_s=20$ and two temperatures for different numbers of electrons $N$. Solid black: ideal Fermi distribution, Eq.~(\ref{eq:Fermi});  Right: Relative difference to $n_0(q)$, Eq.~(\ref{eq:Delta_n0}).
}
\end{figure}

The final property of the UEG to be investigated in the present work is the momentum distribution $n(q)$, which we estimate in our PIMC simulations via the extended ensemble approach introduced in Ref.~\cite{Dornheim_PRB_nk_2021}. As a first step, we have to analyse the dependence on the system size $N$, which is shown in Fig.~\ref{fig:nk_rs20} at $r_s=20$. Specifically, the top and bottom rows correspond to $\theta=8$ and $\theta=32$, and the different symbols distinguish different values of $N$. The left panels depict $n(q)$ itself, and the solid black curves correspond to the ideal Fermi gas, where the momentum distribution is given by
\begin{eqnarray}\label{eq:Fermi}
n_0(\mathbf{q}) = \frac{1}{1+\textnormal{exp}\left(\beta(E_\mathbf{q}-\mu)\right)}\ ,
\end{eqnarray}
with $E_\mathbf{q}=\mathbf{q}^2/2$ and $\mu$ being the chemical potential. 
First and foremost, we note the excellent qualitative agreement of the PIMC data with Eq.~(\ref{eq:Fermi}) over the entire depicted $q$-range. In other words, the impact of electronic exchange--correlation effects is relatively small even at $r_s=20$ and $\theta=8$, which is the most strongly coupled case that have considered in this work.
In addition, no differences between the results for different $N$ can be seen with the naked eye in the plots of $n(q)$.
To resolve any dependence on $N$, we plot the relative deviation to $n_0(q)$,
\begin{eqnarray}\label{eq:Delta_n0}
\frac{\Delta n(q)}{n_0(0)} = \frac{n(q)-n_0(q)}{n_0(0)}\ ,
\end{eqnarray}
in the right column of Fig.~\ref{fig:nk_rs20}. On this scale, we detect small differences between the individual PIMC data sets that, however, do not exceed $0.1\%$ for both $\theta=8$ and $\theta=32$. In addition, we find a small yet significant increase in the occupation at small momenta. This counter-intuitive phenomenon is a direct consequence of the Coulomb coupling, and has been explored in detail in Refs.~\cite{Militzer_PRL_2002,Hunger_PRE_2021,Dornheim_PRB_nk_2021,Dornheim_PRE_2021}. For completeness, we also mention that it is closely associated to an interaction-induced lowering in the kinetic energy at some conditions.

\begin{figure}\centering
\hspace*{-0.06\textwidth}\includegraphics[width=0.55\textwidth]{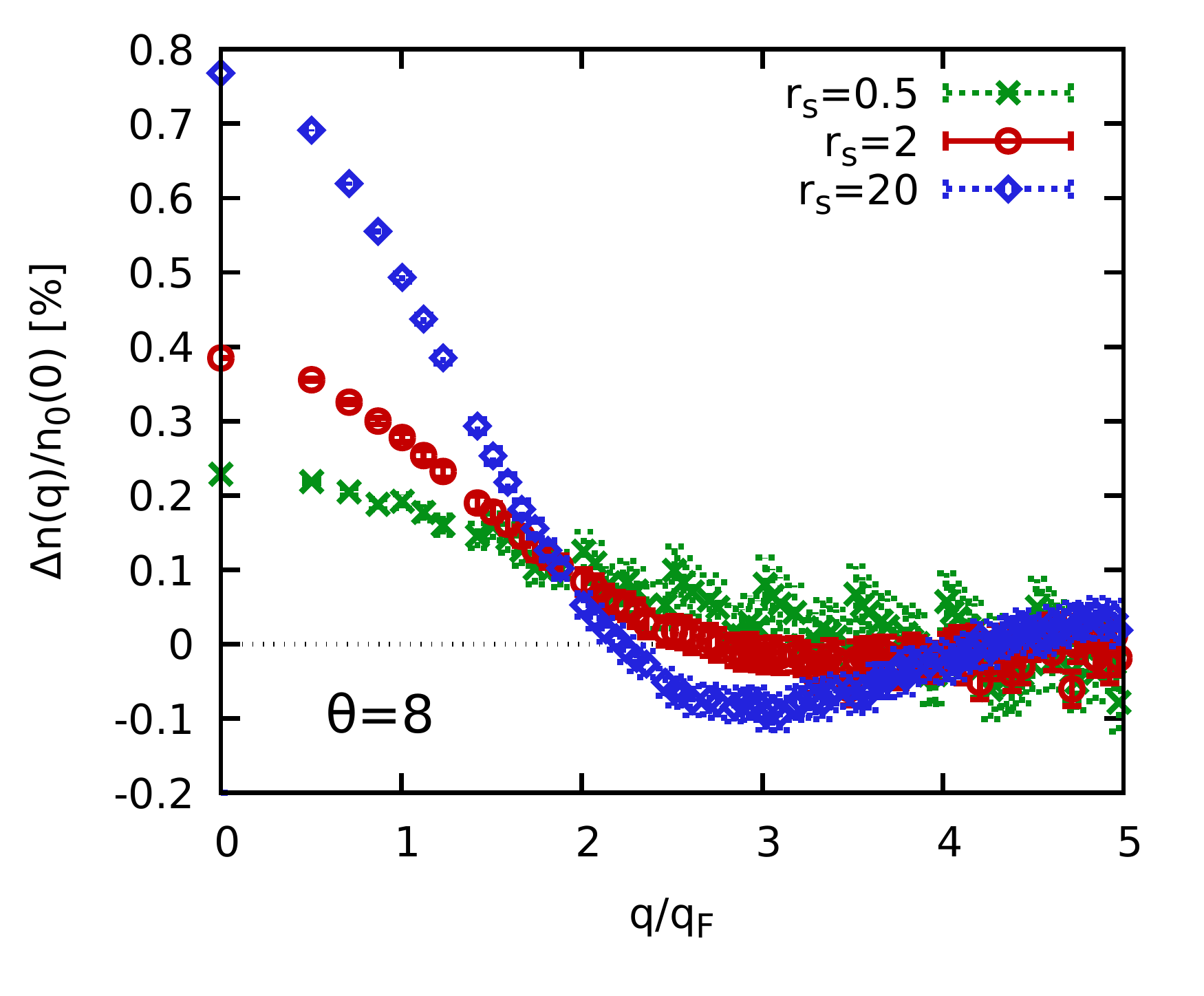}\includegraphics[width=0.55\textwidth]{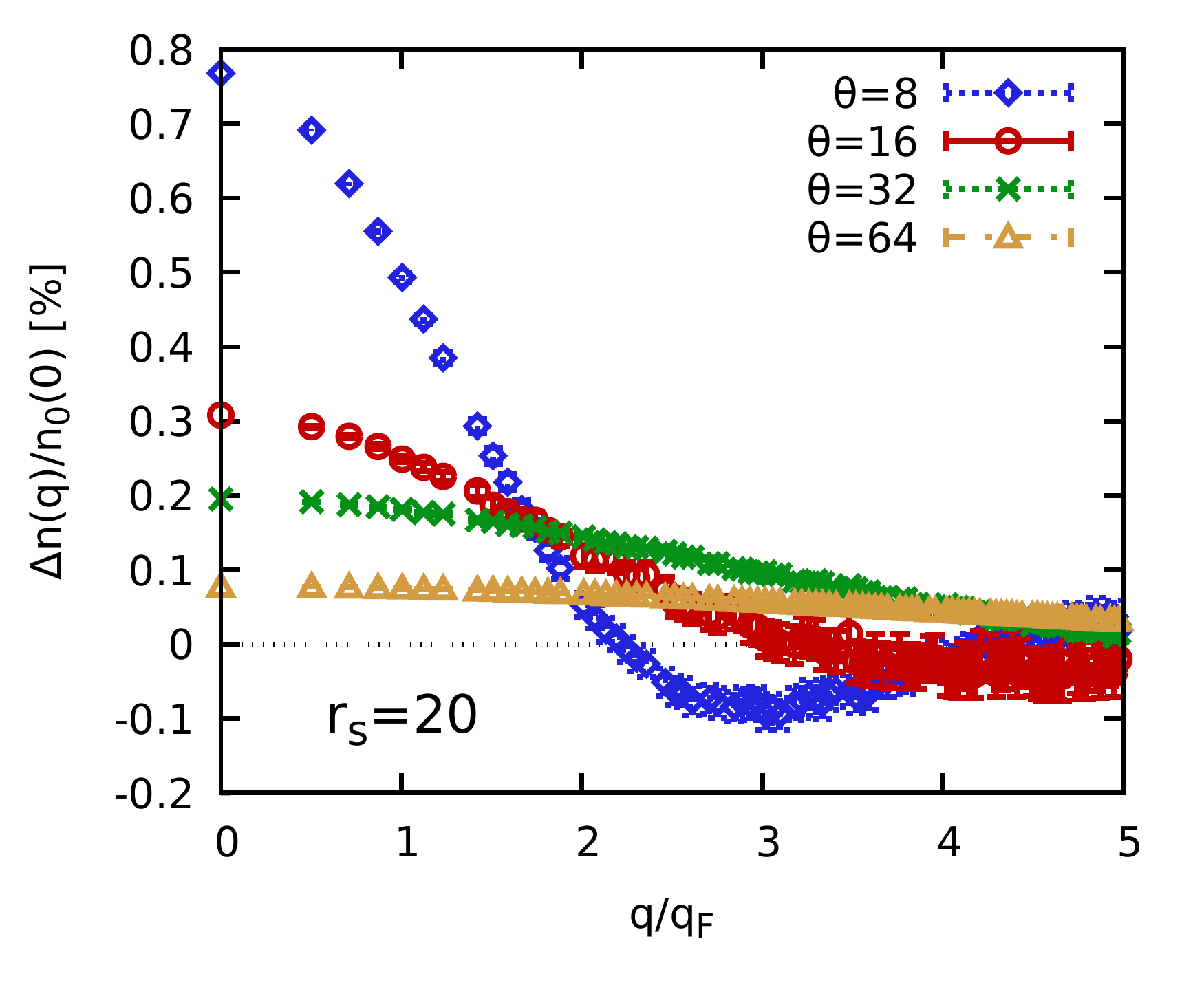}
\caption{\label{fig:nk_delta}
Left: Relative difference between our PIMC results and the ideal Fermi distribution at $\theta=8$ for $r_s=0.5$ (green crosses), $r_s=2$ (red circles), and $r_s=20$ (blue diamonds). Right: Same quantity at $r_s=20$ for $\theta=8$ (blue diamonds), $\theta=16$ (red circles), $\theta=32$ (green crosses), and $\theta=64$ (yellow triangles).
}
\end{figure}

Let us next investigate the dependence of interaction effects on the density parameter $r_s$, which is shown in the left panel of Fig.~\ref{fig:nk_delta} for $\theta=8$. In particular, the green crosses, red circles, and blue diamonds show the evaluation of Eq.~(\ref{eq:Delta_n0}) for $r_s=0.5$, $r_s=2$, and $r_s=20$, respectively. Interestingly, we find the interaction induced increase for all three cases, whereas the negative minimum around $q\sim3q_\textnormal{F}$
only appears for $r_s=20$.

The right panel of the same figure shows results for the $\theta$-dependence at $r_s=20$, and the blue diamonds, red circles, green crosses, and yellow triangles correspond to $\theta=8$, $\theta=16$, $\theta=32$, and $\theta=64$. First and foremost, we find that the overall magnitude of $\Delta n/n_0$ monotonically decreases with $\theta$, as the system becomes increasingly noninteracting, cf.~Fig.~\ref{fig:over}. For $\theta=64$, we find deviations not exceeding $0.1\%$, which is on the same level as any finite-size effects that might be present in the PIMC data.

\section{Summary and Outlook\label{sec:summary}}

In summary, we have presented extensive new \emph{ab initio} PIMC simulations of the UEG in the high-temperature regime, i.e., for $8\leq\theta\leq128$. Our main findings can be summarised as follows. a) We have closely examined the classical relation [Eq.~(\ref{eq:G_classical})] between the static structure factor $S(q)$ and the static LFC $G(q)$, and its validity strongly depends on the density parameter. For $r_s=20$, the system becomes first non-degenerate and only subsequently noninteracting when the temperature is increased. Therefore, Eq.~(\ref{eq:G_classical}) becomes accurate, and the static LFC can be directly estimated from $S(q)$. In stark contrast, the UEG becomes first noninteracting and then non-degenerate with increasing $\theta$ for smaller $r_s$. In this case, $G(q)$ has negligible impact on $S(q)$, and Eq.~(\ref{eq:G_classical}) remains inaccurate for any temperature. b) Our new results for the interaction energy per particle have allowed us to gauge the accuracy of previous parametrizations and analytical models. In particular, the parametrization of $f_\textnormal{xc}$ by Groth \emph{et al.}~\cite{groth_prl} interpolates between PIMC data for $\theta\leq8$ and the Debye-H\"uckel limit for large $T$. This is appropriate for $r_s=0.5$ and $r_s=2$, but leads to a maximum deviation of $\Delta v/v\sim1.5\%$ at $r_s=20$ around $\theta=32$. Nevertheless, we note that this is most likely of no practical relevance, as both $v$ and $f_\textnormal{xc}$ only constitute a small fraction of the total (free) energy in this regime~\cite{status}. Moreover, we have found that the high temperature expansion of the equation of state is in good agreement to our new PIMC data for $r_s=0.5$ and $r_s=2$, whereas it only becomes accurate for $r_s=20$ around $\theta\gtrsim10^2$. We confirm the non-existence of the suggested $\xi/6$-term~\cite{WDK2005,Kraeft_2015}.
c) We have analysed the momentum distribution function $n(q)$, and have detected an interaction-induced increase in the zero-momentum state even for $\theta\gtrsim32$ that is consistent to previous investigations at lower temperature~\cite{Militzer_PRL_2002,Militzer_momentum_HEDP_2019,Hunger_PRE_2021,Dornheim_PRB_nk_2021,Dornheim_PRE_2021}.

Our investigation closes an important gap in the understanding of the UEG---the archetypical system of interacting electrons---which is highly important in its own right~\cite{review,status}.
PIMC simulations and virial expansions are also of fundamental interest for more complex systems like two-component plasmas, and may be used to improve density functional calculations of, e.g., the electrical conductivity where the contribution of electron-electron collisions is not included~\cite{R21}. One may also strife to improve parametrizations like GDSMFB by taking into account not just the Debye-H\"uckel law but higher order terms as well. Similarly, the high-density/low-temperature limit may be checked. In addition, all our PIMC results are freely available online~\cite{repo} and can be used as input for improved parametrizations, and to benchmark semi-classical simulation techniques~\cite{sem}.

\section*{Acknowledgments}

   We gratefully acknowledge computing time on a Bull Cluster at the Center for Information Services and High Performance Computing (ZIH) at Technische Universit\"at Dresden, and at the Norddeutscher Verbung f\"ur Hoch- und H\"ochstleistungsrechnen (HLRN) under grant \emph{shp00026}.
   
  This work was partially funded by the Center for Advanced Systems Understanding (CASUS) which is financed by Germany’s Federal Ministry of Education and Research (BMBF) and by the Saxon Ministry for Science, Culture and Tourism (SMWK) with tax funds on the basis of the budget approved by the Saxon State Parliament.

\appendix
\section{Choice of the short range potential}\label{apa}
In this section, we consider the two component version. This follows the argument in Ref.~\cite{Kraeft_2015}. The free energy reads
\begin{eqnarray}\label{free}
  F-F_{id}=&&\frac{V}{2}\sum_{ab}n_an_b\int \int_0^1\frac{d\lambda}{\lambda}
  [\lambda U_{ab}(r)F_{ab}(r)]dr\,,\nonumber\\
     &&U_{ab}(r)=V_{ab}(r)+{\cal V}_{ab}(r)\,.
   \end{eqnarray} 
Here,  $V$ is the Coulomb potential, ${\cal V}$ is a short range potential having parameter $\lambda$.

The pair distribution is given by
  \begin{eqnarray}\label{distrf}
 &&F_{ab}(r)=\exp(g_{ab}(r)-\beta V_{ab}'(r))\{1+\cdots\}\nonumber\\
&&=\exp(-\beta V_{ab}'(r))\{1+g_{ab}(r)+\frac{1}{2}g_{ab}^2(r)+\cdots\}\,.
\nonumber\\
  \end{eqnarray}
where  $V'$ short range potential, the different notation is for pedagogical reasons. 
$g_{ab}(r)=\frac{e_ae_b}{k_B T r}\exp(-\kappa r)$ is the Debye distribution function, 
$\xi_{ab}=-\frac{e_ae_b}{k_BT \lambda_{ab}}$ is the two-component Born parameter, with the definitions $\lambda_{ab}=\sqrt{2\pi\hbar^2\beta / m_{ab} }$ and $m_{ab}=m_am_b/(m_a+m_b)$.
A decomposition of the free energy gives
\begin{eqnarray}\label{decomp}
 && F- F_{id}=\frac{V}{2}\sum_{ab}n_an_b \int_0^1\frac{d \lambda}{\lambda}
 \{
 \lambda V_{ab}(r)(1+g_{ab}(r))\nonumber\\
 && +\lambda V_{ab}(1+g_{ab})(\exp(-\beta V_{ab}')-1)\nonumber\\
 &&+\lambda V_{ab}[\exp(g_{ab})-1-g_{ab}]\exp(-\beta V_{ab}')\nonumber\\
 &&+\lambda {\cal V}_{ab}F_{ab}\}d{\bf r}\,.
\end{eqnarray}
The first line represents the limiting law, i.e., the $n^{3/2}$ term.
We neglect higher orders. The Coulomb potential is represented according to 
G.Schmitz \cite{schmitz1111} by $V_{ab}=-k_BT g_{ab} 
+\cdots$. The third line of Eq.~(\ref{decomp}) reduces to
\begin{equation}\label{thirdline}
 I_3=\lambda k_B T g_{ab} [F_{ab}-1-g_{ab}]\,.
 \end{equation}
We define the derivative with respect to $\lambda$
 \begin{equation}\label{deriv}
  H'=-k_B T\frac{\partial}{\partial \lambda}\left\{\exp(g_{ab}-\beta 
V_{ab}')-1-g_{ab}-\frac{1}{2}g_{ab}^2\right\}\,.
 \end{equation}
 The result is
 \begin{equation}\label{deriv1}
  H'=-k_B T\left(\frac{\partial g_{ab}}{\partial \lambda}-\beta \frac{\partial 
V_{ab}'}{\partial \lambda}\right)F_{ab}+k_B T\frac{\partial g_{ab}}{\partial 
\lambda}(1+g_{ab})\,.
 \end{equation}
We now  rewrite
 \begin{equation}
  H'=-k_BT\frac{\partial g_{ab}}{\partial 
\lambda}[F_{ab}-1-g_{ab}]+\frac{\partial V_{ab}'}{\partial \lambda}F_{ab}\,.
 \end{equation}
 To avoid contradictions, one now has to make the identification
 ${\cal V}_{ab}(r)=V_{ab}'(r)$.
 Under this condition, we get the formula given in \cite{EHK1967}
 \begin{eqnarray}\label{HoEbKe}
  &&F-F_{id}=-\frac{k_BT V}{12 \pi}\kappa^3-\frac{k_BT 
V}{2}\sum_{ab}n_an_b\nonumber\\
&&\times \int d{\bf r}\{\exp[g_{ab}(r)-\beta 
V_{ab}^{short}(r)]-1-g_{ab}-\frac{1}{2}g_{ab}^2\}.\nonumber\\
 \end{eqnarray}
 This equation leads to the additional term in question, $\xi/6$. The latter 
does not show up iff ${\cal V}=0$, in Eq.~(\ref{free}). The mean value of the Kelbg potential \cite{Kelbg63} delivers the term $\xi/6$, the mean value of the Coulomb potential does 
not.

\section*{References}

\bibliographystyle{unsrt}
\bibliography{bibliography}

\begin{thebibliography}{100}
\bibitem{repo}
A link to a repository containing all PIMC raw data will be made available upon
  publication.
\end{thebibliography}

\end{document}